\documentclass[a4paper, 11pt]{article}
\pdfoutput=1

\usepackage{jheppubnew}
\usepackage{graphicx}
\usepackage{bm}
\usepackage{caption, subcaption}

\renewcommand{\textbullet}{\begin{picture}(1,1)(0,-1.6)\rule{1.6mm}{1.6mm}\end{picture}\,\,\,}

\newcommand{\be}{\begin{equation}}
\newcommand{\ee}{\end{equation}}
\renewcommand{\[}{\begin{equation}}
\renewcommand{\]}{\end{equation}}
\newcommand{\tto}{\rightarrow}

\newcommand{\D}{\mathrm{d}}

\newcommand{\Z}{\mathbb{Z}}

\renewcommand{\O}{\mathcal{O}}

\newcommand{\ep}{\epsilon}

\newcommand{\<}{\langle}
\renewcommand{\>}{\rangle}

\newcommand{\nn}{\nonumber}
\newcommand{\lla}{\langle \! \langle}
\newcommand{\rra}{\rangle \! \rangle}

\newcommand{\lwick}{:\!}
\newcommand{\rwick}{\!:}

\DeclareMathOperator{\re}{Re}

\DeclareMathOperator{\Li}{Li}

\newcommand{\bs}[1]{\boldsymbol{#1}}
\DeclareMathOperator{\K}{K}

\title{Scalar 3-point functions in CFT:\\
 renormalisation, beta functions and anomalies}
\author[a]{Adam Bzowski,}
\author[b]{Paul McFadden}
\author[c]{and Kostas Skenderis}
\affiliation[a]{Institute for Theoretical Physics, K.U.~Leuven, Belgium.}
\affiliation[b]{Theoretical Physics Group, Blackett Laboratory, Imperial College, London, U.K.}
\affiliation[c]{STAG Research Centre and Mathematical Sciences, University of Southampton, U.K.}
\emailAdd{adam.bzowski@fys.kuleuven.be}
\emailAdd{p.mcfadden@imperial.ac.uk}
\emailAdd{k.skenderis@soton.ac.uk}

\begin{document}

\abstract{
We present a comprehensive discussion of renormalisation of 3-point functions of  scalar operators in conformal field theories in general dimension. We have previously shown that conformal symmetry uniquely determines the momentum-space 3-point functions in terms of certain integrals involving a product of three Bessel functions (triple-$K$ integrals). The triple-$K$ integrals diverge when the dimensions of operators satisfy certain relations and we discuss how to obtain renormalised 3-point functions in all cases. There are three different types of divergences: ultralocal, semilocal and nonlocal, and a given divergent triple-$K$ integral may have any combination of them.
Ultralocal divergences may be removed using local counterterms and this results in new conformal anomalies.  Semilocal divergences may be removed by renormalising the sources, and this results in CFT correlators that satisfy Callan-Symanzik equations with beta functions.  In the case of non-local divergences, it is the triple-$K$ representation that is singular, not the 3-point function. Here, the CFT correlator is the coefficient of the leading nonlocal singularity, which satisfies all the expected conformal Ward identities.  Such correlators exhibit enhanced symmetry: they are also invariant under dual conformal transformations where the momenta play the role of coordinates. When both anomalies and beta functions are present the correlators exhibit novel analytic structure containing products of logarithms of momenta. We illustrate our discussion with numerous examples, including free field realisations and AdS/CFT computations.
}

\maketitle

\section{Introduction}

Conformal invariance and its implications for correlation functions is a well-studied subject \cite{DiFrancesco:1997nk}. 
Already from the first works on this topic it was clear that 2- and 3-point functions are fixed by conformal invariance up to constants. For example, the 2-point and 3-point functions of scalar operators are given by \cite{Polyakov:1970xd}
\begin{align}
\< \O(\bs{x}) \O(0) \> &= \frac{C_{\O}}{|\bs{x}|^{2 \Delta}},  \label{2pt} \\
\langle \mathcal{O}_1(\bs{x}_1) \mathcal{O}_2(\bs{x}_2) \mathcal{O}_3(\bs{x}_3) \rangle &= \frac{C_{123}}{|\bs{x}_1 - \bs{x}_2|^{\Delta_1 + \Delta_2 - \Delta_3} |\bs{x}_2 - \bs{x}_3|^{\Delta_2 + \Delta_3 - \Delta_1} |\bs{x}_3 - \bs{x}_1|^{\Delta_3 + \Delta_1 - \Delta_2}}, \label{3pt}
\end{align}
where $\Delta$, $\Delta_1$, $\Delta_2$ and $\Delta_3$ are the conformal dimensions of the operators $\O$, $\O_1$, $\O_2$ and $\O_3$ respectively, and 
$C_\O$ and $C_{123}$ are constants.  These results were obtained using position-space techniques and hold when the operators are at separated points. 

Correlation functions should be well-defined distributions, {\it i.e.}, they should have a Fourier transform.   It is well known 
that when the dimension of the operator is 
\begin{equation} \label{2pt_cond}
\Delta = \frac{d}{2} + k, \qquad k =0, 1, 2, \ldots
\end{equation}
the 2-point function (\ref{2pt}) does not have a Fourier transform because of short-distance singularities. One needs to regularise and renormalise the correlator and this gives rise to new conformal anomalies \cite{Osborn:1991gm,Osborn:1993cr,Petkou:1999fv}.\footnote{2-point functions of tensorial operators ({\it e.g.}, the stress tensor)
also have conformal anomalies and it is in this context that conformal anomalies were first discovered \cite{Deser:1976yx}.}
The renormalised correlators then satisfy anomalous conformal Ward identities. The purpose of this paper is to 
present a renormalised version of the 3-point correlators (\ref{3pt}). In particular, we would like to understand the analogue of the condition (\ref{2pt_cond}), the possible new conformal anomalies that arise, and their structure.

In \cite{Bzowski:2013sza} we initiated a study of conformal field theory in momentum space.\footnote{The initial motivation for this work was the need 
for momentum-space CFT correlators in the context of  holographic cosmology  \cite{McFadden:2009fg,McFadden:2010na,McFadden:2010vh,Bzowski:2011ab,
Bzowski:2012ih}; similar applications of conformal/de Sitter symmetry in cosmology have
been discussed in \cite{Antoniadis:2011ib, Maldacena:2011nz, Creminelli:2012ed,
Kehagias:2012pd,Kehagias:2012td,Coriano:2012hd,Mata:2012bx,Ghosh:2014kba,Arkani-Hamed:2015bza}. Other recent works
that contain explicit computations of CFT correlation functions in momentum
space include \cite{Giannotti:2008cv, Armillis:2009sm, Armillis:2009pq,
Coriano:2012wp,Chowdhury:2012km, Coriano:2013jba, Bonora:2015nqa}.} In particular, we started a systematic analysis of the implications of the conformal Ward identities  and we presented a complete solution of the conformal Ward identities for scalar and tensor 3-point functions. Here we will present a comprehensive discussion of regularisation/renormalisation for scalar 3-point functions. The corresponding discussion for tensorial 3-point function will be discussed in a sequel \cite{tensor}. 

The organisation of this paper, and an overview of our 
plan of attack, is as follows.
We start in section \ref{sec:CWI} with the conformal Ward identities in position space, and derive their corresponding form in momentum space.    
Rather than attempting to construct a well-defined Fourier transform for the correlators \eqref{2pt} and \eqref{3pt} (which, while straightforward for 2-point functions, is very challenging for 3-point functions \cite{Bzowski:2012ih}), we will instead simply solve the conformal Ward identities directly in momentum space.
As preparation for our analysis of 3-point functions, in section \ref{sec:2pt} we first solve the momentum-space Ward identities for 2-point functions, reviewing their renormalisation and the anomalies that arise in cases where the  condition \eqref{2pt_cond} is satisfied.

Our main analysis of CFT 3-point functions then follows in section \ref{sec:3pt}.  In 
section \ref{subsec:Ward1}, we convert the conformal Ward identities from their original tensorial form to a purely scalar form.  The solution for 3-point functions can then be written as an integral of three Bessel-$K$ functions:
\begin{equation} \label{tripleKfirstglimpse}
\lla \O_1(\bs{p}_1) \O_2(\bs{p}_2) \O_3(\bs{p}_3) \rra \propto \int_0^\infty \D x \: x^{d/2-1} \prod_{j=1}^3 p_j^{\Delta_j-d/2} K_{\Delta_j-d/2}(p_j x).
\end{equation}
This is the triple-$K$ integral, and we review its derivation in 
section \ref{subsec:gen_soln}.  (Our double-bracket notation for momentum-space correlators  simply indicates the removal of the overall momentum-conserving delta function.) 
For generic values of the operator dimensions this triple-$K$ integral is well defined, either directly through convergence of the integral or else indirectly through analytic continuation, leading to a correspondingly well-defined 3-point function in momentum space.    
As we will show, however, there are certain special values of the operator dimensions for which the triple-$K$ integral is singular.  In these cases  regularisation and renormalisation are required.  The condition identifying these special values is:
\begin{equation} \label{3pt:cond}
\frac{d}{2} \pm (\Delta_1 - \frac{d}{2}) \pm (\Delta_2 - \frac{d}{2})\pm (\Delta_3 - \frac{d}{2})=-2 k. 
\end{equation}
Here, $d$ is the spacetime dimension (though we work throughout in Euclidean signature for simplicity) and $k$ is any non-negative integer ({\it i.e.}, $k=0,1,2,\ldots$).  Any independent choice of the $\pm$ signs can be made for each of the terms in this expression, and a different value of $k$ is permitted for each choice.  

The remainder of section \ref{sec:3pt} then presents our renormalisation procedure.
First, we discuss the different types of singularities that can arise in the triple-$K$ integral: 
these correspond to the different choices of signs for which the singularity condition \eqref{3pt:cond} can be satisfied.
The different types of singularity are not mutually exclusive and can arise either separately or in various combinations.  Each type of singularity is linked to the existence of a particular type of counterterm that can be added to the CFT action: the nature of these counterterms then reveals how to deal with each of the different types of singularity.

In general, the singularities may be either ultralocal, semilocal or nonlocal, by which we mean that the corresponding position-space expressions have support either only when all three insertion points coincide (ultralocal), only when two insertions coincide (semilocal), or else without any insertions coinciding (nonlocal).  In momentum space, ultralocal singularities correspond to expressions that are purely analytic in the squared momenta ({\it i.e.,} $p_1^2$, $p_2^2$ and $p_3^2$, where each $p_i^2 = \bs{p}_i\cdot\bs{p}_i$), while semilocal singularities are constructed from terms each of which is non-analytic in only a single squared momentum.  Nonlocal singularities, on the other hand, are constructed from terms that are individually non-analytic in two or more squared momenta.
For the triple-$K$ integral to contain such nonlocal singularities, the singularity condition \eqref{3pt:cond} must admit at least one solution with either two or three plus signs.  
If nonlocal singularities are absent but the triple-$K$ integral has semilocal singularities, the singularity condition \eqref{3pt:cond} admits a solution with two minus signs and one plus sign.   If instead only ultralocal singularities are present 
(as was the case for 2-point functions when \eqref{2pt_cond} was satisfied), 
the singularity condition \eqref{3pt:cond} can only be satisfied with three minus signs.

In section \ref{subsec:ren}, we show that ultralocal singularities in the triple-$K$ integral can be removed through the addition of local counterterms constructed from the sources.  The corresponding renormalised 3-point functions then contain single logarithms of momentum divided by the renormalisation scale $\mu$.  This explicit $\mu$-dependence signals the presence of a conformal anomaly.  Interestingly, this anomaly can arise in both odd- and even-dimensional spaces, unlike the more familiar trace anomaly that appears when we put the CFT on a background metric.
Semilocal singularities of the triple-$K$ integral can be removed by a renormalisation of the sources for the scalar operators.  In this case we find a surprising new result: that the corresponding renormalised 3-point correlators contain {\it double} logarithms of momenta.\footnote{Double logs were also observed earlier in \cite{PerezVictoria:2001pa} in the context of AdS/CFT computations. We thank Manuel Perez-Victoria for bringing this paper to our attention.}  These renormalised correlators obey Callan-Symanzik equations with non-trivial beta function terms. There is no contradiction with the theory being a CFT, however, as these beta functions are for sources that couple to composite operators, rather than to operators appearing in the fundamental Lagrangian of the theory.
Finally, nonlocal singularities of the triple-$K$ integral cannot be removed by local counterms: instead it is the triple-$K$ representation that is singular.
In such cases the renormalised correlator is simply given by the leading nonlocal singularity of the triple-$K$ integral, which as we will show directly satisfies the appropriate conformal Ward identities.

Section \ref{subsubsec:reg} discusses our regularisation procedure for the divergent triple-$K$ integral: this is most easily accomplished by infinitesimally shifting the dimensions of operators and of the spacetime itself.    These shifts give rise to corresponding shifts in the indices of the Bessel-$K$ functions that appear in the triple-$K$ integral, as well as in the power of the integration variable.  The advantage of this regularisation scheme is that the regulated triple-$K$ integral preserves conformal invariance, and satisfies a set of regulated conformal Ward identities.  It is also straightforward to extract the divergences of the regulated triple-$K$ integral as the regulator is removed.  
As we will show, the divergences can be read off from a simple series expansion of the integrand about the origin.

In section \ref{subsubsec:change_scheme} we discuss the residual freedom in the regularisation scheme, corresponding to the precise manner in which the operator and spacetime dimensions are shifted. It is straightforward to convert between the different choices of scheme, and we discuss the procedure for doing this.
As the regulated triple-$K$ integrals satisfy regulated Ward identities, by expanding in powers of the regulator one can identify the Ward identities satisfied by the individual divergent terms in the regulated triple-$K$ integral.  These Ward identities contain anomalous terms as we show in section \ref{subsubsec:Ward2}, although we defer a full analysis until section \ref{sec:anomalies_and_betafns}.

In section \ref{subsubsec:1/ep} we illustrate in detail our renormalisation procedure for all cases in which the triple-$K$ integral has only a single pole in the regulator, and present a number of explicit examples.  This case is the simplest that can arise; cases where the regulated triple-$K$ integral contains higher-order singularities are discussed in section \ref{subsubsec:higherorder}, which again presents a number of worked examples, postponing a complete analysis to appendix \ref{sec:gen_results}.

In certain cases the correlation functions we consider can be realised in perturbative conformal field theories or in 
free field theories such as massless scalars or fermions. When this happens, the correlators can be calculated using perturbation theory by means of (typically multi-loop and heavily divergent) Feynman diagrams. The standard renormalisation procedure for Feynman diagrams then proceeds loop by loop, where nested divergences are removed at every step leading to a sequence of momentum integrals, each exhibiting only ultralocal divergences. This renormalisation procedure differs only in execution from our more general procedure, which is valid for any CFT (perturbative or non-perturbative), but is otherwise completely equivalent. In both cases the divergences are removed by the addition of counterterms to the action, and these counterterms have identical form (modulo scheme dependence).  Any possible difference in the final renormalised correlation functions can therefore be removed by introducing finite counterterms, meaning that the two schemes are equivalent. However, since conformal field theories may be not perturbative, the methods we present in this paper are much more general than Feynman diagram-based calculations.

In section \ref{sec:anomalies_and_betafns}, we present a general first-principles discussion of the conformal Ward identities obeyed by the renormalised correlators, including the contributions from both beta functions and conformal anomalies.  As well as confirming the Ward identities found earlier for specific correlators, we obtain a general understanding of 
the relationship between the anomalous terms appearing in the Ward identities for dilatations and for special conformal transformations.
As we show, this relationship sometimes leads to additional constraints on the renormalisation-scheme dependent constants that feature in the renormalised correlators.

In section \ref{sec:dual_conf_sym} we discuss dual conformal invariance: the extraordinary observation that in certain cases the CFT 3-point functions in {\it momentum space} take precisely the form expected for a CFT 3-point function in {\it position space} (namely \eqref{3pt} with $\bs{x}_i\rightarrow \bs{p}_i$).  
For this additional momentum-space conformal symmetry to be present,  the leading divergence of the regulated triple-$K$ integral must be nonlocal.  
We give a number of examples and clarify the origin of dual conformal invariance by relating triple-$K$ integrals to the star-triangle duality of ordinary 1-loop massless Feynman integrals.

We summarise and present our main conclusions in section \ref{sec:discussion}.
Four important appendices then complete our analysis.
In appendix  \ref{sec:gen_results}, 
we derive a complete classification of all possible singularities of the triple-$K$ integral for any 3-point correlator.  Renormalising in a convenient choice of scheme, we arrive 
at explicit expressions for the renormalised 3-point functions wherever these can be read off from the singularities of the triple-$K$ integral.
Changes of renormalisation scheme are related to a corresponding non-uniqueness of the triple-$K$ representation as we discuss.
Appendix \ref{sec:shadow} then elaborates on the curious relations found between correlators of operators with `shadow' dimensions $\Delta$ and $d-\Delta$.
Appendix \ref{sec:freefields} provides independent confirmation of our main results (including the presence of double logarithms of momenta) through explicit free field calculations. Here we also demonstrate that our renormalisation procedure yields results equivalent to those obtained through a conventional perturbation theory analysis. 
Appendix \ref{sec:AdS/CFT} discusses triple-$K$ integrals in a holographic context, explaining  
how they arise in AdS/CFT calculations of 3-point functions.  We present a complete worked example of holographic renormalisation for the 3-point function of a marginal operator in three dimensions.

\section{Conformal Ward identities} \label{sec:CWI}

Let $\mathcal{O}_1, \mathcal{O}_2, \ldots, \mathcal{O}_n$ be conformal primary operators of dimensions $\Delta_1, \Delta_2, \ldots, \Delta_n$. 
The dilatation Ward identity in position space reads 
\begin{equation} \label{e:ward_dil_x}
0 = \sum_{j=1}^n \Big(\Delta_j + x_j^\mu \frac{\partial}{\partial x_j^\mu} \Big) \langle \mathcal{O}_1(\bs{x}_1) \ldots \mathcal{O}_n(\bs{x}_n) \rangle.
\end{equation}
This Ward identity tells us that the correlator is a homogeneous function of the positions of degree $-\Delta_t$, where the total dimension $\Delta_t = \sum \Delta_j$.

The Ward identity associated with special conformal transformations for $n$-point functions is 
\begin{equation} \label{e:ward_cwi_x0}
0 =  \sum_{j=1}^n \left( 2 \Delta_j x_j^\mu + 2 x_j^\mu x_j^\nu \frac{\partial}{\partial x_j^\nu} - x_j^2 \frac{\partial}{\partial x_{j \mu}} \right)  \langle \mathcal{O}_1(\bs{x}_1) \ldots \mathcal{O}_n(\bs{x}_n) \rangle,
\end{equation}
where $\mu$ is a free Lorentz index. For tensorial operators an additional term appears, see \cite{Bzowski:2013sza}.
In position space, the special conformal Ward identity is a first-order linear PDE. 
It can be solved by using the fact that special conformal transformations can be obtained by combining inversions and translations, and then analysing the implications of inversions.
Here we will instead solve the special conformal Ward identity directly.

In momentum space, translational invariance implies that we can pull out a momentum-conserving delta function,
\begin{equation} \label{delta}
\langle \mathcal{O}_1(\bs{p}_1) \cdots  \mathcal{O}_n (\bs{p}_n)
\rangle = (2 \pi)^d \delta( \bs{p}_1 + \cdots  + \bs{p}_n ) \lla
\mathcal{O}_1(\bs{p}_1) \cdots  \mathcal{O}_n(\bs{p}_n) \rra,
\end{equation}
thereby defining the reduced matrix element which we denote with double
brackets.\footnote{In some of the literature, for example in \cite{Maldacena:2011nz, Arkani-Hamed:2015bza}, the reduced 
matrix elements are denoted by  $\<\  \ \>'$.}
The Ward identities  for the reduced matrix elements are  then 
\begin{align}
0 & = \left(- (n - 1) d+ \sum_{j=1}^n \Delta_j  - \sum_{j=1}^{n-1} p_j^\mu \frac{\partial}{\partial p_j^\mu} \right) \lla \mathcal{O}_1(\bs{p}_1) \ldots \mathcal{O}_n(\bs{p}_n) \rra, \label{e:ward_dil} \\
0 & =  \sum_{j=1}^{n-1} \left( 2 (\Delta_j - d) \frac{\partial}{\partial p_{j\mu}} - 2 p_j^\nu \frac{\partial}{\partial p_j^\nu} \frac{\partial}{\partial p_{j\mu}} + p_j^\mu \frac{\partial}{\partial p_j^\nu} \frac{\partial}{\partial p_{j \nu}} \right)  \lla \mathcal{O}_1(\bs{p}_1) \ldots \mathcal{O}_n(\bs{p}_n) \rra, \label{e:ward_cwi0}
\end{align}
where we used the momentum-conserving delta function to express $\bs{p}_n$ in terms of the other momenta. 

The dilatation Ward identity (\ref{e:ward_dil}) is again easy to deal with: it tells us that the reduced matrix elements are homogeneous functions of  degree $\Delta_t-(n-1)d$. The special conformal Ward identity (\ref{e:ward_cwi0}) is now a second-order linear PDE
(while it was first-order in position space), so at first sight going to momentum space appears to make the problem more difficult. 
However, momentum space has one advantage: any tensorial object can be expanded in a basis constructed out of 
momenta and the metric. Let us denote the differential operator on the right-hand side of \eqref{e:ward_cwi0} as $\mathcal{K}^\mu$, so that the conformal Ward identities may be compactly expressed as
\begin{equation} \label{e:cwiK}
\mathcal{K}^\mu \lla \mathcal{O}_1(\bs{p}_1) \ldots \mathcal{O}_n(\bs{p}_n) \rra = 0.
\end{equation}
Since $\mathcal{K}^\mu$ carries one free Lorentz index,  $\mathcal{K}^\mu$ can be decomposed into a basis of independent vectors $p_j^{\mu}$, $j=1,2, \ldots, n-1$, \textit{i.e.},
\begin{equation} \label{e:cwiKdecomp}
\mathcal{K}^\mu = p_1^\mu \mathcal{K}_1 + \ldots + p_{n-1}^\mu \mathcal{K}_{n-1}.
\end{equation}
The Ward identity \eqref{e:ward_cwi0} thus gives rise  to $(n-1)$ scalar equations, 
\begin{equation}
\mathcal{K}_j \lla \mathcal{O}_1(\bs{p}_1) \ldots \mathcal{O}_n(\bs{p}_n) \rra = 0, \qquad j=1,\, 2, \ldots, n-1.
\end{equation}
Altogether the dilatation and special conformal Ward identities constitute $n$ differential equations. A Poincar\'{e}-invariant $n$-point function of scalar operators depends on $n(n-1)/2$ kinematic variables, so after imposing the conformal Ward identities, the correlator should be a function of $n(n-3)/2$ variables.  This agrees with position-space considerations: the number of conformal cross-ratios in $n$ variables in $d>2$ dimensions is $n(n-3)/2$.

\section{2-point functions} \label{sec:2pt}

As a warm-up exercise,
in this section we discuss CFT 2-point functions. We will use this section to establish the benchmarks we 
want to achieve for 3-point functions. 

Poincar\'{e} symmetry implies that the correlator depends only on the magnitude of a single vector $\bs{p}_1=-\bs{p}_2 \equiv \bs{p}$ and both the dilatation and special conformal Ward identities \eqref{e:ward_dil} and \eqref{e:ward_cwi0} simplify to ordinary differential equations.

We start by discussing the implications of special conformal transformations. The special conformal Ward identity is indeed proportional 
to $p^\mu$ (after using $\D/\D p_\mu = (p^\mu/p)\D/\D p$) and the corresponding scalar equation reads
\begin{align} \label{e:cwi_2pt}
0 & = \mathcal{K} \lla \O_1(\bs{p}) \O_2(-\bs{p}) \rra = \left[ \frac{d^2}{d p^2} + \frac{d+1-2 \Delta_1}{p} \frac{d}{d p} \right] \lla \O_1(\bs{p}) \O_2(-\bs{p}) \rra. 
\end{align}
As we shall see, the differential operator $\mathcal{K}$  will reappear later in our discussion of the conformal Ward identities for 3-point functions. Note also that 
\begin{equation}
\mathcal{K} = \frac{1}{p^{d+1-2 \Delta_1}} \frac{d}{d p} \left( p^{d+1-2 \Delta_1} \frac{d}{d p} \right) 
\end{equation}
which, when acting on 
spherically symmetric configurations,  is equal to the box operator in $\mathbb{R}^{d+2-2 \Delta_1}$ with $p$ the radial coordinate.

The general solution of (\ref{e:cwi_2pt}) is 
\begin{equation} \label{e:sol_cwi}
\lla \O_1(\bs{p}) \O_2(-\bs{p}) \rra  = c_{0} p^{2 \Delta_1 -d} + c_1,
\end{equation}
where $c_{0}$ and $c_1$ are integration constants. We still need to impose the dilatation Ward identity,
\begin{equation} \label{e:dil2_2pt}
D \lla \O_1(\bs{p}) \O_2(-\bs{p}) \rra = \left[ d - \Delta_1 - \Delta_2 + p \frac{d}{d p} \right] \lla \O_1(\bs{p}) \O_2(-\bs{p}) \rra=0.
\end{equation}
Inserting (\ref{e:sol_cwi}) we find that\footnote{In the special case $\Delta_1=d/2$ the general solution of 
(\ref{e:cwi_2pt}) is $\lla \O_1(\bs{p}) \O_2(-\bs{p}) \rra  = c_{0} + c_1 \ln p$ and then inserting in (\ref{e:dil2_2pt}) we find 
(\ref{e:sol_dil}).}
\begin{equation} \label{e:sol_dil}
\Delta_1=\Delta_2 \equiv \Delta, \quad c_1=0.
\end{equation}
We thus recover the well-known fact that only operators with the same dimension have non-zero 2-point function in CFT. 
The general form of the 2-point function is
\begin{equation} \label{sec2pt:sol}
\lla \O_1(\bs{p}) \O_2(-\bs{p}) \rra  = c_{\Delta} p^{2 \Delta -d},
\end{equation}
where we renamed $c_0 \to c_{\Delta}$.

For generic dimension $\Delta$ this is the end of the story. Something special happens however when 
\begin{equation} \label{sec2pt: 2pt_cond}
\Delta = \frac{d}{2} + k, \qquad k =0, 1, 2, \ldots
\end{equation}
When this condition holds, 
\begin{equation}
\lla \O_1(\bs{p}) \O_2(-\bs{p}) \rra  = c_{\Delta} p^{2 k}. 
\end{equation}
This correlator is local,\footnote{In position space the problem is that the standard expression, $1/x^{2 \Delta}$, does not have 
a Fourier transform when $\Delta=d/2 +k$. Indeed, using
$$
\int d^d \bs{x} \: e^{-i \bs{p} \cdot \bs{x}} \frac{1}{x^{2 \Delta}} = \frac{\pi^{d/2} 2^{d - 2 \Delta} \Gamma \left( \frac{d - 2 \Delta}{2} \right)}{\Gamma ( \Delta )} p^{2 \Delta - d}
$$
we see that the gamma function has a pole when $\Delta=d/2+k$. One may proceed by differential regularisation to obtain
the renormalised correlator. The final result (upon taking the Fourier transform, which now exists) agrees with
(\ref{e:2ptren}).
} {\it i.e.,} it has support only at $x^2=0$, since if we Fourier transform to position space it is proportional to  (derivatives of) a delta function,
\begin{equation} \label{2pt_special}
\< \O(\bs{x}) \O(0) \> = c_{\Delta} (- \Box)^k \delta(\bs{x}).
\end{equation}
When the dimension of the operator is (\ref{sec2pt: 2pt_cond}) there is something else special: there is a new local term of dimension $d$, namely
\begin{equation} \label{2pt_count}
\phi \,\Box^k \phi,
\end{equation}
where $\phi$ is the source of $\O$. This term can appear as a new counterterm  (and  as we shall see below, as a new contribution to the trace of the energy momentum, {\it i.e.}, a new conformal anomaly \cite{Petkou:1999fv}).  Adding the counterterm (\ref{2pt_count}) with appropriate (finite) coefficient one may arrange to cancel the right-hand side of (\ref{2pt_special}), $\< \O(\bs{x}) \O(0) \>  =0$.
In a unitary theory, this implies that $\O=0$ an an operator. However, we know there are CFTs containing non-trivial operators of 
dimension $\Delta = d/2 +k$. For example, all half-BPS scalar operators of $\mathcal{N}=4$ SYM in $d=4$ have dimensions of this form.

What happens in these special cases is that there are new UV infinities and we need to renormalise theory. As we shall see, 
the renormalised correlators will be non-trivial. However, the theory will now have a conformal anomaly: the conformal Ward identities will be violated by local terms.  Our strategy will be the following. First, we will regularise the theory and  solve the conformal Ward identities in the regulated theory. We will then add counterterms to remove the UV infinities and remove the regulator to obtain renormalised correlators.

To proceed we need to discuss our regularisation. We want to analyse the problem in complete generality, {\it i.e.}, with no reference to any specific model, and the only parameters in our disposal are the space-time dimension and the dimensions of the operators. We proceed by using a dimensional regularisation that also shifts the dimensions of the operators as follows,
\begin{equation} \label{sec2t:dimreg}
d \mapsto \tilde{d}=d + 2 u \epsilon, \qquad\qquad \Delta \mapsto \tilde{\Delta}=\Delta + (u + v)\epsilon,
\end{equation}
where $u$ and $v$ are arbitrary real numbers and $\epsilon$ denotes a regulator. More generally, one may shift each dimension by a different amount but we found that this scheme is sufficient for the discussion up to 3-point functions. We will discuss special 
choices of $u$ and $v$ below.

The solution of the conformal Ward identities in the regulated theory is exactly the same as in (\ref{sec2pt:sol}) (with $d$ and 
$\Delta$ replaced by $\tilde{d}$ and $\tilde{\Delta}$) but the integration constant $c_{\Delta}$ can depend on the regulator,
\begin{equation} \label{sec2pt:sol_reg}
\lla \O(\bs{p}) \O(-\bs{p}) \rra_{\text{reg}} = c_{\Delta}(\epsilon, u, v) p^{2 \tilde{\Delta} - \tilde{d}}=
c_{\Delta}(\epsilon, u, v) p^{2 \Delta - d + 2 v \epsilon}.
\end{equation}
In dimensional regularisation, UV infinities appear as poles in $\epsilon$. In local QFT, UV infinities should be local and this implies 
that $c_{\Delta}$ can have at most a first-order pole,
\begin{equation} 
c_{\Delta}(\epsilon, u, v) = \frac{c_{\Delta}^{(-1)}(u, v)}{\epsilon} + c_{\Delta}^{(0)}(u, v) + O(\epsilon).
\end{equation}
Inserting this in (\ref{sec2pt:sol_reg}) and expanding in $\epsilon$ we find,
\begin{equation} \label{e:2ptreg}
\lla \O(\bs{p}) \O(-\bs{p}) \rra_{\text{reg}} = p^{2 \Delta-d} \left[ \frac{c_{\Delta}^{(-1)}}{\epsilon} + c_{\Delta}^{(-1)} v \ln p^2 + c_{\Delta}^{(0)} + O(\epsilon) \right].
\end{equation}

The generators of dilatations and special conformal transformations in the regulated theory are related to those of the original as follows,
\[ \label{reg_DK}
\tilde{D} = D  - 2 v \epsilon, \qquad
\tilde{\mathcal{K}}=  \mathcal{K} - 2 v \epsilon \frac{1}{p} \frac{d}{d p}.
\]
Notice that in the $v=0$ scheme the generators are not corrected. However, for this scheme the 2-point function itself is not  regulated
so this is not a useful scheme for 2-point functions. This will change when we move to 3-point functions and it will turn out that 
for scalar 3-point functions this is a convenient scheme.  From now on we will stay with a general $(u,v)$ scheme.
The fact that the regulated correlator (\ref{e:2ptreg}) is annihilated by $\tilde{D}$ and $\tilde{\mathcal{K}}$ implies that the 
terms that appear in its $\epsilon$ expansion will satisfy related equations. In particular, the leading-order term in the $\epsilon$ 
expansion should satisfy the Ward identities of the un-regulated theory which we have already solved. 

Let us start with the generic case, $\Delta \neq d/2 +k$. In this case there are no true UV  infinities and our earlier  discussion
shows that (\ref{sec2pt:sol}) is the correct 2-point function. It is instructive however to still discuss it starting from the regulated 
theory.  The regulated 2-point function (\ref{e:2ptreg}) has a $1/\epsilon$ singularity. However, its coefficient is nonlocal and 
thus it cannot be removed by a local counterterm. On the other hand, it satisfies the correct  (non-anomalous) Ward identities,
\begin{equation} \label{2pt_nonlocal}
\tilde{D} \lla \O(\bs{p}) \O(-\bs{p}) \rra_{\text{reg}} =0 \qquad \Rightarrow \qquad D p^{2 \Delta-d} =0
\end{equation}
and the same with $\tilde{D}$ and $D$ replaced by $\tilde{\mathcal{K}}$ and $\mathcal{K}$. 
It follows that $p^{2 \Delta-d}$ is the correct 2-point function.  In a sense the leading-order pole is `fake': we could remove it by multiplying $c_{0}(\epsilon, u, v)$ by $\epsilon$. This discussion may look somewhat superfluous but we will find an exactly analogous situation when we discuss 3-point functions 

Let us now discuss the case $\Delta=d/2 +k$. Here, the leading-order divergence is local and satisfies the Ward identities. 
This is precisely as expected on general grounds: divergences should be local and should be invariant under the original symmetries 
of the theory.  
With $\phi$ again denoting the source for the operator $\O$, the regulated action reads
\begin{equation} \label{e:regS}
S[\phi] = S_{\rm{CFT}} + \int \D^{d + 2 u \epsilon} \bs{x} \: \phi \O.
\end{equation}
If $Z$ denotes the generating functional of the regulated theory,
\begin{equation}
Z[\phi] = \int \mathcal{D} \Phi\, e^{-S[\phi]},
\end{equation}
then
\begin{equation}
\< \O(\bs{x}_1) \O(\bs{x}_2) \>_{\text{reg}} = \left. \frac{\delta^2Z}{\delta \phi(\bs{x}_1) \delta \phi(\bs{x}_2)} \right|_{\phi = 0}.
\end{equation}
The divergence in the 2-point function \eqref{e:2ptreg} can be removed by the addition of the counterterm action
\begin{equation} \label{e:Sct2pt}
S_{\text{ct}} = a_{\text{ct}}(\epsilon, u, v) \int \D^{d + 2 u \epsilon} \bs{x} \: \mu^{2 v \epsilon}\phi \,\Box^k \phi ,
\end{equation}
where $a_{\text{ct}}(\epsilon, u, v)$ is a counterterm constant. As is standard in dimensional regularisation, the renormalisation scale $\mu$ appears for dimensional reasons. In the regularisation scheme \eqref{sec2t:dimreg}, $\phi$ has  scaling dimension
$d - \Delta + (u - v) \epsilon$ and this implies that $\mu$ enters with power $2 v \epsilon$.

The contribution from the counterterm action reads
\begin{equation} \label{e:2ptct}
\lla \O(\bs{p}) \O(-\bs{p}) \rra_{\text{ct}} = - 2 a_{\text{ct}}(\epsilon, u, v) (-p^2)^{k} \mu^{2 v \epsilon}
\end{equation}
and cancels the divergence in \eqref{e:2ptreg} if
\begin{equation}
a_{\text{ct}}(\epsilon, u, v) = \frac{(-1)^{k}}{2} \left[ \frac{c_{\Delta}^{(-1)}(u, v)}{\epsilon} + a_0(u, v) + O(\epsilon) \right]\hspace{-0.5mm},
\end{equation}
where $a_0$ is an arbitrary constant. We can now take the limit $\epsilon \rightarrow 0$ to obtain the renormalised correlation function
\begin{align} \label{e:2ptren}
\lla \O(\bs{p}) \O(-\bs{p}) \rra & = p^{2 k} \left[ c_{\Delta}^{(-1)} v \ln \frac{p^2}{\mu^2} + c_{\Delta}^{(0)} - a_0 \right] 
= p^{2 k} \left[ c_{\Delta} \ln \frac{p^2}{\mu^2} + c_{\Delta}' \right],
\end{align}
where $c_{\Delta}$ is the actual normalisation of the 2-point function and the combination $c_{\Delta}'=(c_{\Delta}^{(0)} - a_0)$ is scheme dependent, since it can be absorbed by a redefinition of the scale $\mu$. The renormalised 2-point function (\ref{e:2ptren})  is however scale dependent,
\begin{equation} \label{sec2pt:an}
\mathcal{A}_2 = \mu \frac{\partial}{\partial \mu} \lla \O(\bs{p}) \O(-\bs{p}) \rra = -2 c_{\Delta} p^{2 k}.
\end{equation}
There is thus a conformal anomaly,
\begin{equation}
\mu \frac{\partial}{\partial \mu} W = A,
\end{equation}
where $W=\ln Z$ is the generating functional of connected correlation functions and $A$ is the conformal anomaly \cite{Petkou:1999fv},
\begin{equation}
A = \int \D^d \bs{x} \, \mathcal{A}_k \phi\, \Box^k \phi + \cdots 
\end{equation}
where $\mathcal{A}_k$ is the anomaly coefficient (which can be read off from \eqref{sec2pt:an}), the sum is over all operators of dimension 
$\Delta=d/2 +k$, and  the dots indicate terms higher order in the sources and terms that are associated with non-scalar operators 
(such as the more often discussed terms that depend only on the background metric). In the next section we will 
compute  the terms cubic in the sources. 

\section{3-point functions}\label{sec:3pt}

We now present the analogue discussion for scalar 3-point functions. We start with the conformal Ward identities and 
their solution for generic conformal dimensions,  then discuss the special cases where renormalisation may 
be required. We illustrate our discussion throughout with explicit examples.

\subsection{Ward identities}\label{subsec:Ward1}

Poincar\'{e} invariance implies that 3-point functions can be expressed in terms of three variables, which we choose to be 
the magnitudes of the three momenta,
\begin{equation}
p_j = | \bs{p}_j |, \quad j=1,2,3.
\end{equation}
Using the chain rule and noting that $\bs{p}_3 = - \bs{p}_1 - \bs{p}_2$, we find
\begin{align}
\frac{\partial}{\partial p_{1 \mu}} 
& = \frac{p_1^\mu}{p_1} \frac{\partial}{\partial p_1} + \frac{p_1^\mu + p_2^\mu}{p_3} \frac{\partial}{\partial p_3}. \label{e:pdiff}
\end{align}
The dilatation Ward identity \eqref{e:ward_dil} may then be processed to become
\begin{equation} \label{e:dil3}
0 = D \lla \O_1(\bs{p}_1) \O_2(\bs{p}_2) \O_3(\bs{p}_3) \rra = \left( 2 d - \Delta_t + \sum_{j=1}^3 p_j \frac{\partial}{\partial p_j} \right) \lla \O_1(\bs{p}_1) \O_2(\bs{p}_2) \O_3(\bs{p}_3) \rra,
\end{equation}
where $\Delta_t = \Delta_1 + \Delta_2 + \Delta_3$. This equation shows that the correlation function is a homogeneous function 
of degree $\Delta_t - 2d$, which implies that
\begin{equation}
\lla \O_1(\bs{p}_1) \O_2(\bs{p}_2) \O_3(\bs{p}_3) \rra = p_1^{\Delta_t - 2 d} F \left( \frac{p_2}{p_1}, \frac{p_3}{p_1} \right),
\end{equation}
where $F$ is a general function of two variables. 

Let us now discuss the special conformal Ward identity \eqref{e:ward_cwi0}.  As noted in section \ref{sec:CWI}, it implies two scalar equations.  The first one reads
\begin{align}
0 & = \mathcal{K}_1 \lla \O_1(\bs{p}_1) \O_2(\bs{p}_2) \O_3(\bs{p}_3) \rra \nn\\
& = \left[ \frac{\partial^2}{\partial p_1^2} + \frac{\partial^2}{\partial p_3^2} + \frac{2 p_1}{p_3} \frac{\partial^2}{\partial p_1 \partial p_3} + \frac{2 p_2}{p_3} \frac{\partial^2}{\partial p_2 \partial p_3} - \frac{2 \Delta_1 - d - 1}{p_1} \frac{\partial}{\partial p_1} \right.\nn\\
& \qquad\qquad \left.  \: - \frac{2 \Delta_1 + 2 \Delta_2 - 3 d - 1}{p_3} \frac{\partial}{\partial p_3} \right] \lla \O_1(\bs{p}_1) \O_2(\bs{p}_2) \O_3(\bs{p}_3) \rra, \label{3pt:k1}
\end{align}
while the second equation,  $\mathcal{K}_2 \lla \O_1(\bs{p}_1) \O_2(\bs{p}_2) \O_3(\bs{p}_3) \rra =0$,
is obtained from this one by substituting $p_1 \leftrightarrow p_2$ and $\Delta_1 \leftrightarrow \Delta_2$.

Let us consider the combinations
\begin{equation}
\K_{13} = \mathcal{K}_1 - \frac{2}{p_3} \frac{\partial}{\partial p_3} D, \qquad\qquad \K_{23} = \mathcal{K}_2 - \frac{2}{p_3} \frac{\partial}{\partial p_3} D.
\end{equation}
The effect of the dilatation terms is to remove the terms with mixed derivatives in  (\ref{3pt:k1}).
In this way we arrive at the particularly simple set of equations discussed in \cite{Bzowski:2013sza},
\begin{equation} \label{SCWI}
0 = \K_{13} \lla \O_1(\bs{p}_1) \O_2(\bs{p}_2) \O_3(\bs{p}_3) \rra = \K_{23} \lla \O_1(\bs{p}_1) \O_2(\bs{p}_2) \O_3(\bs{p}_3) \rra,
\end{equation}
where
\begin{align}
\K_{ij} & = \K_i - \K_j, \\
\K_i & = \frac{\partial^2}{\partial p_i^2} - \frac{2 \Delta_i - d - 1}{p_i} \frac{\partial}{\partial p_i},
\end{align}
for $i,j=1,2,3$. Note that $\K_i$ is the same operator that appeared in our analysis of 2-point functions, see (\ref{e:cwi_2pt}).

\subsection{General solution}\label{subsec:gen_soln}

The system of the dilatation and special conformal Ward identities is equivalent to that defining the generalised hypergeometric function of two variables Appell $F_4$ \cite{Coriano:2013jba,Bzowski:2013sza} and from this fact one can infer general properties such as the uniqueness of the solution. An explicit form of the general solution is given in terms of triple-$K$ integrals \cite{Bzowski:2013sza}, 
\begin{equation} \label{e:corr3}
\lla \O_1(\bs{p}_1) \O_2(\bs{p}_2) \O_3(\bs{p}_3) \rra = c_{123} I_{\alpha \{\beta_1 \beta_2 \beta_3\}}(p_1, p_2, p_3),
\end{equation}
where $c_{123}$ is an integration constant and 
 \begin{equation} \label{e:I}
I_{\alpha \{\beta_1 \beta_2 \beta_3\}}(p_1, p_2, p_3) = \int_0^\infty \D x \: x^\alpha \prod_{j=1}^3 p_j^{\beta_j} K_{\beta_j}(p_j x).
\end{equation}
is the triple-$K$ integral. Here $K_\nu(x)$ denotes the modified Bessel function of the second kind (or the Bessel-$K$ function for short), while the parameters
\begin{equation} \label{e:dDelta}
\alpha = \frac{d}{2} - 1, \qquad\qquad \beta_j = \Delta_j - \frac{d}{2}, \quad j=1,2,3.
\end{equation}

Before we proceed to use this result, let us present an elementary derivation of it. We will start by solving (\ref{SCWI})
using separation of variables, 
\be
\lla \O_1(\bs{p}_1)\O_2(\bs{p}_2)\O_3(\bs{p}_3)\rra = f_1(p_1)f_2(p_2)f_3(p_3).
\ee
Inserting this ansatz in (\ref{SCWI}), we obtain
\be
\frac{K_1 f_1}{f_1} = \frac{K_2 f_2}{f_2}=\frac{K_3 f_3}{f_3}=x^2,
\ee
where $x^2$ is a constant since the equalities hold for arbitrary $p_i$.  The equation $K_i f_i = x^2 f_i$ is equivalent to Bessel's equation and has the general solution
\be    \label{sepsoln}
f_i(p_i) =p_i^{\beta_i} \big(a_K K_{\beta_i}(p_i x)+a_I I_{\beta_i}(p_ix)\big).
\ee
The integrand of the triple-$K$ integral is thus itself a solution of the special conformal Ward identities. 

Now, given a solution of the special conformal Ward identities 
$
f(p_1,p_2,p_3)= \prod_i f_i(p_i),
$ 
we can immediately construct a solution of {\it both} the special conformal {\it and} the dilatation Ward identities by taking the Mellin transform, 
\be
\int_0^\infty \D x\, x^{\alpha-\beta_t} f(p_1 x, p_2 x, p_3 x).
\ee
where $\beta_t = \beta_1+\beta_2+\beta_3$.
To see this, note that 
\be
\sum_{i=1}^3 p_i\frac{\partial}{\partial p_i} f(p_1 x,p_2x,p_3x) = x\frac{\partial}{\partial x} f(p_1 x,p_2x,p_3x)
\ee
and then use integration by parts.  
In order for this Mellin transform to converge, at least one of the $f_i(p_i)$ must be a Bessel-$K$ function, as Bessel-$I$ grows exponentially at large $x$.  A closer analysis \cite{Bzowski:2013sza, Coriano:2013jba} (see also appendix \ref{subsec:non-uniqueness}) reveals that in fact all three $f_i(p_i)$ must be Bessel-$K$ functions, as otherwise the resulting 3-point function becomes singular for collinear momentum configurations ({\it e.g.,} $p_1+p_2=p_3$).

It remains to discuss convergence at $x=0$. As it stands, the triple-$K$ integral converges only if 
\begin{equation} \label{e:conv}
\alpha > |\beta_1| + |\beta_2| + |\beta_3| - 1, \qquad p_1, p_2, p_3 > 0.
\end{equation}
However, one can extend the triple-$K$ integral beyond this region by means of analytic continuation. If one considers the triple-$K$ integral as a function of its parameters with momenta fixed, then analytic continuation can be used in order to define the triple-$K$ 
everywhere, provided 
\begin{equation}\label{e:condition}
\alpha + 1 \pm \beta_1 \pm \beta_2 \pm \beta_3 \neq  - 2 k,
\end{equation}
for any choice (of independent) signs and non-negative integer $k$. When the equality holds we recover \eqref{3pt:cond} and 
the triple-$K$ integral contains poles (as we will discuss in detail shortly).  In such cases a non-trivial renormalisation of the correlation function \eqref{e:corr3} may be required.

In summary, when the dimensions are generic, meaning (\ref{3pt:cond}) is not satisfied for any choice of signs and non-negative integer 
$k$, the solution of the dilatation and special conformal Ward identities is (\ref{e:corr3}). This is then the analogue of 
(\ref{sec2pt:sol}) for 3-point functions. We will shortly discuss in detail the special cases but first a couple of examples.
In these examples, and those we consider later, it will  often be useful to label operators and their sources according to their (bare) dimensions, as indicated in square brackets.  In this notation an operator of dimension $\Delta$ and its corresponding source are thus $\O_{[\Delta]}$ and $\phi_{[d-\Delta]}$.

\bigskip

{\it \textbullet \ Example 1: $d = 4$ and $\Delta_1 = \Delta_2 = \Delta_3 = 5/2$.}

\bigskip

This is an example of a finite correlation function expressible in terms of elementary functions. The 3-point function is represented by a triple-$K$ integral
\begin{equation}
\lla \O_{[5/2]}(\bs{p}_1) \O_{[5/2]}(\bs{p}_2) \O_{[5/2]}(\bs{p}_3) \rra = c\, (p_1 p_2 p_3)^{\frac{1}{2}} \int_0^\infty \D x \: x K_{\frac{1}{2}}(p_1 x) K_{\frac{1}{2}}(p_2 x) K_{\frac{1}{2}}(p_3 x),
\end{equation}
where $c$ is the integration constant. 
All Bessel $K$ functions with half-integral indices are elementary.  
In this case the integral is convergent and evaluates to
\begin{equation}
\lla \O_{[5/2]}(\bs{p}_1) \O_{[5/2]}(\bs{p}_2) \O_{[5/2]}(\bs{p}_3) \rra = \frac{c \pi^2}{2^{\frac{3}{2}}} \frac{1}{\sqrt{p_1 + p_2 + p_3}}.
\end{equation}

\bigskip

{\it \textbullet \ Example 2: $d = 4$ and $\Delta_1 = \Delta_2 = \Delta_3 = 2$.}

\bigskip

In this case the 3-point function is given by
\[
\lla \O_{[2]}(\bs{p}_1) \O_{[2]}(\bs{p}_2) \O_{[2]}(\bs{p}_3) \rra = c I_{1\{000\}} 
\]
It turns that this integral has already been computed in the literature  \cite{Davydychev:1992xr,tHooft:1978xw}
and is given by 
\begin{align}
I_{1 \{000\}} & = \frac{1}{2 \sqrt{-J^2}} \left[ \frac{\pi^2}{6} - 2 \ln \frac{p_1}{p_3} \ln \frac{p_2}{p_3} + \ln \left( - X \frac{p_2}{p_3} \right) \ln \left( - Y \frac{p_1}{p_3} \right) \right. \nn\\
& \qquad \qquad \qquad \left. - \: \Li_2 \left( - X \frac{p_2}{p_3} \right) - \Li_2 \left( - Y \frac{p_1}{p_3} \right) \right]. \label{e:I1000}
\end{align}
where
\begin{align} \label{defs}
J^2 & = (p_1 + p_2 - p_3) (p_1 - p_2 + p_3) (-p_1 + p_2 + p_3) (p_1 + p_2 + p_3), \\
X & = \frac{p_1^2 - p_2^2 - p_3^2 + \sqrt{-J^2}}{2 p_2 p_3}, \qquad Y = \frac{p_2^2 - p_1^2 - p_3^2 + \sqrt{-J^2}}{2 p_1 p_3}.
\end{align}
As will be discussed in \cite{integrals} (see also \cite{Bzowski:2013sza}),
triple-$K$ integrals with integral indices can be obtained from this integral using a recursion method. 

\bigskip

{\it \textbullet \ Example 3: $d = 4$ and $\Delta_1 = \Delta_2 = \Delta_3 = 7/2$.} \label{Example3}

\bigskip

This is an example of a finite correlation function expressible in terms of a triple-$K$ integral which diverges but nevertheless possesses a unique analytic continuation. The 3-point function is represented by
\begin{align} \label{7/2}
&\lla \O_{[7/2]}(\bs{p}_1) \O_{[7/2]}(\bs{p}_2) \O_{[7/2]}(\bs{p}_3) \rra_{\text{bare}} = c\, I_{1,\{3/2, 3/2, 3/2\}} \nn\\&\qquad\qquad\qquad\qquad\qquad\qquad =
c \,(p_1 p_2 p_3)^{\frac{3}{2}} \int_0^\infty \D x \: x K_{\frac{3}{2}}(p_1 x) K_{\frac{3}{2}}(p_2 x) K_{\frac{3}{2}}(p_3 x).
\end{align}
In this case the condition \eqref{e:conv} is violated so the integral does not converge. However, \eqref{e:condition} does hold for 
all choices of signs and therefore the integral can be defined by means of analytic continuation. In such cases the dimensionally regulated integral is actually finite.

We discuss dimensional regularisation below, in section \ref{subsubsec:reg}. The integral (\ref{7/2}) can be regulated in any $(u,v)$-regularisation scheme (see  \eqref{3pt_DimReg}). However, since the Bessel functions are elementary when their orders are half integers, it is convenient to use the $(1,0)$-scheme, 
\begin{align}
&\lla \O_{[7/2]}(\bs{p}_1) \O_{[7/2]}(\bs{p}_2) \O_{[7/2]}(\bs{p}_3) \rra_{\text{reg}}  = c\, (p_1 p_2 p_3)^{\frac{3}{2}} \int_0^\infty \D x \: x^{1+\epsilon} K_{\frac{3}{2}}(p_1 x) K_{\frac{3}{2}}(p_2 x) K_{\frac{3}{2}}(p_3 x) \nn\\
&\quad = - \frac{c \pi^{\frac{3}{2}}(3 - 2 \ep) \Gamma \big( {-} \frac{5}{2} + \ep \big)}{16 \sqrt{2}(p_1 + p_2 + p_3)^{\frac{1}{2} + \ep}}  
\left[ 4 a_{123}^3 - (10 - 4 \ep) a_{123} b_{123} + (5 - 12 \ep + 4 \ep^2) c_{123} \right],
\end{align}
where 
\be
a_{123} = p_1+p_2+p_3, \quad b_{123} = p_1 p_2 + p_1 p_3+p_2 p_3, \qquad c_{123} = p_1 p_2 p_3.
\ee
This expression is valid for a range of $\ep$, not necessarily close to zero. It has a finite $\ep \rightarrow 0$ limit,
\begin{equation}
\lla \O_{[7/2]}(\bs{p}_1) \O_{[7/2]}(\bs{p}_2) \O_{[7/2]}(\bs{p}_3) \rra = \frac{c \pi^2}{10 \sqrt{2}} \frac{4 a_{123}^3 - 10 a_{123} b_{123} + 5 c_{123}}{\sqrt{p_1 + p_2 + p_3}},
\end{equation}
as anticipated. This 3-point function satisfies all conformal Ward identities.

\bigskip

In summary,  if all the beta indices are half-integral the triple-$K$ integrals can be computed in terms of elementary functions and if they are integral they are given in terms of expressions involving dilogarithms. If the beta indices are generic, the triple-$K$ integral does not appear to be reducible to a more explicit expression.

\subsection{Renormalisation}\label{subsec:ren}

We will now focus on the special cases where the triple-$K$ integral is singular, {\it i.e.}, we will consider the cases where the dimensions of operators satisfy one or more of the the following conditions, 
\begin{equation} \label{e:cond_sigma}
\alpha + 1 + \sigma_1 \beta_1 + \sigma_2 \beta_2 + \sigma_3 \beta_3 = - 2 k_{\sigma_1 \sigma_2 \sigma_3}
\end{equation}
where  $\sigma_i \in \{ \pm \}$, $i=1,2,3$  and the $k_{\sigma_1 \sigma_2 \sigma_3}$ are non-negative integers. There are four 
conditions (up to permutations) depending on the relative number of minus and plus signs. We will call these conditions the
$(---)$, $(--+)$, $(-++)$ and $(+++)$ conditions. In general the condition \eqref{e:cond_sigma} can be satisfied in more than one way, 
with a different number of positive and negative signs and different values of associated non-negative integers $k_{\sigma_1 \sigma_2 \sigma_3}$. We will discuss all possibilities below.

When these conditions hold there are new terms of dimension $d$ that appear (as was the case in our discussion of 2-point functions of operators of dimension $d/2 +k$ in section \ref{sec:2pt}), and the nature of these terms gives a hint of how to deal with each of the 
singularities. Let us discuss each case in turn.

\paragraph{$\bm{(---)}$-condition: $\Delta_1 + \Delta_2 + \Delta_3  = 2d +2 k_{---}$.} In this case the new terms of dimension $d$ have the 
following schematic form
\begin{equation} \label{3pt:count}
\Box^{k_1} \phi_1 \Box^{k_2} \phi_2 \Box^{k_3} \phi_3 ,
\end{equation}
where the $\phi_i$ are sources for the operators $\O_i$ of dimension $\Delta_i$, and $k_1+k_2+k_3=k_{---}$. 
Such terms are a direct analogue of \eqref{2pt_count};  they may appear as counterterms and also as new conformal anomalies.
The fact that new conformal anomalies may appear when the theory has operators with dimensions that satisfy this relation was anticipated in 
\cite{Petkou:1999fv, PerezVictoria:2001pa}. We thus expect that when such singularities are present one would have to renormalise  by adding (\ref{3pt:count}) with the appropriate coefficient  and there would be an associated conformal anomaly. As we shall see such singularities 
are linked with logarithmic terms in the renormalised 3-point functions, similar to what we saw for 2-point functions.

\paragraph{$\bm{(--+)}$-condition: $\Delta_1 + \Delta_2 - \Delta_3  = d +2 k_{--+}$.}
In this case (and similarly for its permutations), the new terms of dimension $d$ have the 
following schematic form
\begin{equation} \label{3pt:beta}
\Box^{k_1} \phi_1 \Box^{k_2} \phi_2 \Box^{k_3} \O_{3},
\end{equation}
where $k_1+k_2+k_3=k_{--+}$.  This term can appear as  a counterterm (with appropriate singular coefficient $a_{\rm ct}$)
and thus in this case we renormalise the 
source of $\O_3$,
\[
\phi_3 \to \phi_3 + a_{\rm ct} \Box^{k_3} (\Box^{k_1} \phi_1 \Box^{k_2} \phi_2).
\]
We then expect the renormalised correlators to satisfy a Callan-Symanzik equation with beta function terms.
These beta functions are for sources that couple to composite operators and not for couplings that appear in the Lagrangian of the theory, 
so there is no contradiction here with the fact that we are discussing CFT correlation functions.  As we shall see, 
singularities of this type are linked with double logarithms in correlation functions. The existence of such double-log terms, noted also in \cite{PerezVictoria:2001pa}, is 
one of our most surprising findings, and will be discussed further in the conclusions.

\paragraph{$\bm{(-++)}$-condition: $\Delta_1 - \Delta_2 - \Delta_3  = 2 k_{-++}$.}
In this case\footnote{Note that when $k_{-++}=0$, we have extremal correlators
which were conjectured not to renomalise \cite{D'Hoker:1999ea}.}  (and similarly for its permutations), the following
term has a classical dimension $d$,
\[ \label{-++}
\Box^{k_1} \phi_1 \Box^{k_2} \O_2 \Box^{k_3} \O_3,
\]
where $k_1+ k_2 + k_3 = k_{-++}$. In other words, classically $\O_1$ has the same dimension\footnote{Note however that quantum mechanically the dimension of the product of two operators may not be the sum of their dimensions, as we are using here in asserting that the two operators have the same dimension.}
as  $\Box^{k_1}(\Box^{k_2} \O_2 \Box^{k_3} \O_3)$.
Such a term cannot  act as a counterterm for the 3-point function. As will shall see, in such cases it is the representation of the 3-point function in terms of the triple-$K$ integrals that is singular, not the correlator itself. The conformal Ward identities have a finite non-anomalous solution.

\paragraph{$\bm{(+++)}$-condition: $\Delta_1 + \Delta_2 + \Delta_3  = d-2 k_{+++}$.}  This is similar to the previous case.
The operator $\Box^{k_1} \O_1 \Box^{k_2} \O_2 \Box^{k_3} \O_3$ is classically marginal and the same comments as in the case
of $(-++)$-condition apply. In particular, this  term cannot act as a counterterm and it is again the representation of the 3-point function that is singular. The conformal Ward identities have a finite non-anomalous solution.

\subsubsection{Regularisation} \label{subsubsec:reg}

We will regularise using the dimensional regularisation (\ref{sec2t:dimreg}). In the regulated theory the solution of the conformal Ward identities is again given by \eqref{e:corr3} but with the indices shifted, and the integration constant depends now on the regularisation
parameters,
\begin{equation} \label{reg_KKK}
\lla \O_1(\bs{p}_1) \O_2(\bs{p}_2) \O_3(\bs{p}_3) \rra_{\text{reg}} = c_{123}(\epsilon, u,v) I_{\tilde{\alpha},\{\tilde{\beta}_i\}}(p_1, p_2, p_3),
\end{equation}
where 
\[ \label{3pt_DimReg}
\tilde{\alpha}=\alpha+u\ep, \qquad \tilde{\beta}_i=\beta_i+v\ep, \quad  i=1,2,3.
\]
We see from these expressions that  $v=0$ is special in that the indices of the Bessel functions remain the same. This makes the 
analysis of the singularity structure of the triple-$K$ integral easier, as we discuss in appendix \ref{sec:gen_results}. However, as mentioned in the previous section, this scheme does not regulate 2-point functions and as such it is not a good scheme for regulating {\it tensorial} 3-point functions involving the energy momentum tensor and conserved current. In these cases, the Weyl and diffeomorphism/conservation Ward identities relate 2- and 3-point functions (see for example \cite{Bzowski:2013sza}). For this reason we will continue to work in the general $(u,v)$ scheme. In section \ref{subsubsec:change_scheme} we will discuss how to go from one scheme to another.
 
The regulated triple-$K$ integral $I_{\tilde{\alpha},\{\tilde{\beta}_i\}}$ is well defined since for nonzero $\ep$ the condition \eqref{e:cond_sigma} (with $\alpha \to \tilde{\alpha}, \beta_i \to \tilde{\beta}_i$) does not hold. The integral is 
nevertheless still singular as $\ep \to 0$, however, and our task is to extract the singularities and understand how to deal with them.  This can be achieved in an elementary fashion as follows.\footnote{
In principle, the singularities of the regulated triple-$K$ integral could also be found by converting 
 to the massless triangle Feynman integral representation following appendix A.3 of \cite{Bzowski:2013sza} then using the double Mellin-Barnes representation in equation (2.5) of \cite{Davydychev:1992xr} (see also \cite{Davydychev:1995mq}).} 
Since the integral converges
at infinity even when $\ep \to 0$, all singularities  come from the $x=0$ region. We therefore split the integral 
 into an upper and a lower piece,
\[ \label{I_mu}
I_{\tilde{\alpha},\{\tilde{\beta}_i\}} = \int_{0}^{\mu^{-1}} \D x\, x^{\tilde{\alpha}}\prod_{j=1}^3 p_j^{\tilde{\beta}_j}K_{\tilde{\beta}_j}(p_jx)
+ \int_{\mu^{-1}}^\infty \D x\, x^{\tilde{\alpha}}\prod_{j=1}^3 p_j^{\tilde{\beta}_j}K_{\tilde{\beta}_j}(p_jx),
\]
where $\mu$ is an arbitrary scale which plays the role of the renormalisation scale. Note that by construction the full answer 
for  $I_{\tilde{\alpha},\{\tilde{\beta}_i\}}$ is independent of $\mu$. 

We now focus on the lower part (which contains the UV infinities) and note that for small $x$, the integrand has a Fr{\"o}benius series
\[\label{gen}
x^{\tilde{\alpha}}\prod_{j=1}^3 p_j^{\tilde{\beta}_j}K_{\tilde{\beta}_j}(p_jx) = \sum_\eta c_\eta x^\eta.
\]
The exponents $\eta$ and the coefficients $c_\eta$ follow from the standard series expansions for Bessel functions.
After some manipulation we find
\begin{align}\label{gen_exp}
& x^{\tilde{\alpha}}\prod_{j=1}^3 p_j^{\tilde{\beta}_j}K_{\tilde{\beta}_j}(p_jx) \nn\\ &= 
\sum_{\{\sigma_j=\pm 1\}} 
\sum_{\{k_j\}=0}^\infty  \Big(\prod_{i=1}^3 \frac{(-1)^{k_i} }{2^{\sigma_i\tilde{\beta}_i+2k_i+1} k_i!} \Gamma(-k_i-\sigma_i\tilde{\beta}_i) p_i^{(1+\sigma_i)\tilde{\beta}_i+2k_i}\Big) x^{\tilde{\alpha} + \sum_j (\sigma_j\tilde{\beta}_j+2k_j)},
\end{align}
where we used the fact that we work in a general $(u,v)$ scheme so neither $\tilde{\alpha}$ nor $\tilde {\beta}_i$ are integers.
The sums here run over all values of the $\sigma_j$ and all non-negative integer values of the $k_j$ (where $j=1,2,3$).
It follows that
\[ \label{eta}
\eta  = \tilde{\alpha} + \sum_j (\sigma_j\tilde{\beta}_j+2k_j )
=-1 + 2 \Big({-} k_{\sigma_1 \sigma_2 \sigma_3}+\sum_j k_j \Big) + \ep ( u + v \sum_j \sigma_j),
\]
where in the second equality we used \eqref{e:cond_sigma}.

Recall that in momentum space, 3-point functions are ultralocal if they  depend analytically on all momenta (\textit{i.e.}, they depend on
positive integral powers of all momenta squared),  semilocal if they depend analytically in two of the three momenta (they depend on
positive integral powers of two of the momenta squared) and otherwise they are nonlocal.
From the form of the expansion \eqref{gen_exp}, we see that terms for which $\{\sigma_i\}=\{-,-,-\}$ are ultralocal, terms for which $\{\sigma_i\}=\{-,-,+\}$ (and permutations) are semilocal, while terms for which $\{\sigma_i\}=\{-,+,+\}$ (and permutations) or $\{\sigma_i\}=\{+,+,+\}$ are generically nonlocal.\footnote{Note also that sending $\tilde{\beta}_i \to-\tilde{\beta}_i$ 
is equivalent to sending $\sigma_i \to -\sigma_i$ ({\it i.e.}, exchanging the singularity type), modulo a factor of $p_i^{2\tilde{\beta}_i}$.  This transformation replaces an operator with its shadow, $\Delta_i\rightarrow d-\Delta_i$, see appendix \ref{sec:shadow}.}

Inserting (\ref{gen}) in (\ref{I_mu}) we find,
\[
I_{\tilde{\alpha},\{\tilde{\beta}_i\}} = \sum_\eta c_\eta \frac{\mu^{-(\eta+1)}}{\eta+1} + \int_{\mu^{-1}}^\infty \D x\, x^{\tilde{\alpha}}\prod_{i=1}^3 p_i^{\tilde{\beta}_i}K_{\tilde{\beta}_i}(p_ix),
\]
Note that the lower limit of integration $x=0$ gives a vanishing contribution: the integral $I_{\tilde{\alpha},\{\tilde{\beta}_i\}}$ is defined by means of analytic continuation from the region where it converges (\textit{i.e.}, \eqref{e:conv} with $\alpha \to \tilde{\alpha}$, $\beta_i \to \tilde{\beta}_i$) and in this region the lower limit vanishes (since $\eta>-1$ in this region).  

We will now analyse the structure of singularities using the following two facts: (i) the upper part of the integral is finite and so can only contribute at order $\ep^0$ and higher, and (ii), the divergent terms  cannot have any dependence on $\mu$.  This follows from the fact that the total integral (\textit{i.e.}, upper plus lower part) is independent of the arbitrary scale $\mu$, and this must remain true when the integral is expanded term by term in powers of $\ep$.  

These two facts allow us to determine the form of the divergent terms, as we now discuss. The first implication is that 
the divergent terms are those with $\eta = -1+w\ep$ for some finite $w$. Indeed, suppose $\eta= m+w\ep$ for $m\neq -1$. Then $1/(\eta +1)$ is regular as $\ep \to 0$ and the singularity must come from the coefficients $c_\eta$.
However, such singularities would be $\mu$ dependent since $\mu^{-(\eta+1)}= \mu^{-(m+1)}(1+ O(\ep))$. Cancelling this leading order 
$\mu$ dependence requires $m=-1$.  We thus conclude (using \eqref{eta}) that
\be
\sum_i k_i =k_{\sigma_1 \sigma_2 \sigma_3}, \qquad w = \{ u -3 v, u-v, u+v, u+3v\}.
\ee
In other words there are four possibilities for $w$ depending on the signs required to satisfy \eqref{e:cond_sigma}. 
This condition may be satisfied for different signs (and different integers $k_{\sigma_1 \sigma_2 \sigma_3}$)
and the number of such conditions that are satisfied simultaneously determines the singularity structure of the integral.

Suppose \eqref{e:cond_sigma} has only a single solution. Then 
\[\label{one_w}
c_{-1+w\ep}\frac{\mu^{-w\ep}}{w\ep} = c_{-1+w\ep}\Big(\frac{1}{w\ep}-\ln \mu + O(\ep)\Big).
\]
In this case, the coefficient $c_{-1+w\ep}$ must be finite as the $\ln \mu$ piece cannot be associated with a divergent power of $\ep$.
On the other hand, if the condition is satisfied in multiple ways the coefficients $c_{-1+w\ep}$ may be singular. In fact if there are $s$ conditions satisfied simultaneously, the $c_{-1+w\ep}$ can diverge as $\ep^{-s+1}$, so the triple-$K$ integral can diverge as $\ep^{-s}$.
Since there are at most four different values of $w$ (in this regularisation scheme) the most singular behaviour is $\ep^{-4}$.

Let us first discuss the case where there are two simultaneous solutions to \eqref{e:cond_sigma}. Expanding 
\be
c_{-1+w\ep}=\frac{c^{(-1)}_{-1+w\ep}}{\ep}+c^{(0)}_{-1+w\ep}+O(\ep),
\ee
we find
\begin{align}
\sum_{w}c_{-1+w\ep}\frac{\mu^{-w\ep}}{w\ep} &= \sum_{w} \left(\frac{c^{(-1)}_{-1+w\ep}}{\ep}+c^{(0)}_{-1+w\ep}+O(\ep)\right)
\Big(\frac{1}{w\ep}-\ln \mu + O(\ep)\Big) \\
&=\frac{1}{\ep^2}\Big(\frac{c^{(-1)}_{-1+w_1\ep}}{w_1}+\frac{c^{(-1)}_{-1+w_2\ep}}{w_2}\Big) \nn\\&\quad  + \frac{1}{\ep}\Big[\Big(\frac{c_{-1+w_1\ep}^{(0)}}{w_1}+\frac{c_{-1+w_2\ep}^{(0)}}{w_2}\Big)-(c_{-1+w_1\ep}^{(-1)}+c_{-1+w_2\ep}^{(-1)})\ln \mu \Big] +O(\ep^0). \nn
\end{align}
For the $\mu$-dependence of the divergent terms to cancel then requires $c_{-1+w_1\ep}^{(-1)}+c_{-1+w_2\ep}^{(-1)}=0$.  
The leading $\ep^{-2}$ divergence of the triple-$K$ integral then carries a coefficient $c_{-1+w_1\ep}^{(-1)} (w_2-w_1)/w_1 w_2$.  
This case occurs for example when we have both $\{\sigma_i\}=\{---\}$ and $\{--+\}$ singularities, for which $w$ equals $u-3v$ and $u-v$ respectively.  
Here, the $\ep^{-2}$ divergence of the triple-$K$ integral appears with a coefficient 
\[ \label{e^2}
c_{-1+(u-3v)\ep}^{(-1)}\frac{2v}{(u-v)(u-3v)},
\]
 giving rise to additional divergences at $u=v$ and $u=3v$ if $c_{-1+(u-3v)\ep}^{(-1)}$ is nonzero at these points.
 Of course, it is still possible that $c^{(-1)}_{-1+w\ep}=0$, \textit{i.e.}, the coefficients $c_{-1+w\ep}$ are finite even when multiple conditions
 hold. In such cases the singularity is of first order. 

In the general case where solutions of \eqref{e:cond_sigma} exist for multiple values of $w$, expanding
\[
\sum_{w}c_{-1+w\ep}\frac{\mu^{-w\ep}}{w\ep} =
\sum_{w}c_{-1+w\ep}\Big(\frac{1}{w\ep}-\ln \mu+\frac{1}{2}w\ep\ln^2\mu+\ldots\Big)
\]
we see that for the divergent part of the triple-$K$ integral to be $\mu$-independent requires
\[
\sum_{w}c_{-1+w\ep} (w\ep)^m=O(\ep^0),
\]
for all $m\ge 0$, in order for the coefficient of $(\ln\mu)^{m+1}$ to vanish.  
Expanding the coefficients as 
\be \label{c_notation}
c_{-1+w\ep} = \sum_{s}c_{-1+w\ep}^{(-s)}\frac{1}{\ep^{s}}, 
\ee
we obtain the nontrivial equations
\[
\sum_{w}c_{-1+w\ep}^{(-s)}w^m = 0, \qquad 0\le m\le s-1.
\]
As there are $s$ equations for $c_{-1+w\ep}^{(-s)}$, to obtain a nontrivial solution requires that there are at least $s+1$ coefficients $c_{-1+w\ep}^{(-s)}$.  Thus, if the leading divergence of the $c_{-1+w\ep}$ is $\ep^{-(s-1)}$, there must be at least $s$ different values of $w$.

With all $\mu$-dependent divergences cancelling, the remaining divergent part of the triple-$K$ integral is then simply
\[\label{Idiv}
I_{\tilde{\alpha},\{\tilde{\beta}_i\}}^{\rm{div}}= \sum_{w} \frac{c_{-1+w\ep}}{w\ep} + O(\ep^0).
\]
For a specific triple-$K$ integral, \eqref{Idiv} is straightforward to evaluate.  
In particular, there is no need to evaluate the triple-$K$ integral itself, only the series expansion of its integrand. We can therefore compute the divergent part of {\it any} triple-$K$ integral, in {\it any} $(u,v)$-scheme, through this procedure.

Before we proceed, we illustrate how to compute \eqref{Idiv} using an example.

\bigskip

{\it \textbullet \, Example 4: Divergence of regulated triple-$K$ integral for $\Delta_1=4$, $\Delta_2=\Delta_3=3$ in $d=4$.} \label{Example4label}

\bigskip

Here $\alpha = 1$ while $\beta_1 = 2$ and $\beta_2=\beta_3=1$.  Thus, the $(---)$-condition is satisfied with 
$k_{---}=1$ and the $(--+)$ and $(-+-)$ conditions are satisfied with $k_{--+}=k_{-+-}=0$.

Expanding the integrand of the regulated triple-$K$ integral, $I_{1+u\ep,\{2+v\ep,1+v\ep,1+v\ep\}}$, the terms of the form $x^{-1+w\ep}$ are
\begin{align}\label{433}
&x^{1+u\ep}p_1^{2+v\ep}p_2^{1+v\ep}p_3^{1+v\ep}K_{2+v\ep}(p_1 x)K_{1+v\ep}(p_2 x)K_{1+v\ep}(p_3 x) = \nn\\&\qquad\quad  
2^{-1 + v\ep} \Gamma(-1 -v\ep) \Gamma(1 + v\ep) \Gamma(2 + v\ep)(p_2^{2 + 2v\ep} + p_3^{2 + 2v\ep}) x^{-1 + (u - v) \ep} \nn\\&\qquad - 
 \frac{ 2^{-1 + 3 v\ep}\Gamma^2(1 + v\ep)\Gamma(2 + v\ep)}{v \ep(1 + v\ep ) }  \big(p_2^2 + p_3^2 + v\ep(p_1^2 + p_2^2 + 
     p_3^2)\big)  x^{-1 + (u - 3 v) \ep}
+\ldots
\end{align}
The divergent part of the regulated triple-$K$ integral is then
\begin{align}\label{433div}
I^{\rm{div}}_{1+u\ep,\{2+v\ep,1+v\ep,1+v\ep\}}
&=\frac{2^{-1 + v\ep} \Gamma(-1 -v\ep) \Gamma(1 + v\ep) \Gamma(2 + v\ep)}{(u-v)\ep}(p_2^{2 + 2v\ep} + p_3^{2 + 2v\ep})\nn\\&\quad
- 
 \frac{ 2^{-1 + 3 v\ep}\Gamma^2(1 + v\ep)\Gamma(2 + v\ep)}{v (1 + v\ep )(u-3v)\ep^2 }  \big(p_2^2 + p_3^2 + v\ep(p_1^2 + p_2^2 + 
     p_3^2)\big) +O(\ep^0) \nn\\&
=-\frac{(p_2^2+p_3^2)}{(u-3v)(u-v)\ep^2}
+\frac{1}{2\ep}\Big(\frac{1}{(u-v)}(p_2^2\ln p_2^2+p_3^2\ln p_3^2)-\frac{1}{(u-3v)}p_1^2\nn\\&\quad
+\frac{v-u(1-2\gamma_E+\ln 4)}{(u-3v)(u-v)}(p_2^2+p_3^2)\Big)+O(\ep^0).
\end{align}
The coefficient of the leading order term is ultralocal while the coefficient of the subleading singularity is semilocal.\footnote{
The same conclusion can be reached using  differential regularisation in position space, see
\cite{PerezVictoria:2001pa}.} 

\subsubsection{Changing the regularisation scheme} \label{subsubsec:change_scheme}

Some regularisation schemes ({\it i.e.,} choices of $u$ and $v$) may be more convenient than others. For example, there may be a scheme in which one can 
compute the regulated integrals exactly.  More generally, different schemes come with different advantages and disadvantages. As discussed earlier (see also appendix \ref{subsec:zero_v_scheme}), the choice $u=1$, $v=0$ is particularly convenient because the indices of the Bessel functions are unchanged.  However, this scheme is unsuitable for tensorial correlators involving conserved currents and/or stress tensors, since these are related via the diffeomorphism Ward identity to 2-point functions which are not regulated by this scheme. The scheme with $u = v$, on the other hand,  has the attractive property that $\Delta$ and $d$ are each shifted by the same amount.  The dimensions of conserved currents and the stress tensor in the regulated theory are thus still correlated with the dimension of the regulated spacetime, as required by conservation. In some cases, however, divergences may have poles in $1/(u-v)$, as we saw in \eqref{433div}. A third useful scheme is to set $u=-v$: here only the spacetime dimension is shifted, and as will be discussed in \cite{integrals}, many regulated integrals can be computed exactly.

Given the different choices of scheme available, we would like to understand the dependence of the renormalised correlators on the scheme used.
In this subsection we discuss how to change from one regularisation $(u_0, v_0)$-scheme to another $(u,v)$-scheme.
Let us consider a divergent triple-$K$ integral, $I_{\alpha,\{\beta_i\}}$ and consider the difference in its value in the two different schemes,
\begin{equation} \label{scheme}
I^{\text{(scheme)}}_{(u_0, v_0) \mapsto (u,v)} = I_{\alpha + u \epsilon\{\beta_1 + v \epsilon, \beta_2 + v \epsilon, \beta_3 + v \epsilon\}} - I_{\alpha + u_0 \epsilon\{\beta_1 + v_0 \epsilon, \beta_2 + v_0 \epsilon, \beta_3 + v_0 \epsilon\}}.
\end{equation}
Note now that triple-$K$ integrals satisfy the following relations:
\begin{align}
L_1 I_{\alpha \{\beta_1, \beta_2, \beta_3\}} &= I_{\alpha +1\{\beta_1-1, \beta_2, \beta_3\}},  \\
M_1 I_{\alpha \{\beta_1, \beta_2, \beta_3\}} &= I_{\alpha +1\{\beta_1+1, \beta_2, \beta_3\}}, 
\end{align}
where 
\begin{equation}
 L_i = - \frac{1}{p_i} \frac{\partial}{\partial p_i}, \qquad M_i = 2 \beta_i - p_i \frac{\partial}{\partial p_i},
 \end{equation}
as can be shown by using the definition of the triple-$K$ integral and the standard properties of Bessel functions (complete proofs will be given in \cite{integrals}).

Suppose that we start with a divergent triple-$K$ integral (an integral where one or more of the conditions \eqref{e:cond_sigma}
hold). Then acting with $L_1$ on its regulated version will decrease $k_{-\sigma_2 \sigma_3}$ by one and leave $k_{+\sigma_2\sigma_3}$ unchanged,
while acting with $M_1$ will decrease $k_{+\sigma_2 \sigma_3}$ by one and leave $k_{-\sigma_2\sigma_3}$ unchanged.
Thus, by acting a sufficient number of times with $L_i$ and/or $M_i$, we will end up with a convergent integral in all cases.
Let $\{D^r\}$ be the set of such differential operators, where $r$ labels each operator in the set. Then
\begin{equation}
D^r I_{\alpha \{\beta_1, \beta_2, \beta_3\}} = I_{\alpha +m^r_1 \{\beta_1+m^r_2, \beta_2 +m^r_3, \beta_3+m^r_4\}},
\end{equation}
where $m^r_1, m^r_2, m^r_3, m^r_4$ are integers, are convergent integrals.  It follows that 
\begin{equation} \label{difeq_scheme}
D^r I^{\text{(scheme)}}_{(u_0, v_0) \mapsto (u,v)}  = 0 + O(\ep).
 \end{equation}
The equations \eqref{difeq_scheme} are a set of differential equations that may be used to determine the momentum dependence of $I^{\text{(scheme)}}_{(u_0, v_0) \mapsto (u,v)}$, which on general grounds should be a sum of local and semilocal terms.  The coefficients of the different terms
are constants that depend on $u,v, u_0, v_0$ and $\ep$, and can be determined by expanding $I_{\alpha + u \epsilon\{\beta_1 + v \epsilon, \beta_2 + v \epsilon, \beta_3 + v \epsilon\}}$ for small $p_i$, extracting all terms up to finite order in $\ep$, then inserting in \eqref{scheme} and comparing with the solution of \eqref{difeq_scheme}.

We will now illustrate this procedure with a simple example. Consider the integral $I_{2\{111\}}$. In this case the $(---)$ condition holds with $k_{---}=0$, and thus it suffices to act once with $L_i$ in order to obtain a convergent integral. We then have $\{D^r\}=\{L_1, L_2, L_3\}$, and 
\eqref{difeq_scheme} reads
\begin{equation}
L_1 I^{\text{(scheme)}}_{(u_0, v_0) \mapsto (u,v)} = L_2 I^{\text{(scheme)}}_{(u_0, v_0) \mapsto (u,v)}=L_3 I^{\text{(scheme)}}_{(u_0, v_0) \mapsto (u,v)} = 0 + O(\ep),
\end{equation}
which implies that $I^{\text{(scheme)}}_{(u_0, v_0) \mapsto (u,v)}$ is independent of momenta,
\begin{equation}
I^{\text{(scheme)}}_{(u_0, v_0) \mapsto (u,v)}  = C(u,v;u_0,v_0;\ep)+ O(\ep).
\end{equation}
We therefore need to compute the momentum-independent terms in $I_{2 + u \epsilon\{1 + v \epsilon, 1 + v \epsilon, 1 + v \epsilon\}}$, up to finite terms in $\epsilon$.  Since we want the momentum-independent part of this integral, we may wish to take first the zero-momentum limit in the integrand and then compute the integral. One has to be careful, however, as taking the limit inside the integral is not always allowed.  Moreover, $I_{2 + u \epsilon\{1 + v \epsilon, 1 + v \epsilon, 1 + v \epsilon\}}$ may diverge in this limit. What we are guaranteed is that $I^{\text{(scheme)}}_{(u_0, v_0) \mapsto (u,v)}$ is independent of momentum. In other words, any IR divergence must be independent of $(u,v)$. 

In the case at hand, we may safely take two momenta to zero, say $p_1$ and $p_2$, but we need to keep the third momentum non-zero,
\begin{align}
\lim_{p_1 \to 0, p_2 \to 0} I_{2+u \epsilon \{1 + v \epsilon, 1 + v \epsilon, 1 + v \epsilon\}} & = 4^{v \ep} \Gamma^2(1 + v \ep) \int_0^\infty \D x \: x^{(u - 2 v) \epsilon} p_3^{1+v \ep} K_{1 + v \ep}(p_3 x). 
\end{align}
This integral can computed using the result
\begin{equation} \label{e:I1K}
\int_0^\infty \D x \: x^{\alpha - 1} K_\nu(c x) = \frac{2^{\alpha - 2}}{c^\alpha} \Gamma \left( \frac{\alpha + \nu}{2} \right) \Gamma \left( \frac{\alpha - \nu}{2} \right),
\end{equation}
with the integral defined outside its domain of convergence $\re \alpha > | \re \nu |$ and $\re c > 0$ through analytic continuation.
Expanding the answer in $\ep$, we find
\begin{equation}
I_{2+u \epsilon \{1 + v \epsilon, 1 + v \epsilon, 1 + v \epsilon\}} = \frac{1}{(u - 3v) \: \epsilon} + \left[ - \ln p_3 + \frac{u}{u - 3v} (\ln 2 - \gamma_E) \right] + O(\epsilon).
\end{equation}
This is divergent as $p_3 \to 0$, but the coefficient is $(u,v)$ independent and we obtain
\begin{equation}
I^{\text{(scheme)}}_{(u_0, v_0) \mapsto (u,v)} = \frac{1}{\ep} \left(\frac{1}{(u - 3v)} -\frac{1}{(u_0 - 3v_0)} \right) +
(\ln 2 - \gamma_E) \left(\frac{u}{u - 3v} - \frac{u_0}{u_0 - 3v_0}\right) + O(\ep),
\end{equation}
which is what we wanted to derive. This allows us to obtain $I_{2+u \epsilon \{1 + v \epsilon, 1 + v \epsilon, 1 + v \epsilon\}} $ in any $(u,v)$ scheme.
More generally, using this method we can convert a triple-$K$ integral evaluated in one scheme to its counterpart in any other scheme.

\subsubsection{Ward identities}\label{subsubsec:Ward2}

The regulated correlators satisfy the original Ward identities by construction
\be
\tilde{D} \lla \O_1(\bs{p}_1) \O_2(\bs{p}_2) \O_3(\bs{p}_3) \rra_{\text{reg}} =0, \quad 
\tilde{K}_{ij} \lla  \O_1(\bs{p}_1) \O_2(\bs{p}_2) \O_3(\bs{p}_3) \rra_{\text{reg}} =0,
\ee
where
\begin{align}
\tilde{D} &= 2 \tilde{d} - \tilde{\Delta}_t + \sum_{j=1}^3 p_j \frac{\partial}{\partial p_j} = D + (u-3 v) \ep , \label{tD} \\
\tilde{K}_{ij} &= \tilde{K}_i - \tilde{K}_j = K_{ij} -2 v \ep 
\left(\frac{1}{p_i} \frac{\partial}{\partial p_i} - \frac{1}{p_j} \frac{\partial}{\partial p_j}\right).
\end{align}
This implies that the coefficient of the leading-order divergence
is also annihilated by $K_{ij}$ and $D$, 
\begin{align}
&D \left(\sum_{w} \frac{c_{-1 + w\ep}^{(-s_{max})}}{w}\right) =0, \qquad  K_{ij}  \left(\sum_{w} \frac{c_{-1 + w\ep}^{(-s_{max})}}{w}\right) =0,  \label{d-cond1} 
\end{align}
while sub-leading coefficients satisfy inhomogeneous equations,
\begin{align}
&D \left(\sum_{w} \frac{c_{-1 + w\ep}^{(-s+1)}}{w} \right)=- (u-3 v) \left(\sum_{w} \frac{c_{-1 + w\ep}^{(-s)}}{w}\right),  \qquad s < s_{max}, \\
&K_{ij} \left(\sum_{w} \frac{c_{-1 + w\ep}^{(-s+1)}}{w} \right)= 2 v 
\left(\frac{1}{p_i} \frac{\partial}{\partial p_i} - \frac{1}{p_j} \frac{\partial}{\partial p_j}\right)\left(\sum_{w} \frac{c_{-1 + w\ep}^{(-s)}}{w}\right), 
\qquad s < s_{max},
\label{d=cond2}
\end{align}
where $s_{max}$ is the power of the most singular behaviour in $c_{-1+w \ep} \sim 1/\ep^{s_{max}}$.

Equations \eqref{d-cond1} imply in particular that if the leading divergence is nonlocal then its coefficient satisfies the 
non-anomalous Ward identities and is therefore the sought-for answer for the 3-point function. We have seen that 
the divergences are nonlocal in the cases of $(-++)$ and $(+++)$ singularities. In other words, in these cases 
it is the representation of the 3-point function in terms of triple-$K$ integral that is singular, not the correlator itself. 
To obtain the correlators it suffices to multiply the triple-$K$ integral by $\ep^{s_{max}}$ and take the limit $\ep \to 0$.
(See below \eqref{2pt_nonlocal} for the analogous discussion for 2-point functions.) On the other hand, if the leading order singularity is local or semilocal, then one needs to renomalise.  This is again exactly analogous to what we saw when we discussed 2-point functions: the solution of the non-anomalous Ward identities is (semi)-local and as such not acceptable as a 3-point function (because one can add finite local counterterms in the action and set these correlators to zero). Instead, after renormalisation one obtains renormalised correlators, which now satisfy anomalous Ward identities to which we will return in section \ref{sec:anomalies_and_betafns}. 

In the following we will organise our discussion according to the degree of singularity of the triple-$K$ integral.

\subsubsection{Triple-$K$ integrals with $1/\ep$ singularity} \label{subsubsec:1/ep}

In this case only one of the conditions \eqref{e:cond_sigma} holds. The analysis then depends on which condition this is.

\paragraph{$\bm{(+++)}$ or  $\bm{(++-)}$ singularities.}  In this case, as discussed above, the correlator can be read off from the leading-order singularity.   We will present the general case in appendix \ref{sec:gen_results} and focus our attention here on a few illustrative examples:

\bigskip

{\it \textbullet \, Example 5: $\Delta_1=\Delta_2=1/2$, $\Delta_3=1$ in $d=3$.}

\bigskip

This is an example of a $(++-)$ singularity: $\alpha=1/2$, $\beta_1=\beta_2=-1$ and $\beta_3=-1/2$ and $k_{++-} =0$. Expanding the triple-$K$ integrand we find
\[
c_{-1+(u+v)\ep}=2^{-3/2-\ep v}(p_1 p_2)^{-2+2\ep v}\Gamma^2(1-v\ep)\Gamma(-1/2+v\ep).
\]
Extracting the leading term as $\ep\tto 0$ we obtain
\[
\lla \O_{[1/2]}(\bs{p}_1)\O_{[1/2]}(\bs{p}_2)\O_{[1]}(\bs{p}_3)\rra \propto (p_1 p_2)^{-2}.
\]
One may easily verify that this 3-point function satisfies the (non-anomalous) conformal Ward identities.
This example may be realised using a free scalar $\Phi$ as $\O_{[1/2]}=\Phi$ and $\O_{[1]}= \,\lwick \Phi^2\rwick$.

\bigskip

{\it \textbullet \, Example 6: $\Delta_1=\Delta_2=\Delta_3=1$ in $d=3$.}

\bigskip

This is an example of a $(+++)$ singularity: $\alpha=1/2$, $\beta_i=-1/2$ and $k_{+++}=0$.  Expanding the triple-$K$ integrand we have
\[
c_{-1+(u+3v)\ep} = 2^{-3/2-3v\ep}\Gamma^3(1/2-v\ep)(p_1p_2p_3)^{-1+2v\ep},
\]
and extracting the leading term as $\ep\tto 0$ we obtain
\[\label{dualex1}
\lla \O_{[1]}(\bs{p}_1)\O_{[1]}(\bs{p}_2)\O_{[1]}(\bs{p}_3)\rra \propto \frac{1}{p_1 p_2 p_3}.
\]
This example may be realised using a free scalar $\Phi$ setting $\O_{[1]}=\,\lwick \Phi^2\rwick$, as in the previous example.

It is also instructive to also analyse this case in the $(1,0)$ scheme:
\begin{align}
\lla \O_{[1]}(\bs{p}_1) \O_{[1]}(\bs{p}_2) \O_{[1]}(\bs{p}_3) \rra_{\text{reg}} & = c_1 (p_1 p_2 p_3)^{- \frac{1}{2}} \int_0^\infty \D x \: x^{\frac{1}{2}+\ep} K_{\frac{1}{2}}(p_1 x) K_{\frac{1}{2}}(p_2 x) K_{\frac{1}{2}}(p_3 x),  \label{o1o1o1} 
\end{align}
The advantage of this scheme is that the index of the Bessel function does not change and since 
$ K_{\frac{1}{2}}(x) = \sqrt{\pi/2 x} \exp(-x)$ the integral is elementary leading to  
\begin{align}
\lla \O_{[1]}(\bs{p}_1) \O_{[1]}(\bs{p}_2) \O_{[1]}(\bs{p}_3) \rra_{\text{reg}} & = \frac{c_1}{p_1 p_2 p_3} \left( \frac{\pi}{2} \right)^{\frac{3}{2}} \left[ \frac{1}{\epsilon} - \ln (p_1 + p_2 + p_3) - \gamma_E + O(\epsilon) \right].
\end{align}
Thus,
\begin{equation}
\lla \O_{[1]}(\bs{p}_1) \O_{[1]}(\bs{p}_2) \O_{[1]}(\bs{p}_3) \rra \propto \frac{1}{p_1 p_2 p_3}.
\end{equation}
One may easily verify that this 3-point function satisfies the (non-anomalous) conformal Ward identities.

\paragraph{$\bm{(---)}$ singularities and new anomalies.}  In this case the divergence is ultralocal and satisfies the conformal Ward identities, as one expects on general grounds. Using (\ref{gen_exp}) and \eqref{one_w} we find the divergent terms are\footnote{When deriving this expression we can set $\tilde{\beta}_i\rightarrow\beta_i$ since the gamma functions are finite: for example, $-k_1+\beta_1 \ge -k_{---}+\beta_1 = -k_{+--}$,  but the assumed absence of a $(+--)$ singularity means that $k_{--+}$ as defined in \eqref{e:cond_sigma} is either non-integer or else a negative integer.}
\begin{align} \label{div}
&\lla \O_1(\bs{p}_1) \O_2(\bs{p}_2) \O_3(\bs{p}_3) \rra_{\text{div}} = c_{123}
\frac{\mu^{(3 v-u) \ep}}{(u-3 v) \ep}  (-1)^{k} \hspace{-1mm} \sum_{k_1+k_2+k_3 =k}\, \prod_{i=1}^3 
\frac{\Gamma(-k_i + \beta_i)} {2^{2 k_i - \beta_i+1} k_i!} \,p_i^{2 k_i},
\end{align}
where $c_{123}$ is a constant and here and in the following we have shortened $k_{---}$ to $k$.

To proceed we add a counterterm to remove the infinity and then remove the regulator to obtained the renormalised 3-point function.  The counterterm  takes form
\begin{equation} \label{e:cts1}
S_{\text{ct}} = \sum_{k_1+k_2+k_3=k} a_{k_1 k_2k_3} (\epsilon, u, v) \int \D^{d + 2 u \epsilon} \bs{x} \:\mu^{(3 v - u) \epsilon}\, \Box^{k_1} \phi_1 \Box^{k_2} \phi_2 \Box^{k_3} \phi_3 ,
\end{equation}
where the renormalisation scale $\mu$ was introduced on dimensional grounds and
\begin{equation}
a_{k_1 k_2k_3} (\epsilon, u, v)  = \frac{a_{k_1 k_2k_3}^{(-1)} (u, v)}{\ep} + a_{k_1 k_2k_3}^{(0)} (u, v). 
\end{equation}
As we shall see, the constant $a_{k_1 k_2k_3}^{(-1)} (u, v)$ is uniquely fixed by requiring the cancellation of infinities, while 
$a_{k_1 k_2k_3}^{(0)} (u, v)$ parametrises the scheme dependence.
 Note that all terms with different contraction of derivatives can be always rearranged in the form of \eqref{e:cts1}. Indeed, using integration by parts,
\begin{equation}
\int \D^{d + 2 u \epsilon} \bs{x}\,\phi_1 \partial_\mu \phi_2 \partial^\mu \phi_3 = \frac{1}{2} \int \D^{d + 2 u \epsilon} \bs{x}\left[ \Box \phi_1 \: \phi_2 \phi_3 - \phi_1 \Box \phi_2 \: \phi_3 - \phi_1 \phi_2 \Box \phi_3 \right],
\end{equation}
which can be used recursively to end up with the expression \eqref{e:cts1}. The counterterm contribution is
\begin{equation} \label{e:cts1_contrib}
\lla \O_1(\bs{p}_1) \O_2(\bs{p}_2) \O_3(\bs{p}_3) \rra_{\text{ct}} = (-1)^{k} \sum_{k_1 + k_2 + k_3 = k} a_{k_1 k_2 k_3} p_1^{2 k_1} p_2^{2 k_2} p_3^{2 k_3} \mu^{(3 v - u) \epsilon}.
\end{equation}
where $k_1+k_2+k_3 = k$ (we assume that all three operators are pairwise different -- otherwise there are additional symmetry factors). Thus with appropriate choice of the coefficients $a_{k_1k_2k_3}$ we may cancel the divergence \eqref{div}
in the 3-point function.  We then define the renormalised correlator as
\be
\lla \O_1(\bs{p}_1) \O_2(\bs{p}_2) \O_3(\bs{p}_3) \rra_{\text{ren}} =
\lim_{\ep \to 0} 
\big[\lla \O_1(\bs{p}_1) \O_2(\bs{p}_2) \O_3(\bs{p}_3) \rra_{\text{reg}} +\lla \O_1(\bs{p}_1) \O_2(\bs{p}_2) \O_3(\bs{p}_3) \rra_{\text{ct}}\big].
\ee

This renormalised correlator depends on the scale $\mu$:
\begin{align} \label{muOOO1}
\mu \frac{\partial}{\partial \mu}  \lla \O_1(\bs{p}_1) \O_2(\bs{p}_2) \O_3(\bs{p}_3) \rra_{\text{ren}} &= (-1)^{k}(3 v-u) \sum_{k_1 + k_2 + k_3 = k}  a_{k_1 k_2 k_3}^{(-1)} p_1^{2 k_1} p_2^{2 k_2} p_3^{2 k_3} \nn\\
&=   (-1)^{k+1}c_{123}\hspace{-1mm}\sum_{{k_1+k_2+k_3 =k}} \, \prod_{i=1}^3 
\frac{\Gamma(-k_i + \beta_i)} {2^{2 k_i - \beta_i+1} k_i!} p_i^{2 k_i}
\end{align}
where in the first equality we used the fact that the regulated 3-point function does not depend on $\mu$, and in the second the fact that the counterterm cancels the infinity in \eqref{div}. This implies that there is a new conformal anomaly $\mathcal{A}_{123}$ associated with this 3-point function. 

The existence of the anomaly implies that the generating functional of correlators $W[\phi_i]$ depends on the mass scale $\mu$,
\begin{equation} \label{anomaly}
\mu \frac{\partial}{\partial \mu} W = A.
\end{equation}
Indeed, differentiating \eqref{anomaly} with respect to $\phi_1, \phi_2$ and $\phi_3$ and comparing with \eqref{muOOO1} we find
\begin{equation}
A =    \int \D^d \bs{x}  \sum_{{k_1+k_2+k_3 =k}}{\cal A}_{k_1k_2k_3} \Box^{k_1} \phi_1 \Box^{k_2} \phi_2 \Box^{k_3} \phi_3,
\end{equation}
where 
\begin{equation} \label{an_3pt}
{\cal A}_{k_1k_2k_3} = c_{123}  \prod_{i=1}^3 
\frac{\Gamma(-k_i + \beta_i)} {2^{2 k_i - \beta_i+1} k_i!}
\end{equation}
and the ratio ${\cal A}_{k_1k_2k_3}/c_{123}$ is universal. 

One may integrate the anomaly equation \eqref{muOOO1} to obtain
\begin{align}
\lla \O_1(\bs{p}_1) \O_2(\bs{p}_2) \O_3(\bs{p}_3) \rra_{\text{ren}} = & \sum_{k_1+k_2+k_3 =k} p_1^{2 k_1} p_2^{2 k_2} p_3^{2 k_3}
(-1)^{k}  {\cal A}_{k_1k_2k_3}  
 \ln \frac{p_1 +p_2 +p_3}{\mu}  \nonumber \\
&\qquad + p_3^{\Delta_t-2d} f(\frac{p_1}{p_3}, \frac{p_2}{p_3}),
\end{align}
where $\Delta_t = \sum_j\Delta_j$ and $f(x,y)$ is an arbitrary function of two variables (which is of course uniquely fixed by the conformal Ward identities).
The argument of logarithm must be linear in momenta and changing the specific combination amounts to redefining $f(x,y)$.
We thus conclude that conformal anomalies lead to terms linear in $\ln p_i$.

\bigskip

{\it \textbullet \, Example 7: $\Delta_1=\Delta_2=\Delta_3=2$ in $d=3$.}

\bigskip

This example is closely related to the example of three operators of dimension one in $d=3$ we discussed earlier.
The correlator in the $(1,0)$-scheme is given by 
\begin{align}
\lla \O_{[2]}(\bs{p}_1) \O_{[2]}(\bs{p}_2) \O_{[2]}(\bs{p}_3) \rra_{\text{reg}} & = -c_{222} (p_1 p_2 p_3)^{\frac{1}{2}} \int_0^\infty \D x \: x^{\frac{1}{2} + \ep} K_{\frac{1}{2}}(p_1 x) K_{\frac{1}{2}}(p_2 x) K_{\frac{1}{2}}(p_3 x), 
\end{align}
where the overall minus is for later convenience.
Notice that this is the same triple-$K$ integral that appeared in (\ref{o1o1o1}). Nevertheless, we will deal with the divergence in a very different way.  The regulated correlator is given by
\begin{align}
\lla \O_{[2]}(\bs{p}_1) \O_{[2]}(\bs{p}_2) \O_{[2]}(\bs{p}_3) \rra_{\text{reg}} &=c_{222}
 \left( \frac{\pi}{2} \right)^{\frac{3}{2}} \left[ -\frac{1}{\epsilon} +\ln (p_1 + p_2 + p_3) + \gamma_E + O(\epsilon) \right], \label{e:O2a}  
\end{align}
In this case the divergence is local and it can be cancel by a local counterterm
\begin{equation}
S_{{\rm ct}} = a(\ep)  \int \D^{3 + 2 \epsilon} \bs{x} \: \phi^3 \mu^{-  \epsilon},
\end{equation}
where $\phi$ is the source of $\O_2$.
Choosing 
\begin{equation}
a(\ep) = \frac{1}{6}c_{222} \left( \frac{\pi}{2} \right)^{\frac{3}{2}} \left(\frac{1}{\ep} + a_0\right),
\end{equation}
where $a_0$ is an arbitrary constant parametrising the scheme dependence, we find for the renormalised correlator
\begin{align}
\lla \O_{[2]}(\bs{p}_1) \O_{[2]}(\bs{p}_2) \O_{[2]}(\bs{p}_3) \rra_{\text{ren}} &=c_{222}
 \left( \frac{\pi}{2} \right)^{\frac{3}{2}} \left[ \ln\Big( \frac{p_1 + p_2 + p_3}{\mu}\Big) + a_1 \right] \label{e:O2b}  
\end{align}
where $a_1=a_0 +\gamma_E$. 

The renormalised correlator correlator now depends on a scale,
\begin{equation}
\mu \frac{\partial}{\partial \mu}  \lla \O_{[2]}(\bs{p}_1) \O_{[2]}(\bs{p}_2) \O_{[2]}(\bs{p}_3) \rra_{\text{ren}} = -c_{222}
 \left( \frac{\pi}{2} \right)^{\frac{3}{2}},
 \end{equation}
 so ${\cal A}_{222} = -c_{222}
 \left( \frac{\pi}{2} \right)^{\frac{3}{2}}$ in agreement with \eqref{an_3pt}. Correspondingly, there is a new conformal anomaly
 \begin{equation}
 \langle T \rangle = \frac{1}{3!} {\cal A}_{222}\, \phi^3,
 \end{equation} 
 and the ratio ${\cal A}_{222}/c_{222}$ indeed does not renomalise.

\paragraph{$\bm{(--+)}$ singularities and beta functions.} In this case the divergence is semilocal and satisfies the conformal Ward identities, as one expects on general grounds. The analysis is identical for the three cases, $(--+)$, $(-+-)$ and $(+--)$, and for concreteness we discuss the case of a $(--+)$ singularity.
Using (\ref{gen_exp}) and \eqref{one_w} we find the divergent terms are\footnote{In deriving this expression we can set $\tilde{\beta}_i\rightarrow\beta_i$ since the gamma functions are finite: for example, $-k_3-\beta_3 \ge -k_{--+}-\beta_3 = -k_{---}$, but the assumed absence of a $(---)$ singularity means that $k_{---}$ as defined in \eqref{e:cond_sigma} is either non-integer or else a negative integer.}
\begin{align} \label{div--+}
&\lla \O_1(\bs{p}_1) \O_2(\bs{p}_2) \O_3(\bs{p}_3) \rra_{\text{div}} = c_{123} \frac{\mu^{(v-u) \ep}}{(u-v) \ep}   \nonumber \\
&\qquad\qquad \times(-1)^{k}  \sum_{{k_1+k_2+k_3 =k}} 
\frac{\Gamma(-k_1 + \beta_1)}{2^{2 k_1 - \beta_1+1} k_1!}
 \frac{\Gamma(-k_2 + \beta_2)}{2^{2 k_2 - \beta_2+1} k_2!}
 \frac{\Gamma(-k_3 - \beta_3)}{2^{2 k_3 + \beta_3+1} k_3!}
 p_1^{2 k_1} p_2^{2 k_2} p_3^{2 \beta_3 +2 k_3}, 
\end{align}
where $k = k_{--+}$ denotes the integer appearing in the defining condition \eqref{e:cond_sigma}. Since this expression is analytic in $p_1^2$ and $p_2^2$ it is semilocal. 

When the dimensions of operators satisfy the $(--+)$ condition there is a possible counterterm given by
\begin{equation} \label{e:cts2}
S_{\text{ct}} = \sum_{k_1+k_2+k_3 =k} a_{k_1 k_2 k_3}(\ep, u,v) \int \D^{d+2 u \ep} \bs{x} \: \mu^{(v-u)\epsilon}\, \Box^{k_1} \phi_1 \Box^{k_2} \phi_2 \Box^{k_3} \O_3 ,
\end{equation}
where 
\begin{equation}
a_{k_1 k_2 k_3}(\ep, u,v) = \frac{a_{k_1 k_2 k_3}^{(-1)}(u,v)}{\ep} + a^{(0)}_{k_1 k_2 k_3}(u,v).
\end{equation}
The coefficient $a^{(0)}_{k_1 k_2 k_3}(u,v)$ parametrises the (finite) scheme-dependent contribution of this counterterm. The counterterm contribution reads
\begin{align} \label{e:cts2_contrib}
&\lla \O_1(\bs{p}_1) \O_2(\bs{p}_2) \O_3(\bs{p}_3) \rra_{\text{ct}} \nn\\[1ex]
&\qquad = (-1)^{k + 1} \sum_{k_1 + k_2 + k_3 = k} a_{k_1 k_2 k_3} p_1^{2 k_1} p_2^{2 k_2} p_3^{2 k_3} \mu^{(v - u) \epsilon} \lla \O_3(\bs{p}_3) \O_3(-\bs{p}_3) \rra_{\text{reg}} \nn\\
& \qquad
= (-1)^{k + 1}  \sum_{k_1 + k_2 + k_3 = k} a_{k_1 k_2 k_3} p_1^{2 k_1} p_2^{2 k_2} p_3^{2 k_3} (c_{\Delta_3} p_3^{2 \Delta_3 - d + 2 v \epsilon})  \mu^{(v - u) \epsilon},
\end{align}
where $c_{\Delta_3}$ is the normalisation of the 2-point function (see \eqref{sec2pt:sol_reg}). Recalling that $\beta_3 = \Delta_3-d/2$ we see that the momentum dependence of (\ref{e:cts2_contrib}) exactly matches that of (\ref{e:cts2}) and therefore we may cancel the infinity by an appropriate choice of $a_{k_1 k_2 k_3}$. The renormalised correlator is then
\be
\lla \O_1(\bs{p}_1) \O_2(\bs{p}_2) \O_3(\bs{p}_3) \rra_{\text{ren}} =
\lim_{\ep \to 0} 
\big[\lla \O_1(\bs{p}_1) \O_2(\bs{p}_2) \O_3(\bs{p}_3) \rra_{\text{reg}} +\lla \O_1(\bs{p}_1) \O_2(\bs{p}_2) \O_3(\bs{p}_3) \rra_{\text{ct}}\big].
\ee  

The renormalised correlator depends on the scale $\mu$,
\begin{equation} \label{muOOO}
\mu \frac{\partial}{\partial \mu}  \lla \O_1(\bs{p}_1) \O_2(\bs{p}_2) \O_3(\bs{p}_3) \rra_{\text{ren}} = (v-u) 
  \sum_{k_1 + k_2 + k_3 = k} (-1)^{k + 1} a^{(-1)}_{k_1 k_2 k_3} p_1^{2 k_1} p_2^{2 k_2} p_3^{2 k_3} (c_{\Delta_3} p_3^{2 \Delta_3 - d}),
\end{equation}
where we used the fact that the regulated 3-point function does not depend on $\mu$.
To understand this result, note that the counterterm amounts to a renormalisation of the source that couples to $\O_3$.  The source $\phi_3$ is in fact the {\it renormalised} coupling, since functionally differentiating with respect to it yields the renormalised correlator, while the bare source is
\begin{equation}
\phi_3^{\rm{bare}}\equiv\phi_3 + \sum_{k_1+k_2+k_3 =k} a_{k_1 k_2 k_3}(\ep, u,v) (-1)^{k_3} \Box^{k_3} (\Box^{k_1} \phi_1 \Box^{k_2} \phi_2) \mu^{(v-u)\epsilon}.
\end{equation}
Inverting perturbatively, to quadratic order we find
\begin{equation}
\phi_3=\phi_3^{\rm{bare}} - \sum_{k_1+k_2+k_3 =k} a_{k_1 k_2 k_3}(\ep, u,v) (-1)^{k_3} \Box^{k_3} (\Box^{k_1} \phi_1^{\rm{bare}} \Box^{k_2} \phi_2^{\rm{bare}}) \mu^{(v-u)\epsilon},
\end{equation}
where we have defined $\phi_1^{\rm{bare}}=\phi_1$ and $\phi_2^{\rm{bare}}=\phi_2$ since these sources are unrenormalised.
As the bare couplings are independent of the renormalisation scale $\mu$, we then obtain the beta function
\begin{equation} \label{beta}
\beta_{\phi_3}\equiv \lim_{\ep \to 0} \mu \frac{\partial \phi_3}{\partial \mu} = - (v-u) \sum_{k_1+k_2+k_3 =k} a^{(-1)}_{k_1 k_2 k_3}(u,v) (-1)^{k_3} \Box^{k_3} (\Box^{k_1} \phi_1 \Box^{k_2} \phi_2). 
\end{equation}
Comparing \eqref{muOOO} and \eqref{beta} we find\footnote{For ease of presentation we assume $\Delta_3 \neq d/2 +k$.} 
\begin{equation} \label{mudmu3pt}
\mu \frac{\partial}{\partial \mu}  \lla \O_1(\bs{p}_1) \O_2(\bs{p}_2) \O_3(\bs{p}_3) \rra_{\text{ren}} =  \frac{\partial^2 \beta_{\phi_3}}{\partial\phi_1 \partial \phi_2} \lla \O_3(\bs{p}_3) \O_3(-\bs{p}_3) \rra_{\text{ren}}. 
\end{equation}

We thus find that in this case the correlators depend on $\mu$ through the implicit $\mu$-dependence of the renormalised source $\phi_3$. In terms of the generating function $W$ we now have
\begin{equation} \label{Dmu}
\mu \frac{\D}{\D \mu} W[\phi_i]=0,
\end{equation}
where the total variation is given by
\begin{equation}
\mu \frac{\D}{\D \mu} = \mu \frac{\partial}{\partial \mu} + \sum_i \int \D^d \bs{x}\, \beta_{\phi_i}\frac{\delta}{\delta \phi_i(\bs{x})}.
\end{equation}
Indeed, differentiating (\ref{Dmu}) with respect to the renormalised sources we recover \eqref{mudmu3pt}.

Integrating \eqref{mudmu3pt} we find that the renormalised correlator will contains terms proportional to either $\ln p_i$, if $\Delta_3 \neq d/2 +k$, or $\ln p_i \ln p_j$ terms if $\Delta_3 = d/2 +k$.  Thus, single logs are not only associated with conformal anomalies but also with beta functions and (perhaps more surprisingly) double logs may also appear in conformal correlators. In the case of double logs, one of the logs is due to the conformal anomaly in 2-point functions and the other is due to the beta function. 

\bigskip

{\it \textbullet \, Example 8: $\Delta_1=\Delta_2=\Delta_3=3$ in $d=3$.}\label{example8}

\bigskip

We will now illustrate this case by discussing the computation of the 3-point function of three marginal operators in $d=3$.
In this case, $\alpha=1/2$, $\beta_1=\beta_2=\beta_3=3/2$ and the $(--+)$, $(-+-)$, $(+--)$ conditions are satisfied with $k_{--+}=k_{-+-}=k_{+--}=0$. 

The bare 3-point function,
\begin{equation} 
\lla \O_{[3]}(\bs{p}_1) \O_{[3]}(\bs{p}_2) \O_{[3]}(\bs{p}_3) \rra_{\text{bare}} = c_{333}\, I_{\frac{1}{2} \{ \frac{3}{2} \frac{3}{2} \frac{3}{2} \}}
\end{equation}
is divergent. As we are in $d=3$ it is most convenient to work in the $(1,0)$-scheme (since then the integral is elementary).
Extracting the divergences as discussed earlier we obtain
\begin{equation} \label{e:O3a}
I_{\frac{1}{2} + \ep \{ \frac{3}{2}, \frac{3}{2}, \frac{3}{2}\}} = \left( \frac{\pi}{2} \right)^{\frac{3}{2}} \frac{p_1^3 + p_2^3 + p_3^3}{3 \ep} + O(\ep^0).
\end{equation}
This divergence is semilocal because it is a sum of terms each of which is analytic in two momenta and non-analytic in one.

To remove this divergence we add the counterterm,
\begin{equation} \label{e:Sct1}
S_{{\rm ct}} = a(\epsilon) \int \D^{3 + 2 \epsilon} \bs{x} \: \mu^{-\epsilon}\phi^2 \O .
\end{equation}
This counterterm does not contribute to 2-point functions and 
its contribution to the 3-point function reads
\begin{align}
\lla \O_{[3]}(\bs{p}_1) \O_{[3]}(\bs{p}_2) \O_{[3]}(\bs{p}_3) \rra_{\text{ct}} & = - 2 a \mu^{-\epsilon} \left[ \lla \O_{[3]}(\bs{p}_1) \O_{[3]}(-\bs{p}_1) \rra_{\text{reg}} + 2 \text{ perms.} \right] \nn\\
& = - 2 a\, c_{3} \mu^{-\epsilon} ( p_1^{3} + p_2^{3} + p_3^{3} ).  \label{3ptcount}
\end{align}
Therefore, the counterterm removes the divergence from the 3-point function provided 
\begin{equation}
a(\ep) = \frac{c_{333}}{c_{3}} \left( \frac{\pi}{2} \right)^{\frac{3}{2}} \left[ \frac{1}{6 \ep} + a^{(0)}+ O(\epsilon^0)\right],
\end{equation}
where $a^{(0)}$ is an undetermined $\epsilon$-independent constant that parametrises scheme dependence.
The renormalised source $\phi$ is related to the bare source via
\begin{equation}
\phi^{\rm{bare}} = \phi + \phi^2 \mu^{-\epsilon} \frac{c_{333}}{c_{3}} \left( \frac{\pi}{2} \right)^{\frac{3}{2}} \left[ \frac{1}{6 \ep} + a^{(0)}+ O(\epsilon^0)\right],
\end{equation}
which after inverting leads to a beta function
\begin{equation}
\beta_{\phi} \equiv \lim_{\ep \to 0} \mu \frac{\partial \phi}{\partial \mu} = 
 \frac{1}{6}  \frac{c_{333}}{c_{3}}  \left(\frac{\pi}{2} \right)^{\frac{3}{2}}  \phi^2. 
\end{equation}

The triple-$K$ integral $I_{\frac{1}{2} \{ \frac{3}{2} \frac{3}{2} \frac{3}{2} \}}$ can easily be calculated in the $(1,0)$-regularisation scheme and reads
\begin{align}
I_{\frac{1}{2} + \ep \{ \frac{3}{2} \frac{3}{2} \frac{3}{2} \}} & = \frac{1}{3} \left( \frac{\pi}{2} \right)^{\frac{3}{2}} \Big[ \frac{p_1^3 + p_2^3 + p_3^3}{\ep}  - \: p_1 p_2 p_3 + (p_1^2 p_2 + p_2^2 p_1 + p_1^2 p_3 + p_3^2 p_1 + p_2^2 p_3 + p_3^2 p_2) \nn\\
& \qquad\qquad\qquad  - \: ( p_1^3 + p_2^3 + p_3^3 ) \ln (p_1 + p_2 + p_3)  + \tfrac{4}{3} ( p_1^3 + p_2^3 + p_3^3 ) \Big].
\end{align}
Adding the contribution from the counterterm \eqref{3ptcount} and sending $\ep \to 0$ we obtain the renormalised correlator,
\begin{align}\label{rencorr333}
 \lla \O_{[3]}(\bs{p}_1) \O_{[3]}(\bs{p}_2) \O_{[3]}(\bs{p}_3) \rra_{\text{ren}} &=  \frac{1}{3} \left( \frac{\pi}{2} \right)^{\frac{3}{2}} c_{333} \Big[- p_1 p_2 p_3 + (p_1^2 p_2+\text{5 perms.})  \nn\\
& \ \  -  ( p_1^3 + p_2^3 + p_3^3 ) \ln \frac{p_1 + p_2 + p_3}{\mu}  - 6 a^{(0)}  (p_1^3 + p_2^3 + p_3^3) \Big].
\end{align}
Note that changing the renormalisation scale $\mu$ amounts to changing $a^{(0)}$, \textit{i.e.}, the scheme-dependent part of the correlator.
Acting with $\mu (\partial/\partial \mu)$ we find
\begin{align}\label{CSeqn333}
\mu \frac{\partial}{\partial \mu} \lla \O_{[3]}(\bs{p}_1) \O_{[3]}(\bs{p}_2) \O_{[3]}(\bs{p}_3) \rra_{\text{ren}} & =  \frac{1}{3} \left( \frac{\pi}{2} \right)^{\frac{3}{2}} c_{333} (p_1^3 + p_2^3 + p_3^3) \nn \\
& = \frac{\partial^2 \beta_{\phi_0}}{\partial \phi_0^2}  \big(\lla \O_{[3]}(\bs{p}_1) \O_{[3]}(-\bs{p}_1) \rra_{\text{ren}} + \text{ perms.}\big),
\end{align}
confirming our earlier general analysis.

\subsubsection{Triple-$K$ integrals with higher-order singularities} \label{subsubsec:higherorder}

Higher-order singularities are associated with multiple conditions holding simultaneously. The analysis of the general case is analogous to what we have discussed already: if there are $(---)$ singularities there are new conformal anomalies while if there are $(--+)$ singularities we have beta functions. The renormalised correlators in such cases depend on the renormalisation scale $\mu$.  The form of this $\mu$-dependence may be found  by functionally differentiating 
\begin{equation} \label{anomaly2}
\mu \frac{\D}{\D \mu} W[\phi_i] = A
\end{equation}
with respect to the renormalised sources $\phi_i$ and noting that
\begin{equation}
\mu \frac{\D}{\D \mu} = \mu \frac{\partial}{\partial \mu} + \sum_i \int \D^d \bs{x}\, \beta_{\phi_i}\frac{\delta}{\delta \phi_i(\bs{x})}.
\end{equation}
If there are additional singularities of type $(++-)$ and/or $(+++)$ then one needs to multiply the triple-$K$ integral by an appropriate power of $\ep$ before removing the regulator. The classification and analysis of all possible cases is discussed in appendix \ref{sec:gen_results}. Here we will discuss two examples that illustrate the general case.

\bigskip

{\it \textbullet \, Example 9: $\Delta_1=4$, $\Delta_2=\Delta_3=3$ in $d=4$.}

\bigskip

In this case $\alpha = 1$, $\beta_1=2$, $\beta_2= \beta_3=1$ and thus both a $(---)$ condition (with $k_{---}=1$) and  $(-+-)$ and $(--+)$ conditions (with $k_{-+-}=k_{--+}=0$) hold simultaneously. The bare 3-point function is given by
\begin{equation}
\lla \O_{[4]}(\bs{p}_1)\O_{[3]}(\bs{p}_2)\O_{[3]}(\bs{p}_3)\rra_{{\rm bare}} = c_{433}\, I_{1,\{211\}}.
\end{equation}
We have already discussed the computation of the divergent terms at the end of section\ \ref{subsubsec:reg} (see example 4 on page \pageref{Example4label}), where we saw that the regulated triple-$K$ integral,
$ I_{1+u \ep,\{2+v \ep,1+v \ep,1+v \ep\}}$, diverges as $\ep^{-2}$.  This leading order singularity is ultralocal while the subleading singularity at order $\ep^{-1}$ is semilocal.

To cancel the infinities we introduce the counterterm action 
\[
\label{e:Sct3}
S_{{\rm ct}}  = \int \D^{d+2u\ep} \bs{x} \big[a_0 \mu^{(v-u)\ep}\phi_{[0]}\phi_{[1]} {\O}_{[3]} +a_1\mu^{(3v-u)\ep} \phi_{[0]}\phi_{[1]}\Box\phi_{[1]}+a_2\mu^{(3v-u)\ep} {\phi_{[1]}}^2\Box\phi_{[0]}\big],
\]
where $\phi_{[0]}$ is the source of $\O_{[4]}$ and $\phi_{[1]}$ is the source of $\O_{[3]}$.  (To reduce clutter here we have used the bare rather than regulated dimensions in our notation, writing $\phi_{[0]}$ as shorthand for $\phi_{[0+(u-v)\ep]}$, {\it etc.})
This generates the following contribution to the 3-point function,
\begin{align} \label{433ct}
&\lla \O_{[4]}(\bs{p}_1)\O_{[3]}(\bs{p}_2)\O_{[3]}(\bs{p}_3)\rra_{{\rm ct}} \nn\\[1ex] &\qquad = -a_1 (p_2^2+p_3^2)\mu^{(3v-u)\ep}-2a_2 p_1^2 \mu^{(3v-u)\ep} 
\nn\\&\qquad\quad 
-a_0 \mu^{(v-u)\ep}[\lla \O_{[3]}(p_2)\O_{[3]}(-p_2)\rra_{{\rm reg}}+\lla\O_{[3]}(p_3)\O_{[3]}(-p_3)\rra_{{\rm reg}}], 
\end{align}
where $a_0$, $a_1$ and $a_2$ have series expansions in $\ep$, and the regulated 2-point function is
\[\label{433reg2pt}
\lla \O_{[3]}(\bs{p})\O_{[3]}(-\bs{p})\rra_{{\rm reg}} = \Big(\frac{c_3^{(-1)}}{\ep}+c_3^{(0)} + O(\ep)\Big)p^{2+2v\ep}.
\]
When \eqref{433ct} is expanded in $\ep$, the divergent terms must match $I^{\rm{div}}_{1+u\ep,\{2+v\ep,1+v\ep,1+v\ep\}}$ as evaluated in \eqref{433div}.   This procedure fixes the coefficients in the counterterm action as
\begin{align}\label{a0eqn}
\frac{a_0}{c_{433}} &= \frac{1}{2v(u-v)c_3^{(-1)}\ep}+a_0^{(0)} +O(\ep),\\ \label{a1eqn}
\frac{a_1}{c_{433}} &= -\frac{1}{2v(u-3v)\ep^2}+\Big(-a_0^{(0)}c_3^{(-1)}+\frac{v-u(1-2\gamma_E+\ln 4)}{2(u-v)(u-3v)}\Big)\frac{1}{\ep}+ a_1^{(0)}+ O(\ep),\\ \label{a2eqn}
\frac{a_2}{c_{433}} &=-\frac{1}{4(u-3v)\ep}+a_2^{(0)} + O(\ep),
\end{align}
where for simplicity we set $c_3^{(0)}=0$. The constants $a_0^{(0)}, a_1^{(0)}$ and $a_2^{(0)}$ parametrise the scheme dependence.
Due to the $a_0$ counterterms, the renormalised source $\phi_{[1]}$ is related to the bare source $\phi_{[1]}^{\rm{bare}}$ by 
\begin{equation}
\phi_{[1]}^{\rm{bare}} = \phi_{[1]} + a_0 \mu^{(v-u)\ep} \phi_{[0]} \phi_{[1]} ,
\end{equation}
leading to a beta function
\begin{equation}\label{beta433}
\beta_{\phi_{[1]}} = \lim_{\ep \to 0} \mu \frac{\partial \phi_{[1]}}{\partial \mu} = \frac{c_{433}}{2 v c_3^{(-1)}} \phi_{[0]} \phi_{[1]}.
\end{equation}

The triple-$K$ integral $ I_{1+u \ep,\{2+v \ep,1+v \ep,1+v \ep\}}$ can be computed exactly using recursion relations \cite{Bzowski:2013sza,integrals}, and after adding the contribution of the counterterm contribution, the limit $\epsilon \rightarrow 0$ may be taken leading to the renormalised correlator
\begin{align} \label{e:O433}
& \lla \O_{[4]}(\bs{p}_1) \O_{[3]}(\bs{p}_2) \O_{[3]}(\bs{p}_3) \rra_{{\rm ren}} = c_{433} \Big( 2 - p_1 \frac{\partial}{\partial p_1} \Big) 
\Big(\frac{1}{4} J^2 I_{1\{000\}} \Big)\nn\\
& \qquad\qquad\qquad + \: \frac{c_{433}}{8} \left[ (p_2^2 - p_3^2) \ln \frac{p_1^2}{\mu^2} \Big( \ln \frac{p_3^2}{\mu^2} - \ln \frac{p_2^2}{\mu^2} \Big) - (p_2^2 + p_3^2) \ln \frac{p_2^2}{\mu^2} \ln \frac{p_3^2}{\mu^2} \right.\nn\\
& \qquad\qquad\qquad\qquad\qquad \left. +(p_1^2 - p_2^2) \ln \frac{p_3^2}{\mu^2} + (p_1^2 - p_3^2) \ln \frac{p_2^2}{\mu^2} + p_1^2 \right] \nn\\
& \qquad\qquad\qquad + \: a_0' \Big( p_2^2 \ln \frac{p_2^2}{\mu^2} + p_3^2 \ln \frac{p_3^2}{\mu^2} \Big)  + \: a'_1 ( p_2^2 + p_3^2 ) + a'_2 p_1^2.
\end{align}
Here, $I_{1\{000\}}$ and $J^2$ are given in  \eqref{e:I1000} and \eqref{defs}, and $a'_0$, $a'_1$ and $a'_2$ are scheme-dependent constants linearly related to $a_0^{(0)}$, $a_1^{(0)}$ and $a_2^{(0)}$.  (In fact, as we will see later in section \ref{sec:anomalies_and_betafns}, the special conformal Ward identities further fix $a'_2+a'_0 = -c_{433}/2$.)
Acting with $\mu (\partial/\partial\mu)$, we find
\begin{align}\label{433muderiv}
&\mu\frac{\partial}{\partial \mu}
\lla \O_{[4]}(\bs{p}_1) \O_{[3]}(\bs{p}_2) \O_{[3]}(\bs{p}_3) \rra_{{\rm ren}} 
\nn\\&\qquad 
=
\frac{c_{433}}{2}\Big({-}p_1^2+\frac{1}{2}(p_2^2+p_3^2)+p_2^2\ln\frac{p_2^2}{\mu^2}+p_3^2\ln\frac{p_3^2}{\mu^2}\Big)
-2a'_0(p_2^2+p_3^2)\nn\\&\qquad
=  \frac{\partial^2\beta_{\phi_{[1]}}}{\partial\phi_{[0]}\partial\phi_{[1]}} \Big(
\lla \O_{[3]}(\bs{p}_2)\O_{[3]}(-\bs{p}_2)\rra_{{\rm ren}}+\lla\O_{[3]}(\bs{p}_3)\O_{[3]}(-\bs{p}_3)\rra_{{\rm ren}}\Big)+\mathcal{A}_{433},
\end{align}
where the anomaly
\[\label{433anomaly}
\mathcal{A}_{433} = -\frac{c_{433}}{2} p_1^2 + \big(\frac{c_{433}}{4}-2a'_0\big)(p_2^2+p_3^2).
\]
In this case, only the coefficient of $p_1^2$ (divided by the overall normalisation of the 3-point function $c_{433}$) is physically meaningful: the remainder of the anomaly is scheme-dependent and can be adjusted by adding finite counterterms to change $a'_0$.

Note that the dimensions of the operators $\O_{[3]}$  and $\O_{[4]}$ are such that $f(\phi_{[0]}) \phi_{[1]}\Box\phi_{[1]}$
has dimension four for any function $f(\phi_{[0]})$ of the dimensionless sources $\phi_{[0]}$.  As discussed in section \ref{sec:2pt},
the 2-point function of the operator $\O_{[3]}$ also requires renormalisation and a counterterm of the form 
$S_{{\rm ct}}  \propto  \int \D^{4} \bs{x} \phi_{[1]}\Box\phi_{[1]}$. This counterterm and the second counterterm in \eqref{e:Sct3} maybe considered as the expansion of $f(\phi_{[0]})$ around $\phi_{[0]} \approx 0$.  Similarly, the conformal anomaly may contain a term proportional to $g(\phi_{[0]}) \phi_{[1]}\Box\phi_{[1]}$ for some function $g$ of $\phi_{[0]}$, and we have found that 
 the part associated with the 3-point function is scheme dependent.

\bigskip

{\it\textbullet\, Example 10: $\Delta_1=4$ and $\Delta_2=\Delta_3=2$ in $d=4$.}

\bigskip

In this case $\alpha=1$, $\beta_1=2$, $\beta_2=\beta_3 = 0$ and so
we have $(-++)$, $(--+)$, $(-+-)$ and $(---)$ singularities with $k_{-++}=k_{--+}=k_{-+-}=k_{---}=0$.

The divergent part of the regulated triple-$K$ integral is 
\begin{align}
I^{\rm{div}}_{\tilde{\alpha},\{\tilde{\beta}\}}&=\sum_w \frac{c_{-1+w\ep}}{w\ep}+O(\ep^0)
= \frac{c_{-1+(u+v)\ep}}{(u+v)\ep}+\frac{c_{-1+(u-v)\ep}}{(u-v)\ep}+\frac{c_{-1+(u-3v)\ep}}{(u-3v)\ep}+O(\ep^0),
\end{align}
where
\begin{align}
c_{-1+(u+v)\ep}&=2^{-1-v\ep}\Gamma^2(-v\ep)\Gamma(2+v\ep)(p_2p_3)^{2v\ep},\nn\\
c_{-1+(u-v)\ep}&=2^{-1+v\ep}\Gamma(-v\ep)\Gamma(v\ep)\Gamma(2+v\ep)(p_2^{2v\ep}+p_3^{2v\ep}),\nn\\
c_{-1+(u-3v)\ep} &= 2^{-1+3v\ep}\Gamma^2(v\ep)\Gamma(2+v\ep).
\end{align}
Expanding out, we find 
\begin{align}
\frac{c_{-1+(u+v)\ep}}{(u+v)\ep} &= \frac{1}{2v^2(u+v)\ep^3}+\frac{1+\gamma_E-\ln 2+2\ln(p_2p_3)}{2v(u+v)\ep^2}\nn\\&\quad
+\frac{1}{4(u+v)\ep}\Big[
\frac{\pi^2}{2}-1+\big(1+\gamma_E-\ln 2+2\ln (p_2p_3)\big)^2\Big]
+O(\ep^0), \nn\\[1ex]
\frac{c_{-1+(u-v)\ep}}{(u-v)\ep} &=  -\frac{1}{(u-v)v^2\ep^3}-\frac{(1-\gamma_E+\ln (2p_2p_3))}{(u-v)v\ep^2} \nn\\&\quad
-\frac{1}{2(u-v)\ep}\Big[
\frac{\pi^2}{2}-1+\big(1-\gamma_E+\ln(2p_2p_3)\big)^2+\ln^2(p_2/p_3)\Big]+O(\ep^0), \nn\\[1ex]
\frac{c_{-1+(u-3v)\ep}}{(u-3v)\ep} &= \frac{1}{2v^2(u-3v)\ep^3}+\frac{1-3\gamma_E+3\ln 2}{2v(u-3v)\ep^2} \nn\\&\quad
+\frac{1}{4(u-3v)\ep}\Big[
\frac{\pi^2}{2}-1+\big(1-3\gamma_E+3\ln 2\big)^2\Big]
+O(\ep^0).
\end{align}
The leading $\ep^{-3}$ divergence of 
$I^{\rm{div}}_{\tilde{\alpha},\{\tilde{\beta}\}}$ is therefore ultralocal while the subleading $\ep^{-2}$ divergence is semilocal.   Only the sub-subleading order $\ep^{-1}$ divergence is nonlocal, and it is this that is proportional to the renormalised correlator once the $\ep^{-3}$ and $\ep^{-2}$ divergences have been removed. 
We therefore write
\[
\lla \O_{[4]}(\bs{p}_1)\O_{[2]}(\bs{p}_2)\O_{[2]}(\bs{p}_3)\rra_{\rm{ren}} = \lim_{\ep\rightarrow 0}\Big[2\ep(u+v)\,c_{422}\, 
I^{\rm{div}}_{\tilde{\alpha},\{\tilde{\beta}\}}
+\lla \O_{[4]}(\bs{p}_1)\O_{[2]}(\bs{p}_2)\O_{[2]}(\bs{p}_3)\rra 
_{\rm{ct}}\Big],
\]
where $c_{422}$ is a theory-dependent constant that is independent of $\ep$ and represents the overall normalisation of the 3-point function.  (The additional factor of $2(u+v)$ is purely for convenience.)
The counterterm contribution follows from the action
\[\label{ct422}
S_{\rm{ct}} = \int \D^{4+2u\ep} x \sqrt{g}\big[a_0\mu^{(3v-u)\ep} \phi_{[0]}\phi_{[2]}^2+
a_1 \mu^{(v-u)\ep}\phi_{[0]}\phi_{[2]}\O_{[2]}\big],
\]
namely,
\begin{align}\label{ctcontr}
& \lla \O_{[4]}(\bs{p}_1)\O_{[2]}(\bs{p}_2)\O_{[2]}(\bs{p}_3)\rra 
_{\rm{ct}}
=2a_0\mu^{(3v-u)\ep} \nn\\
&\qquad\qquad -a_1\mu^{(v-u)\ep}\big[\lla\O_{[2]}(\bs{p}_2)\O_{[2]}(-\bs{p}_2)\rra_{\rm{reg}}+\lla\O_{[2]}(\bs{p}_3)\O_{[2]}(-\bs{p}_3)\rra_{\rm{reg}}\big]
\end{align}
where
\[
\lla\O_{[2]}(\bs{p})\O_{[2]}(-\bs{p})\rra_{\rm{reg}}= C_2 p^{2v\ep}, \qquad
C_2 = \frac{C_2^{(-1)}}{\ep}+C_2^{(0)}+C_2^{(1)}\ep+O(\ep^2). 
\]
(Once again, to reduce clutter we are labelling operators and sources through their bare rather than their regulated dimensions.)
Working in the most compact scheme where $C_2^{(0)}=C_2^{(1)}=0$, to obtain a finite renormalised correlator we require 
\begin{align}
\frac{a_0}{2(u+v)c_{422}} &= \frac{-1}{v(u-3v)(u+v)\ep^2}
+\frac{1}{\ep}\Big(a_1^{(0)}C_2^{(-1)}+\frac{2\big((\gamma_E-\ln 2)u-v\big)}{(u-3v)(u-v)(u+v)}\Big)+ a_0^{(0)}+O(\ep),\\
\frac{a_1}{2(u+v)c_{422}} &= \frac{-1}{C_2^{(-1)} v (u-v)(u+v)\ep}
+a_1^{(0)}+a_1^{(1)}\ep+O(\ep^2).
\end{align}
(Note we must keep $a_1^{(1)}$ here as the regulated 2-point function is proportional to $\ep^{-1}$.)
The counterterms \eqref{ct422} mean the renormalised source $\phi_{[2]}$ is related to the corresponding bare source according to
\[
\phi_{[2]}^{\rm{bare}} = \phi_{[2]} +a_1\phi_{[0]}\phi_{[2]}\mu^{(v-u)\ep},
\]
generating a beta function
\[
\beta_{\phi_{[2]}} \equiv \lim_{\ep\rightarrow 0}\mu\frac{\partial\phi_{[2]}}{\partial \mu}=-(v-u)\phi_{[0]}\phi_{[2]}a_1^{(-1)} = -\frac{2c_{422}}{C_2^{(-1)}v}\phi_{[0]}\phi_{[2]}.
\]
The renormalised correlator is then 
\begin{align}\label{rencorr422}
\lla \O_{[4]}(\bs{p}_1)\O_{[2]}(\bs{p}_2)\O_{[2]}(\bs{p}_3)\rra_{\rm{ren}} 
&=
c_{422}\ln\frac{p_2^2}{\mu^2}\ln\frac{p_3^2}{\mu^2}+a'_1 \Big(\ln\frac{p_2^2}{\mu^2}+\ln\frac{p_3^2}{\mu^2}\Big)+a'_0,
\end{align}
where  $a'_1$ and $a'_0$ are ($\ep$-independent) scheme-dependent constants linearly related to the $a_1^{(0)}$, $a_1^{(1)}$ and $a_0^{(0)}$ above.  Specifically, the relation is
\begin{align}
a'_1 &=  - a_1^{(0)}C_2^{(-1)}v +\frac{2c_{422}}{(u-v)}[u(\gamma_E-\ln 2)-v], \\
a'_0 &= 2a_0^{(0)}-2a_1^{(1)}C_2^{(-1)} +\frac{2c_{422}}{(u-3v)(u-v)}
[2(\ln2-\gamma_E)^2u^2+4(\ln 2-\gamma_E)u v+\pi^2 v^2].
\end{align}
Notice also that since $\Delta_1=\Delta_2+\Delta_3$ this correlator is extremal. As we expect, the momentum dependence of the nonlocal part of \eqref{rencorr422} then matches that of the product of 2-point functions $\lla \O_{[2]}(\bs{p}_2)\O_{[2]}(-\bs{p}_2)\rra \lla \O_{[2]}(\bs{p}_3)\O_{[2]}(-\bs{p}_3)\rra$.\footnote{Note however the semi- and ultralocal terms in
the correlator ({\it i.e.,} the terms proportional to $a_1'$ and $a_0'$) can be adjusted arbitrarily through finite counterterms, as can be seen from (4.145) and (4.146).}

Under a change of renormalisation scale, 
\begin{align}
&\mu\frac{\partial}{\partial\mu}\, \lla \O_{[4]}(\bs{p}_1)\O_{[2]}(\bs{p}_2)\O_{[2]}(\bs{p}_3)\rra_{\rm{ren}} 
= -2c_{422} \Big(\ln\frac{p_2^2}{\mu^2}+\ln\frac{p_3^2}{\mu^2}\Big)-4a'_1 \nn\\&\qquad\qquad
= \frac{\partial\beta_{\phi_{[2]}}}{\partial\phi_{[2]}\partial\phi_{[0]}}\Big(\lla\O_{[2]}(\bs{p}_2)\O_{[2]}(-\bs{p}_2)\rra_{\rm{ren}}+\lla\O_{[2]}(\bs{p}_3)\O_{[2]}(-\bs{p}_3)\rra_{\rm{ren}}\Big)+\mathcal{A}_{422},
\end{align}
where the anomaly $\mathcal{A}_{422} = -4a'_1$.
In this example, then, the anomaly is purely scheme-dependent and can be adjusted arbitrarily through the addition of finite counterterms.

As in the case of the previous example, the existence of a dimensionless source implies that we can consider counterterms and anomalies of the form $f(\phi_{[0]}) \phi_{[2]}^2$, where $f$ is a function of $\phi_{[0]}$.  The Taylor expansion of this function is fixed by the $n$-point function and we have determined the terms up to linear order.  As in the previous example, the corresponding conformal anomaly due to the 3-point function is again scheme dependent.

\section{Beta functions and anomalies} 
\label{sec:anomalies_and_betafns}

In this section we examine more closely the anomaly and beta function terms that appear in the conformal Ward identities.  Since these terms break conformal symmetry, we will start from the diffeomorphism and Weyl Ward identities that hold for a general quantum field theory.  We will restrict our considerations to scalar operators; for a more complete discussion we refer the reader to \cite{Osborn:1991gm, Baume:2014rla}.

First, let us consider the variation of the generating functional for renormalised correlators under a variation of the renormalised sources $\phi_i$,  
\[
\delta W = -\int \D^d \bs{x}\,\sqrt{g}\,\big(\tfrac{1}{2} \<T_{\mu\nu}\>_s \delta g^{\mu\nu} + \sum_{i}\<\O_i\>_s\delta \phi_i \big).
\]
Here, the quantum field theory lives on an arbitrary background metric $g_{\mu\nu}$, the background source profiles $\phi_i$ are also arbitrary, as indicated by the subscript $s$ (for source) on the 1-point functions.
The index $i$ labels the different scalar operators, and is distinct from the spatial indices $\mu,\nu$.  
Under a diffeomorphism, $x^\mu \rightarrow x^\mu+\xi^\mu$, we have
\[\label{diffvar}
\delta g_{\mu\nu} = 2\nabla_{(\mu}\xi_{\nu)}, \qquad
\delta \phi_i= \xi^\mu\partial_\mu\phi_i,\qquad
\delta W=0,
\]
giving rise to the Ward identity
\[
0=\nabla^\mu \<T_{\mu\nu}\>_s + \sum_i \<\O_i\>_s \partial_\nu\phi_i.
\]
The corresponding Ward identities for correlators, if required, can then be derived by functionally differentiating this relation with respect to the sources $\phi_i$ before setting them to zero and returning to a flat metric.

Under a Weyl transformation of the background metric $g_{\mu\nu} \rightarrow e^{2\sigma(\bs{x})}g_{\mu\nu}$, we have instead
\[
\delta g_{\mu\nu} = 2\sigma g_{\mu\nu}, \qquad
\delta \phi_i = \sigma \mathcal{B}_{\phi_i}, \qquad \delta W = A = \int \D^d\bs{x}\,\sqrt{g}\, \sigma \mathcal{A},
\] 
where the $\mathcal{B}_{\phi_i}$ and the anomaly density $\mathcal{A}$ are local functions of dimension $d-\Delta_i$ and $d$ respectively, constructed from the set of sources $\{\phi_i,\, g_{\mu\nu}\}$ and their derivatives.
According to our present conventions where $\phi_i$ has a bare dimension $d-\Delta_i$, 
\[
\mathcal{B}_{\phi_i} = (\Delta_i-d)\phi_i + \beta_{\phi_i},
\]
where $\beta_{\phi_i}$ is the beta function for $\phi_i$.  (We could alternatively regard $\mathcal{B}_{\phi_i}$ as $\mu^{d-\Delta_i}$ times the beta function for the dimensionless coupling $\phi_i^{\rm{dimless}} = \phi_i\mu^{\Delta_i-d}$.)
Note also that since $W$ is the generating functional of renormalised correlators, $\mathcal{A}$ is a finite quantity.  
Writing the trace of the stress tensor as $T^\mu_\mu=T$, the corresponding Ward identity is then
\[\label{WeylWI}
\<T\>_s = \sum_i \mathcal{B}_{\phi_i} \<\O_i\>_s + \mathcal{A}.
\]

Let us now proceed to conformal transformations, which are diffeomorphisms mapping flat space to itself up to a Weyl transformation, 
\[
\delta g_{\mu\nu}=2\partial_{(\mu}\xi_{\nu)} = \frac{2}{d}(\partial\cdot\xi) \delta_{\mu\nu}.
\]
We therefore specialise to a flat background metric $g_{\mu\nu}=\delta_{\mu\nu}$ and write all indices henceforth in the lowered position, although we keep the scalar source profiles $\phi_i$ arbitrary. 
To undo the action of this diffeomorphism on the metric we can use an opposing Weyl transformation with $\sigma = -\frac{1}{d}(\partial\cdot \xi)$.  The net variation of the sources is then
\[
\delta g_{\mu\nu}=0, \qquad \delta \phi_i = \xi_\mu\partial_\mu\phi_i-\frac{1}{d}(\partial\cdot\xi)\mathcal{B}_{\phi_i}, \qquad
\delta W = -\frac{1}{d} \int\D^d\bs{x}\,(\partial\cdot\xi)\, \mathcal{A},
\]
which after integrating by parts yields the  conformal Ward identity
\begin{align}
0 = \int\D^d \bs{x} \,\Big[\frac{1}{d}(\partial\cdot\xi)\, \mathcal{A}+\sum_i \Big(\frac{1}{d}(\partial\cdot\xi)(\Delta_i\phi_i+\beta_{\phi_i})+\phi_i\xi_\mu\partial_\mu\Big)\<\O_i\>_s\Big].
\end{align}

To obtain the corresponding identities for correlators we must now functionally differentiate with respect to the sources before restoring them to zero.
Since we assume that the theory with all sources switched off (denoted by a subscript zero) is a CFT, $\beta_{\phi_i}$ begins at quadratic order in the sources as we saw in previous sections, hence
\[
\beta_i|_0 = 0, \qquad \frac{\partial \beta_{\phi_i}}{\partial \phi_j}\Big|_0 = 0.
\]
We will also assume all 1-point functions vanish once the sources are switched off, {\it i.e.,} conformal symmetry is not spontaneously broken.
Functionally differentiating three times with respect to the sources, we then obtain
\begin{align}\label{genCWI3}
0&=\sum_{i=1}^3 
\left[\Big(\Delta_i(\partial\cdot\xi)_{\bs{x}_i}+d\xi_\mu(\bs{x}_i)\frac{\partial}{\partial x_{i\mu}}\Big)\<\O_1(\bs{x}_1)\O_2(\bs{x}_2)\O_3(\bs{x}_3)\>\right.
\nn\\&\qquad\quad
\left.-\Big(\frac{\partial \beta_{\phi_i}(\bs{x}_1)}{\partial\phi_1(\bs{x}_1)\partial\phi_2(\bs{x}_2)}\Big|_0(\partial\cdot\xi)_{\bs{x}_1}\<\O_i(\bs{x}_1)\O_3(\bs{x}_3)\> + \mathrm{cyclic\,permutations\,of\,(123)}\Big) \right]
\nn\\&\quad
+ (\partial\cdot\xi)_{\bs{x}_1}\frac{\delta^3 A}{\delta\phi_1(\bs{x}_1)\delta\phi_2(\bs{x}_2)\delta\phi_3(\bs{x}_3)}\Big|_0.
\end{align}
This is the general 3-point  conformal Ward identity including all beta function and anomalous contributions.  The beta function contributions are semilocal, arising only when the dimensions of the operators in the 2-point functions coincide, while the anomaly contribution is ultralocal.

More generally, we see that the existence of a beta function contribution requires a nonzero quadratic term
in the expansion of the beta function about the origin: on dimensional grounds, for $\beta_{\phi_i}$ to contain a term $\sim\Box^m\phi_j\Box^n\phi_k$ requires $-\Delta_i +\Delta_j+\Delta_k =d + 2(m+n)$ or equivalently $\alpha+1+\beta_i-\beta_j-\beta_k=-2(m+n)$.  The corresponding triple-$K$ integral therefore has a singularity of $(+--)$ type with $k_{+--}=m+n$.  (For $k_{+--}>0$, note also that the second derivative of the beta function in \eqref{genCWI3} leads to boxes acting on delta functions.)
 Similarly, to have an anomalous contribution requires $\mathcal{A}$ to contain a term $\sim \Box^l\phi_1\Box^m\phi_2\Box^n\phi_3$ hence $\Delta_1+\Delta_2+\Delta_3 = 2(d+l+m+n)$ or $\alpha+1-\beta_1-\beta_2-\beta_3 = -2(l+m+n)$.  The triple-$K$ integral then has a singularity of $(---)$ type with $k_{---}=l+m+n$.  These conditions, while necessary, are not always sufficient as we will see in example 13 below.
 
To obtain specifically the dilatation Ward identity we must set $\xi_\mu = x_\mu$, meaning $\partial\cdot\xi=d$, while to obtain the special conformal Ward identity we set $\xi_\mu=x^2b_\mu-2(b\cdot x)x_\mu$ for some vector $b_\mu$, whereupon $\partial\cdot\xi = -2d(b\cdot x)$.
Let us now consider a few examples to illustrate this discussion.

\bigskip

{\it\textbullet\, Example 11: $\Delta_1=4$ and $\Delta_2=\Delta_3=3$ in $d=4$.}

\bigskip

Here, the $(+--)$ and the $(---)$ conditions are  satisfied with $k_{+--}=0$ and $k_{---}=1$, leading us to expect both a beta function and an anomaly. 
On purely dimensional grounds, the possible contributions to the beta functions are 
\[
\beta_{\phi_{[0]}} = B_{000}\phi_{[0]}^2+O(\phi_{[0]}^3), \qquad
\beta_{\phi_{[1]}} = B_{110}\phi_{[1]}\phi_{[0]} + O(\phi_{[1]}\phi_{[0]}^2),
\]
labelling sources by their bare dimensions for compactness. 

The dilatation Ward identity then reads
\begin{align}
0 &= \Big(10+\sum_{i=1}^3 x_{i\mu}\frac{\partial}{\partial x_{i\mu}}\Big)\<\O_{[4]}(\bs{x}_1)\O_{[3]}(\bs{x}_2)\O_{[3]}(\bs{x}_3)\>
\nn\\&\quad
-B_{110}\<\O_{[3]}(\bs{x}_2)\O_{[3]}(\bs{x}_3)\>\Big(\delta(\bs{x}_1-\bs{x}_2)
+\delta(\bs{x}_1-\bs{x}_3)\Big)
+\frac{\delta^3 A}{\delta\phi_{[0]}(\bs{x}_1)\delta\phi_{[1]}(\bs{x}_2)\delta\phi_{[1]}(\bs{x}_3)}\Big|_0,
\end{align}
while the special conformal Ward identity is
\begin{align}
0 &= \Big[-2\bs{b}\cdot(4\bs{x}_1+3\bs{x}_2+3\bs{x}_3)
+\sum_{i=1}^3 \Big(x_i^2 b_\mu-2(\bs{b}\cdot \bs{x}_i)x_{i\mu}\Big)\frac{\partial}{\partial x_{i\mu}}\Big]\<\O_{[4]}(\bs{x}_1)\O_{[3]}(\bs{x}_2)\O_{[3]}(\bs{x}_3)\>\nn\\&\qquad
+2(\bs{b}\cdot \bs{x}_1)B_{110}\<\O_{[3]}(\bs{x}_2)\O_{[3]}(\bs{x}_3)\>\big(\delta(\bs{x}_1-\bs{x}_2)
+\delta(\bs{x}_1-\bs{x}_3)\big)
\nn\\&\qquad
-2(\bs{b}\cdot \bs{x}_1)\frac{\delta^3 A}{\delta\phi_{[0]}(\bs{x}_1)\delta\phi_{[1]}(\bs{x}_2)\delta\phi_{[1]}(\bs{x}_3)}\Big|_0.
\end{align}
We therefore have both beta functions and an anomalous contribution as anticipated.
Extracting the factor of $b_\mu$
and converting to momentum space, these two identities become
\begin{align}
0&= \Big[-2+\sum_{i=2}^3 p_{i\mu}\frac{\partial}{\partial p_{i\mu}}\Big]\lla \O_{[4]}(\bs{p}_1)\O_{[3]}(\bs{p}_2)\O_{[3]}(\bs{p}_3)\rra
\nn\\&\qquad
+B_{110}\Big(\lla \O_{[3]}(\bs{p}_2)\O_{[3]}(-\bs{p}_2)\rra +\lla \O_{[3]}(\bs{p}_3)\O_{[3]}(-\bs{p}_3)\rra\Big) +\mathcal{A}_{433}, \\
0 &= \sum_{i=2}^3 \Big[-2\frac{\partial}{\partial p_{i\mu}}-2p_{i\nu}\frac{\partial}{\partial p_{i\nu}}\frac{\partial}{\partial p_{i\mu}}+p_{i\mu}\frac{\partial}{\partial p_{i\nu}}\frac{\partial}{\partial p_{i\nu}}\Big]\lla\O_{[4]}(\bs{p}_1)\O_{[3]}(\bs{p}_2)\O_{[3]}(\bs{p}_3)\rra,
\end{align}
where 
\[
\frac{\delta^3 A}{\delta\phi_{[0]}(\bs{p}_1)\delta\phi_{[1]}(\bs{p}_2)\delta\phi_{[1]}(\bs{p}_3)}\Big|_0
= (2\pi)^d \delta(\bs{p}_1+\bs{p}_2+\bs{p}_3)\mathcal{A}_{433}(p_1,p_2,p_3).
\]
Decomposing these vector equations into a scalar basis, the dilatation Ward identity is
\begin{align}
0&=D\,\lla \O_{[4]}(\bs{p}_1)\O_{[3]}(\bs{p}_2)\O_{[3]}(\bs{p}_3)\rra 
\nn\\[1ex]&\qquad
+B_{110}\big(\lla \O_{[3]}(\bs{p}_2)\O_{[3]}(-\bs{p}_2)\rra +\lla \O_{[3]}(\bs{p}_3)\O_{[3]}(-\bs{p}_3)\rra\big) +\mathcal{A}_{433}, 
\end{align}
where
\[
D = -2+\sum_{i=1}^3 p_i \frac{\partial}{\partial p_i},
\]
while the special conformal Ward identities are
\begin{align}\label{CWI433v4}
0&= \Big[\frac{2}{p_1}\frac{\partial}{\partial p_1}D+K_{31}\Big]\lla \O_{[4]}(\bs{p}_1)\O_{[3]}(\bs{p}_2)\O_{[3]}(\bs{p}_3)\rra, \\ \label{CWI433v3}
0&= \Big[\frac{2}{p_1}\frac{\partial}{\partial p_1}D+K_{21}\Big]\lla \O_{[4]}(\bs{p}_1)\O_{[3]}(\bs{p}_2)\O_{[3]}(\bs{p}_3)\rra, 
\end{align}
or equivalently,
\begin{align}\label{CWI433v1}
K_{23}\lla \O_{[4]}(\bs{p}_1)\O_{[3]}(\bs{p}_2)\O_{[3]}(\bs{p}_3)\rra &=0, \\ \label{CWI433v2}
K_{31}\lla \O_{[4]}(\bs{p}_1)\O_{[3]}(\bs{p}_2)\O_{[3]}(\bs{p}_3)\rra
&=\frac{2}{p_1}\frac{\partial}{\partial p_1}\mathcal{A}_{433},
\end{align}
where $K_{ij}=K_i-K_j$ with
\begin{align}
K_1 = \frac{\partial^2}{\partial p_1^2}-\frac{3}{p_1}\frac{\partial}{\partial p_1},\qquad
K_2 = \frac{\partial^2}{\partial p_2^2}-\frac{1}{p_2}\frac{\partial}{\partial p_2}, \qquad
K_3 = \frac{\partial^2}{\partial p_3^2}-\frac{1}{p_3}\frac{\partial}{\partial p_3}.
\end{align}
While \eqref{CWI433v1} follows trivially from permutation symmetry, \eqref{CWI433v2} is non-trivial and relates the anomalous contributions appearing in the dilatation and the special conformal Ward identities.
In fact, we can use this identity 
(or equivalently \eqref{CWI433v4}) 
to eliminate a scheme-dependent term in our earlier result \eqref{e:O433} for the renormalised correlator.
Under dilatations
\begin{align}
D\,\lla \O_{[4]}(\bs{p}_1) \O_{[3]}(\bs{p}_2) \O_{[3]}(\bs{p}_3) \rra &= -\mu\frac{\partial}{\partial \mu}
\lla \O_{[4]}(\bs{p}_1) \O_{[3]}(\bs{p}_2) \O_{[3]}(\bs{p}_3) \rra \nn\\
&=
\frac{c_{433}}{2}\Big[p_1^2-\frac{1}{2}(p_2^2+p_3^2)-p_2^2\ln\frac{p_2^2}{\mu^2}-p_3^2\ln\frac{p_3^2}{\mu^2}\Big]
+2a'_0(p_2^2+p_3^2),
\end{align}
hence
\[
\frac{2}{p_1}\frac{\partial}{\partial p_1}D\,\lla \O_{[4]}(\bs{p}_1) \O_{[3]}(\bs{p}_2) \O_{[3]}(\bs{p}_3) \rra = 2 c_{433}.
\]
One can likewise show that
\[
K_{31}\lla \O_{[4]}(\bs{p}_1) \O_{[3]}(\bs{p}_2) \O_{[3]}(\bs{p}_3) \rra = (a'_2+a'_0) K_{31} p_1^2=4(a'_2+a'_0),
\]
and hence the special conformal Ward identity \eqref{CWI433v2} fixes
\[\label{433a2a0reln}
a'_2+a'_0 = -\frac{c_{433}}{2}.
\]
There are therefore only two, rather than three scheme-dependent coefficients in \eqref{e:O433}.

\bigskip

{\it\textbullet\, Example 12: $\Delta_1=4$ and $\Delta_2=\Delta_3=2$ in $d=4$.}

\bigskip

Here, we have $(-++)$, $(--+)$, $(-+-)$ and $(---)$ singularities with $k_{-++}=k_{--+}=k_{-+-}=k_{---}=0$.   
On purely dimensional grounds, the possible contributions to the beta functions are
\[
\beta_{\phi_{[0]}} = B_{000}\phi_{[0]}^2+O(\phi_{[0]}^3),\qquad
\beta_{\phi_{[2]}} = B_{220}\phi_{[2]}\phi_{[0]}+B_{200}\phi_{[0]}\Box\phi_{[0]}+O(\phi_{[2]}\phi_{[0]}^2, \phi_{[0]}^2\Box\phi_{[0]}),
\]
labelling sources by their bare dimensions once again.

The dilatation Ward identity is 
\begin{align}
0 &= \Big(8+\sum_{i=1}^3 x_{i\mu}\frac{\partial}{\partial x_{i\mu}}\Big)\<\O_{[4]}(\bs{x}_1)\O_{[2]}(\bs{x}_2)\O_{[2]}(\bs{x}_3)\> \nn\\&\quad
-B_{220}\<\O_{[2]}(\bs{x}_2)\O_{[2]}(\bs{x}_3)\>\Big(\delta(\bs{x}_1-\bs{x}_2)
+\delta(\bs{x}_1-\bs{x}_3)\Big)
+\frac{\delta^3 A}{\delta\phi_{[0]}(\bs{x}_1)\delta\phi_{[2]}(\bs{x}_2)\delta\phi_{[2]}(\bs{x}_3)}\Big|_0,
\end{align}
while the special conformal Ward identity reads
\begin{align}
0 &= \left[-2\bs{b}\cdot(4\bs{x}_1+2\bs{x}_2+2\bs{x}_3)
+\sum_{i=1}^3\Big(x_i^2 b_\mu-2(\bs{b}\cdot \bs{x}_i)x_{i\mu}\Big)\frac{\partial}{\partial x_{i\mu}}\right]\<\O_{[4]}(\bs{x}_1)\O_{[2]}(\bs{x}_2)\O_{[2]}(\bs{x}_3)\>\nn\\&\quad
+2(\bs{b}\cdot \bs{x}_1) B_{220}\<\O_{[2]}(\bs{x}_2)\O_{[2]}(\bs{x}_3)\>\big(\delta(\bs{x}_1-\bs{x}_2)
+\delta(\bs{x}_1-\bs{x}_3)\big)\nn\\&\quad
-2(\bs{b}\cdot \bs{x}_1)\frac{\delta^3 A}{\delta\phi_{[0]}(\bs{x}_1)\delta\phi_{[2]}(\bs{x}_2)\delta\phi_{[2]}(\bs{x}_3)}\Big|_0.
\end{align}
The remainder of the analysis then closely mirrors that of the previous example.
The momentum-space dilatation Ward identity reads
\begin{align}\label{CWI422no2}
0&= D\,\lla \O_{[4]}(\bs{p}_1)\O_{[2]}(\bs{p}_2)\O_{[2]}(\bs{p}_3)\rra \nn\\[1ex]&\qquad
+B_{220}\big(\lla \O_{[2]}(\bs{p}_2)\O_{[2]}(-\bs{p}_2)\rra +\lla \O_{[2]}(\bs{p}_3)\O_{[2]}(-\bs{p}_3)\rra\big) +\mathcal{A}_{422}, 
\end{align}
where 
\[
D = \sum_{i=1}^3 p_i \frac{\partial}{\partial p_i}.
\]
This is consistent with \eqref{rencorr422} above since
$(D+\mu(\partial/\partial\mu))\, \lla \O_{[4]}(\bs{p}_1)\O_{[2]}(\bs{p}_2)\O_{[2]}(\bs{p}_3)\rra=0$.

Meanwhile, the special conformal Ward identities are
\begin{align}\label{CWI422no1}
0&= \Big[\frac{2}{p_1}\frac{\partial}{\partial p_1}D+K_{31}\Big]\lla \O_{[4]}(\bs{p}_1)\O_{[2]}(\bs{p}_2)\O_{[2]}(\bs{p}_3)\rra, \\
0&= \Big[\frac{2}{p_1}\frac{\partial}{\partial p_1}D+K_{21}\Big]\lla \O_{[4]}(\bs{p}_1)\O_{[2]}(\bs{p}_2)\O_{[2]}(\bs{p}_3)\rra, 
\end{align}
or equivalently,
\begin{align}
K_{23}\lla \O_{[4]}(\bs{p}_1)\O_{[2]}(\bs{p}_2)\O_{[2]}(\bs{p}_3)\rra &=0, \\ 
K_{31}\lla \O_{[4]}(\bs{p}_1)\O_{[2]}(\bs{p}_2)\O_{[2]}(\bs{p}_3)\rra
&=\frac{2}{p_1}\frac{\partial}{\partial p_1}\mathcal{A}_{422},
\end{align}
where $K_{ij}=K_i-K_j$ with
\begin{align}
K_1 = \frac{\partial^2}{\partial p_1^2}-\frac{3}{p_1}\frac{\partial}{\partial p_1},\qquad
K_2 = \frac{\partial^2}{\partial p_2^2}+\frac{1}{p_2}\frac{\partial}{\partial p_2}, \qquad
K_3 = \frac{\partial^2}{\partial p_3^2}+\frac{1}{p_3}\frac{\partial}{\partial p_3}.
\end{align}
The renormalised correlator indeed satisfies these identities, since \eqref{rencorr422}  obeys
\[
\frac{2}{p_1}\frac{\partial}{\partial p_1} D\,
\lla \O_{[4]}(\bs{p}_1)\O_{[2]}(\bs{p}_2)\O_{[2]}(\bs{p}_3)\rra 
=0, \qquad K_{31}\lla \O_{[4]}(\bs{p}_1)\O_{[2]}(\bs{p}_2)\O_{[2]}(\bs{p}_3)\rra 
=0.
\]

Note that in this case (unlike the previous), the special conformal Ward identities provide no additional constraints on the scheme-dependent constants in  \eqref{rencorr422}.

\bigskip

{\it\textbullet\, Example 13: $\Delta_1=\Delta_2=2$ and $\Delta_3=4$ in $d=3$.}\label{Example13label}

\bigskip

While it is necessary for the triple-$K$ integral to have a $(---)$ singularity in order to have an anomaly, the presence of such a singularity is not sufficient to guarantee the anomaly is nonzero.  In fact, whenever we have only $(++-)$ and $(---)$ singularities, 
the anomaly vanishes and the renormalised correlator obeys the {\it homogeneous} conformal Ward identities, as we saw in section \ref{subsubsec:Ward2}.
(Given the absence of a $(+--)$ singularity, beta function contributions are also clearly forbidden.)
In the present example, which falls into this category, we have $k_{++-}=0$ and $k_{---}=1$.  The correlator is extremal, $\Delta_1+\Delta_2=\Delta_3$, and can be realised in terms of a free scalar $\Phi$ as $\O_1=\O_2=\,\lwick\Phi^4\rwick$ and $\O_3=\,\lwick\Phi^8\rwick$.

The leading divergence of the regulated triple-$K$ integral occurs at $\ep^{-1}$ order and is nonlocal in the momenta.  The renormalised correlator then follows by multiplying through by an overall constant of order $\ep$ and sending $\ep\rightarrow 0$, yielding
\[
\lla \O_1(\bs{p}_1)\O_2(\bs{p}_2)\O_3(\bs{p}_3)\rra \propto F_{++-} + a F_{---}
\] 
where\footnote{See also \eqref{case7result} in appendix \ref{subsec:zero_v_scheme}.}
\[
F_{++-}=  p_1p_2,\qquad 
F_{---} = 3(p_1^2+p_2^2)- p_3^2.
\]  
The nonlocal piece $F_{++-}$  is equal to the product $\lla \O_1(\bs{p}_1)\O_1(-\bs{p}_1)\rra \lla\O_2(\bs{p}_2)\O_2(-\bs{p}_2)\rra$ as we would expect for an extremal correlator.
The finite constant $a$ multiplying the ultralocal $F_{---}$ piece can be adjusted arbitrarily through the addition of a finite counterterm 
\[\label{3mct}
S_{\rm{ct}}= a \int \D^{3+2u\ep}\bs{x}\,\mu^{-(u-3v)\ep}
\big[ 3(\Box \phi_1\phi_2\phi_3+\phi_1\Box\phi_2\phi_3)-\phi_1\phi_2\Box\phi_3\big].
\]

Both $F_{++-}$ and $F_{---}$ independently satisfy the homogeneous dilatation and special conformal Ward identities, $DF=0$ and $K_{ij}F=(K_i-K_j)F=0$, as is easily verified noting that 
\[
D = -2+\sum_{i=1}^3 p_i \frac{\partial}{\partial p_i}, \qquad K_1 =\frac{\partial^2}{\partial p_1^2}, \qquad K_2 =\frac{\partial^2}{\partial p_2^2}, \qquad K_3 = \frac{\partial^2}{\partial p_3^2} -\frac{4}{p_3}\frac{\partial}{\partial p_3}.
\] 
Indeed this makes sense, as the finite counterterm \eqref{3mct} fails to generate a nonzero anomaly:
\[
\mu\frac{\partial}{\partial\mu}
\lla\O_1(\bs{p}_1)\O_2(\bs{p}_2)\O_3(\bs{p}_3)\rra=
\lim_{\ep\rightarrow 0}(\ep a) (3(p_1^2+p_2^2)-p_3^2) =0.
\]
The point here is that we {\it only} have a finite counterterm: there are no counterterms with divergent coefficients, since the renormalised correlator is given by multiplying the leading $\ep^{-1}$ divergence of the triple-$K$ integral through by an overall constant of order $\ep$.  (This must be the case as there are no counterterms to remove the $(++-)$ singularity.)
To have a nonzero anomaly would instead require a $(---)$ counterterm whose coefficient has an $\ep^{-1}$ pole.

 \section{Dual conformal symmetry}
\label{sec:dual_conf_sym}

Several of the renormalised 3-point functions we have met thus far have the curious property of dual conformal symmetry: their {\it momentum-space} expressions take the form expected of a CFT 3-point function in {\it position space}.\footnote{Early hints of dual conformal symmetry emerged in \cite{Broadhurst:1993ib, Drummond:2006rz}, and were later developed in the context of scattering amplitudes in $\mathcal{N}=4$ SYM, see {\it e.g.,} \cite{Drummond:2008vq, Drummond:2010km, Henn:2011xk}. Dual conformal symmetry is known to be connected to the existence of a Yangian algebra.}
One example is when solely the $(+++)$ condition is satisfied with $k_{+++}=0$.  In this case, $\Delta_1+\Delta_2+\Delta_3=d$, and we find ({\it e.g.,} from the general formula \eqref{case1_result} in appendix \ref{sec:gen_results}) 
\begin{align}
\lla\O_1(\bs{p}_1)\O_{2}(\bs{p}_2)\O_{3}(\bs{p}_3)\rra
 \propto p_1^{2\Delta_1-d}p_2^{2\Delta_2-d}p_3^{2\Delta_3-d} = p_1^{\Delta_1-\Delta_2-\Delta_3}p_2^{\Delta_2-\Delta_3-\Delta_1}p_3^{\Delta_3-\Delta_1-\Delta_2}.
\end{align}
Defining
\[\label{ydef}
\bs{p}_1=\bs{y}_{23}=\bs{y}_2-\bs{y}_3, \qquad
\bs{p}_2=\bs{y}_{31}=\bs{y}_3-\bs{y}_1, \qquad
\bs{p}_3=\bs{y}_{12}=\bs{y}_1-\bs{y}_2,
\]
to ensure momentum conservation $\sum_i\bs{p}_i=0$, we then have
\begin{align}
\lla\O_{1}(\bs{y}_{23})\O_{2}(\bs{y}_{31})\O_{3}(\bs{y}_{12})\rra \propto \frac{1}{|y_{23}|^{\Delta_2+\Delta_3-\Delta_1}|y_{31}|^{\Delta_3+\Delta_1-\Delta_2}|y_{12}|^{\Delta_1+\Delta_2-\Delta_3}}.
 \end{align}
The 3-point function thus has exactly the form imposed by conformal symmetry acting on the $\bs{y}$ coordinates.   This dual momentum-space conformal symmetry is present {\it in addition} to the position-space conformal symmetry we started with, which acts on the original $\bs{x}$ coordinates. 

In the example above, the operator dimensions associated with the dual conformal symmetry are the same as for the original conformal symmetry.  This is not always the case, however, as can be seen from the following example.  Consider the case where solely the condition $(++-)$ is satisfied, with $k_{++-}=0$.  Now we have $\Delta_1+\Delta_2=\Delta_3$
and the correlator is extremal.  From \eqref{case2_result} in appendix \ref{sec:gen_results}, the renormalised correlator is
\[
\lla\O_1(\bs{p}_1)\O_{2}(\bs{p}_2)\O_{3}(\bs{p}_3)\rra
 \propto p_1^{2\Delta_1-d}p_2^{2\Delta_2-d}.
\]
Defining 
\[
\bar{\Delta}_1 = d/2-\Delta_2, \qquad \bar{\Delta}_2=d/2-\Delta_1, \qquad 
\bar{\Delta}_3 = d-\Delta_3,
\]
we see that
\begin{align}
\lla\O_{1}(\bs{y}_{23})\O_{2}(\bs{y}_{31})\O_{3}(\bs{y}_{12})\rra \propto \frac{1}{|y_{23}|^{\bar{\Delta}_2+\bar{\Delta}_3-\bar{\Delta}_1}|y_{31}|^{\bar{\Delta}_3+\bar{\Delta}_1-\bar{\Delta}_2}|y_{12}|^{\bar{\Delta}_1+\bar{\Delta}_2-\bar{\Delta}_3}}.
 \end{align}
The dimensions $\bar{\Delta}_i$ associated with the dual conformal symmetry are therefore in general different from those associated with the position-space conformal symmetry.
(Note however the modified dimensions still satisfy the extremality condition $\bar{\Delta}_1+\bar{\Delta}_2=\bar{\Delta}_3$.)

A third case where dual conformal symmetry can arise is when both $(+++)$ and $(++-)$ conditions are simultaneously satisfied (see case (5) in appendix \ref{sec:gen_results}).
This requires $\beta_3$ to be an integer: if $\beta_3\in\Z^+$ and $k_{+++}=0$, then the 3-point function is the same as in the first example above, while if $\beta_3\in\Z^-$ and $k_{++-}=0$, the 3-point function is the same as in the second example.

In all the examples above we had either $k_{+++}=0$ or $k_{++-}=0$.  To understand what happens more generally, consider 
for example
the case where only the $(+++)$ condition is satisfied with $k_{+++}=1$.  If all the $\beta_i\ge 0$ say, from \eqref{case1_result} the renormalised correlator is 
\[\label{kppp1}
\lla\O_{1}(\bs{y}_{23})\O_{2}(\bs{y}_{31})\O_{3}(\bs{y}_{12})\rra \propto y_{23}^{2\beta_1}y_{31}^{2\beta_2}y_{12}^{2\beta_3}\Big(\frac{y_{23}^2}{\beta_1+1}+\frac{y_{31}^2}{\beta_2+1}+\frac{y_{12}^2}{\beta_3+1}\Big).
\]
Now, in order to have dual conformal symmetry, it is necessary for the correlator to transform appropriately under inversions $\bs{y}_i\rightarrow \bs{y}_i/y_i^2$, namely
\[
\lla\O_{1}(\bs{y}_{23})\O_{2}(\bs{y}_{31})\O_{3}(\bs{y}_{12})\rra \rightarrow
y_1^{\bar{\Delta}_1}y_2^{\bar{\Delta}_2}y_3^{\bar{\Delta}_3}
\lla\O_{1}(\bs{y}_{23})\O_{2}(\bs{y}_{31})\O_{3}(\bs{y}_{12})\rra,
\]
where the $\bar{\Delta}_i$ denote generic dual conformal dimensions.
Since under inversions,
\[
y_{12}^2 \rightarrow y_{12}^2\,(y_1 y_2)^{-2}, 
\]
we see that \eqref{kppp1} transforms as a {\it sum} of 3-point functions of  different conformal dimensions, rather than as a single 3-point function.  This behaviour occurs whenever the renormalised correlator is purely the sum of products of momenta raised to various powers, without any logarithms being present.\footnote{According to the general classification scheme in appendix \ref{sec:gen_results}, this happens in cases (1), (2), (5) and (7); in these cases the leading divergence of the regulated triple-$K$ integral is nonlocal as per table \ref{sing_table}.}

As dual conformal symmetry is more typically encountered in the context of massless Feynman diagrams \cite{Drummond:2010km}, it is interesting to analyse the triple-$K$ integral from this perspective.
As shown in appendix A.3 of \cite{Bzowski:2013sza}, we can rewrite the regulated triple-$K$ integral as a massless 1-loop Feynman integral,
\[\label{FeynI}
I_{\tilde{\alpha},\{\tilde{\beta}_i\}}
= 
2^{-4}(2/\pi)^{\tilde{d}/2}
\Gamma(\tilde{\delta}_1)
\Gamma(\tilde{\delta}_2)\Gamma(\tilde{\delta}_3)\Gamma(\tilde{d}-\tilde{\delta}_t)
\int \D^{\tilde{d}} \bs{p} \,\frac{1}{|\bs{p}|^{2\tilde{\delta}_3}|\bs{p}-\bs{p}_1|^{2\tilde{\delta}_2}|\bs{p}+\bs{p}_2|^{2\tilde{\delta}_1}}.
\]
In this formula $\tilde{\delta}_i=\tilde{\beta}_i-\tilde{\beta}_t/2+\tilde{d}/4$, where $\tilde{\delta}_t = \sum_i \tilde{\delta}_i$ and $\tilde{\beta}_t=\sum_i\tilde{\beta}_i$, and we regulate in our usual manner so that $\tilde{\beta}_i=\beta_i+v \ep$ and $\tilde{d} =d+2u\ep$.  
Setting $\bs{p}=\bs{y}-\bs{y}_3$, this 1-loop triangle integral is then related to an equivalent star integral in which we integrate over the position $\bs{y}$ of a central vertex,
\[\label{starint}
\int \D^{\tilde{d}} \bs{p} \,\frac{1}{|\bs{p}|^{2\tilde{\delta}_3}|\bs{p}-\bs{p}_1|^{2\tilde{\delta}_2}|\bs{p}+\bs{p}_2|^{2\tilde{\delta}_1}}
= \int \D^{\tilde{d}}\bs{y}\,\frac{1}{|\bs{y}-\bs{y}_1|^{2\tilde{\delta}_1}|\bs{y}-\bs{y}_2|^{2\tilde{\delta}_2}|\bs{y}-\bs{y}_3|^{2\tilde{\delta}_3}} \equiv J_{\tilde{d},\{\tilde{\delta}_i\}}(\{\bs{y}_i\}).
\]
For this star integral to possess dual conformal symmetry, it must transform under inversions $\bs{y}_i\rightarrow \bs{y}_i/y_i^2$ in the same manner as a CFT 3-point function, namely
\[\label{inv_condn}
J_{\tilde{d},\{\tilde{\delta}_i\}}(\{\bs{y}_i/y_i^2\}) = y_1^{2\tilde{\delta}_1}y_2^{2\tilde{\delta}_2}y_3^{2\tilde{\delta}_3}
J_{\tilde{d},\{\tilde{\delta}_i\}}(\{\bs{y}_i\}).
\]
To achieve this requires 
\[\label{nec_condns}
\tilde{\delta}_t = \tilde{d} \qquad \Rightarrow \qquad
\tilde{\delta}_i = \tilde{\Delta}_i, \qquad \tilde{\Delta}_t =  \tilde{d},
\]
as can be seen by inverting the integration variable $\bs{y}\rightarrow \bs{y}/y^2$. 
When this condition is satisfied, however, the relation between the star integral and the triple-$K$ integral in \eqref{FeynI} is singular, due to the factor of $\Gamma(\tilde{d}-\tilde{\delta}_t)$.  Indeed, this makes sense as the  regulated triple-$K$ integral does not by itself possess dual conformal symmetry.  (One can verify directly that the triple-$K$ integral fails to transform correctly under inversions  $\bs{y}_i\rightarrow \bs{y}_i/y_i^2$.)  Dual conformal symmetry therefore cannot exist for CFT 3-point functions for which renormalisation is not required, \textit{i.e.}, cases where the singularity condition \eqref{e:cond_sigma} is not satisfied and the triple-$K$ integral can be defined through analytic continuation alone.

How then can dual conformal symmetry arise in certain of the remaining cases for which renormalisation {\it is} required?
The answer is that, in order for the renormalised correlator to possess dual conformal symmetry, we need not require that the star integral possesses {\it exact} dual conformal symmetry: it is sufficient that this holds simply to leading order in $\ep$.

In the first example above, where the $(+++)$ condition alone held with $k_{+++}=0$, we had $\Delta_t=d$ and so $\beta_t=-d/2$.  We then find 
\[
\tilde{\delta}_i = \Delta_i+(u-v)\ep/2, \qquad
\tilde{d}-\tilde{\delta}_t = (u+3v)\ep/2.
\]
The star integral \eqref{starint} now only satisfies \eqref{inv_condn} at order $\ep^0$, since after inverting we pick up a net factor of $y^{2(\tilde{\delta}_t-\tilde{d})} = y^{-(u+3v)\ep} = 1+O(\ep)$ in the numerator of the integral.  In addition, the factor of $\Gamma(\tilde{d}-\tilde{\delta}_t) = \Gamma((u+3v)\ep/2)$ in \eqref{FeynI} contributes an $\ep^{-1}$ pole.   Consequently, only the leading $\ep^{-1}$ divergence of the regulated triple-$K$ integral possesses dual conformal invariance.  As we have already seen, however, this leading $\ep^{-1}$ divergence is  precisely the renormalised correlator: since there are no counterterms when the $(+++)$ condition alone is satisfied, the renormalised correlator is obtained by multiplying the regulated triple-$K$ integral through by an overall constant of order $\ep$ before sending $\ep\rightarrow 0$.

In the second example, where the $(++-)$ condition alone was satisfied with $k_{++-}=0$, 
the emergence of dual conformal symmetry is less obvious as the star integral \eqref{starint} does not satisfy the condition \eqref{nec_condns}.
One can show, however, that the gamma function prefactors in \eqref{FeynI} are all finite as $\ep\rightarrow 0$, and as we know the triple-$K$ integral has an $\ep^{-1}$ divergence, the star integral must therefore diverge as $\ep^{-1}$.  This leading $\ep^{-1}$ divergence of the star integral, which is proportional to the renormalised correlator, does then possess dual conformal symmetry.

 \section{Discussion}\label{sec:discussion}

We have presented a comprehensive discussion of the renormalisation of 3-point functions of primary operators in conformal field theory.
Our results were obtained by solving the conformal Ward identities and as such they apply to all CFTs, perturbative or non-perturbative, and in any dimension. Renormalisation is required when the dimensions of operators involved in the 
3-point function satisfy specific relations. 

Our discussion 
is analogous to that 
for 2-point functions, where renormalisation is required when the operators involved have dimension such that $\Delta -d/2$ is integral.  Correspondingly, there is a conformal anomaly, and (like the more familiar conformal anomaly that depends on the background metric) the coefficients of these anomalies 
are part of the CFT data.  Operators with such dimensions are common in CFTs, and also in supersymmetric CFTs as BPS operators typically have such dimensions (for example, $1/2$-BPS operators in $\mathcal{N}=4$ SYM). A recent application of the
anomalies related to 2-point functions may be found in \cite{Gomis:2015yaa}.

In the case of 3-point functions, renormalisation leads to a richer structure: new conformal anomalies arise and beta functions appear.
The   generating functional of CFT connected correlators satisfies
 \begin{equation}  \label{conc:gen}
\mu \frac{\D}{\D \mu} W[\phi_i] = A,
\end{equation}
where $\phi_i$ are the renormalised sources and 
\begin{equation}
\mu \frac{\D}{\D \mu} = \mu \frac{\partial}{\partial \mu} + \sum_i \int \D^d \bs{x}\, \beta_{\phi_i}\frac{\delta}{\delta \phi_i(\bs{x})}.
\end{equation}
Anomalies arise when 
\begin{equation} \label{conc:an}
\Delta_1 + \Delta_2 + \Delta_3  = 2d +2 k_{---},
\end{equation}
while a beta function for the source that couples to $\O_{3}$
will appear when 
\begin{equation} \label{conc:beta}
\Delta_1 + \Delta_2 - \Delta_3  = d +2 k_{--+}, 
\end{equation}
where $k_{---}$ and $k_{---}$ are non-negative integers (and similarly for permutations).
The beta functions are due to renormalisation of the sources.\footnote{Note that the fact that renormalisation requires the sources of composite operators to renormalise is not new: for example, BRST renormalisation of Yang-Mills theory requires renormalisation of the sources that couple to the BRST variation of the Yang-Mills field and  of the ghost fields, see for example \cite{Itzykson:1980rh}.}

If either \eqref{conc:an} or \eqref{conc:beta} holds, \eqref{conc:gen} implies that the 3-point function will depend logarithmically on the renormalisation scale $\mu$, and thus it will contain logarithms of momenta. If both conditions hold simultaneously, $\Delta_3-d/2$ must be integral and thus $\O_{3}$  is one of the operators that have anomalies already at the level of 2-point functions. In this case,  \eqref{conc:gen} implies that the 3-point functions contain double logarithms. The fact that 3-point functions 
can exhibit such analytic structure is one of the most surprising results to emerge from this work.

A further special case arises when one of the other two operators is marginal.  The coefficient of the conformal anomaly due to the 2-point function of $\O_{3}$ may now become a function of the source of the marginal operator, 
and indeed we find such an anomaly does arise at the level of 3-point functions. This anomaly however is scheme-dependent and 
the corresponding $\mu$-dependence of the 3-point function may be set to zero by a choice of scheme.

A different set of special cases arises when the operators have dimensions that satisfy one (or both) of the following conditions:
\begin{equation} \label{conc: dualConf}
\Delta_1 - \Delta_2 - \Delta_3  = 2 k_{-++}, \qquad  \Delta_1 + \Delta_2 + \Delta_3  = d-2 k_{+++}
\end{equation}
(along with permutations), where $k_{-++}$ and $k_{+++}$ are non-negative integers. In such cases, the triple-$K$ representation of the 3-point functions is singular, not the correlators themselves. The corresponding 3-point functions may be extracted from the singular part of the triple-$K$ integral and satisfy non-anomalous conformal Ward identities. Actually, these correlators exhibit enhanced symmetry.
If $k_{-++}=0$ and/or $k_{+++}=0$, the correlators take the form of position-space correlators but with differences in position replaced by momenta, {\it i.e.,} 
these correlators are dual conformal invariant. If $k_{-++} \neq 0$ and/or $k_{+++}\neq 0$ the correlators are instead equal to a sum of terms, each of which is individually dual conformal invariant (albeit with different conformal weights).  It would interesting to understand the implications of dual conformal invariance.

We emphasise that we are considering the theory at the fixed point and the correlation functions we derive are those of the CFT.
If we were to promote the source of  $\O_{3}$ to a new coupling, however, then the deformed theory would run.  
A corollary of our analysis is a {\it necessary condition for a marginal operator $\O_{[d]}$ to be exactly marginal: its 3-point  function 
$\< \O_{[d]} \O_{[d]} \O_{[d]}\>$ should vanish}.  If this 3-point function is non-vanishing there will be a beta function (see {\it e.g.,} example 8), and the deformed theory will not be conformal. A similar argument (in $d=2$) based on OPEs was made in \cite{Kadanoff, Kadanoff:1978pv}.\footnote{We thank Adam Schwimmer and Stefan Theisen for bringing these references to our attention.}

In this paper we discussed 
the renormalisation of 3-point functions of {\it scalar} operators.  The same techniques also apply to 
tensorial 3-point functions, but there are new issues that arise. More specifically, since the diffeomorphism and Weyl Ward identities 
relate 2- and 3-point functions, we need a regulator that regulates both. For this reason the $(1,0)$-scheme which proved so useful here cannot be used there. Moreover, conservation requires that in $d$ dimensions  the stress
tensor has dimension $d$ and conserved currents have dimension $d-1$.
This condition requires a $u=v$ scheme, however the regulated expressions appear to have singularities when $u=v$.  We will discuss in detail how to overcome these problems and renormalise tensorial correlators in a sequel to this work \cite{tensor}.

It would be interesting to extend our discussion to higher-point functions. Correlators higher than 3-point functions are not uniquely determined by the conformal Ward identities: conformal invariance allows for an arbitrary function of cross-ratios in position space. One would first need to understand what is the analogue of the cross-ratio in momentum space. The singularity structure is also richer since there are different short distance behaviours depending on how many points are coincident. One would anticipate obtaining new anomalies 
when \cite{Petkou:1999fv}
\begin{equation}
\sum_{i=1}^n \Delta_i = (n-1) d + 2 k_1,
\end{equation}
and new contributions to beta functions, which are of order $(n-1)$ in the sources, when
\begin{equation}
\sum_{i=1}^{n-1} \Delta_i - \Delta_n = (n-2)   d + 2 k_2
\end{equation}
and permutations, where where $k_1, k_2$  are non-negative integers. These  two cases should correspond to ultralocal divergences 
and divergences where all but one point is coincident. All other divergences should already be accounted for by the counterterms introduced to renormalise lower point functions. Based on the case of 3-point functions studied here, one may anticipate that 
correlators with dimensions that satisfy the analogue of \eqref{conc: dualConf} should also be special.\footnote{These would be cases where one can construct dimension $d$ combinations obtained by products of operators, as well as sources and derivatives.} It would be interesting to see whether such correlators are dual conformal invariant. 

Anomalies have provided invaluable insights into quantum field theory and have led to many important results. In this paper, we uncovered a new set of conformal anomalies that originate from divergences in 3-point functions of scalar operators, and we  saw that even without anomalies CFT correlators can depend on a scale (via the scale-dependence of the renormalised sources).  Moreover,  CFT 3-point functions may depend quadratically on logarithms of momenta.  
It will be exciting to explore the implications and applications of these results.

\section*{Acknowledgements} We would like to thank Adam Schwimmer and James Drummond for discussions.
AB and KS would like to thank the Galileo Galilei Institute in Florence for support and hospitality during the workshop ``Holographic Methods for Strongly Coupled Systems'' and
PM thanks the Centre du Recherches Mathematiques, Montreal. 
KS gratefully acknowledges support from the Simons Center for Geometry and Physics, Stony Brook University and the 
``Simons Summer Workshop 2015:  New advances in Conformal Field Theories'' during which some of the research for this paper was performed. 
AB is supported by the Interuniversity Attraction Poles Programme
initiated by the Belgian Science Policy (P7/37) and the European
Research Council grant no.~ERC-2013-CoG 616732 HoloQosmos.
AB would like to thank COST for partial support via the STMS grant COST-STSM-MP1210-29014.  PM is supported by the STFC Consolidated Grant ST/L00044X/1.
AB and PM would like to thank the University of Southampton for hospitality during parts of this work.

\appendix

\section{General results} 
\label{sec:gen_results}

The triple-$K$ integral is singular whenever the condition \eqref{e:cond_sigma}, namely
\begin{equation} \label{e:cond_sigma2}
\alpha+1+ \sigma_1 \beta_1 + \sigma_2 \beta_2 + \sigma_3 \beta_3 = - 2 k_{\sigma_1 \sigma_2 \sigma_3},
\end{equation}
is satisfied for some non-negative integers $k_{\sigma_1 \sigma_2 \sigma_3}$ and (independent) choice of signs $\sigma_i \in \{ \pm \}$, $i=1,2,3$, with $\alpha=d/2-1$ and $\beta_i = \Delta_i-d/2$.
In the main text, we focused on cases where only a single solution of this condition exists.  In general, however, this condition may have multiple solutions, each with a different number of positive and negative signs, and potentially different values of $k_{\sigma_1 \sigma_2 \sigma_3}$.  When such multiple solutions exist, the regulated triple-$K$ integral typically has higher-order poles in $\ep$, with the maximum permitted being $\ep^{-s}$ where $s$ is the number of different solutions of \eqref{e:cond_sigma2} (not counting simple permutations).
Our purpose in this appendix is to classify  all the cases that can arise, including those where multiple solutions of \eqref{e:cond_sigma2} exist, and to understand their singularity structure.  We will also give explicit results for the renormalised 3-point function wherever this can be determined purely from the singularities of the regulated triple-$K$ integral.

\subsection{Classification of cases}\label{subsec:classification}

Let us call a solution of \eqref{e:cond_sigma2} associated with some $\sigma_i\in\{\pm\}$  a solution of type $(\sigma_1 \sigma_2 \sigma_3)$.
To classify the cases where \eqref{e:cond_sigma2} admits multiple solutions, 
we first observe that certain types of solution are mutually incompatible, since on physical grounds 
\begin{equation} \label{e:dimasum}
d > 0, \qquad\qquad \Delta_i > 0.
\end{equation}
(Note the latter condition is a weaker restriction than unitarity which requires $\Delta_i\ge (d-2)/2$.)
The types of solution that cannot appear simultaneously are therefore:
\begin{equation} 
\begin{array}{lll}
\left\{ \begin{array}{l} +++ \\ --- \end{array} \right. &
\left\{ \begin{array}{l} ++- \\ --+ \end{array} \right. &
\left\{ \begin{array}{l} +++ \\ +-- \end{array} \right.
\end{array}
\end{equation}
In the first two cases, we would violate the condition $d>0$, while in the third we would violate the condition $\Delta_1>0$.  For example, to have solutions of both type $(+++)$ and $(---)$ requires
\begin{align}
d/2 + \beta_1+\beta_2+\beta_3 &= -2 k_{+++},\nn\\
d/2 - \beta_1-\beta_2-\beta_3 &= -2 k_{---},
\end{align}
but on adding these equations we find  $d=-2(k_{+++}+k_{---})\le 0$.  Similarly, to have both $(+++)$ and $(+--)$ solutions requires
\begin{align}
d/2 + \beta_1+\beta_2+\beta_3 &= -2 k_{+++},\nn\\
d/2 + \beta_1-\beta_2-\beta_3 &= -2 k_{+--},
\end{align}
but on adding we find $d+2\beta_1 = 2\Delta_1 = -2(k_{+++}+k_{+--}) \le 0$.

Excluding cases with incompatible solution types, the remaining allowed cases are:
\begin{equation} \label{e:full_class}
\begin{array}{llll}
+++ & ++- & +-- & --- \\
& & & \\
\left\{ \begin{array}{l} +++ \\ ++- \end{array} \right. &
\left\{ \begin{array}{l} ++- \\ +-- \end{array} \right. &
\left\{ \begin{array}{l} ++- \\ --- \end{array} \right. &
\left\{ \begin{array}{l} +-- \\ --- \end{array} \right. \\
& & & \\
\left\{ \begin{array}{l} ++- \\ +-- \\ --- \end{array} \right. & & & 
\end{array}
\end{equation}
For ease of reference we have numbered these cases (1)--(9) as listed in table \ref{sing_table}.
We will also need to keep track of which permutations of the $(+--)$ type solution are present, subdividing cases (3), (6), (8) and (9) into further subcases accordingly (see later).
Fortunately, we do not need to do the same for the type $(++-)$ solutions as these can only arise in a single permutation due to the condition $\Delta_i>0$.  For example, if we had both $(++-)$ and $(+-+)$ solutions of \eqref{e:cond_sigma2}, then on adding we would find $\Delta_1 = -k_{++-}-k_{+-+} \le 0$.

\begin{table}[t]
\begin{center}
\begin{tabular}{|c|c|c|c|}
\hline
Case & Solution types present & Leading divergence 
& First nonlocal divergence\\ \hline
1 & $(+++)$ & $\ep^{-1}$ & $\ep^{-1}$ \\ \hline
2 & $(++-)$ & $\ep^{-1}$ & $\ep^{-1}$ \\ \hline
3 & $(+--)$ & $\ep^{-1}$ & $\ep^0$ \\ \hline
4 & $(---)$ & $\ep^{-1}$ & $\ep^0$ \\ \hline
5 & $(+++)$ and $(++-)$ & $\ep^{-2}$ & 
$\ep^{-2}$ \\ \hline
6 & $(++-)$ and $(+--)$ & $\ep^{-2}$ & $\ep^{-1}$ \\ \hline
7 & $(++-)$ and $(---)$ & $\ep^{-1}$ & $\ep^{-1}$ \\ \hline
8 & $(+--)$ and $(---)$ & $\ep^{-2}$ & $\ep^0$
\\ \hline
9 & $(++-)$, $(+--)$ and $(---)$ & $\ep^{-3}$& $\ep^{-1}$ \\ \hline
\end{tabular}
\end{center}
\caption{Singular cases consistent with $d>0$ and $\Delta_i>0$, including those where  \eqref{e:cond_sigma2} admits multiple solutions.  The third and fourth columns refer to the divergence of the corresponding regulated triple-$K$ integral, as discussed in section \ref{subsec:zero_v_scheme}.  The third column lists the maximum leading divergence of the regulated triple-$K$ integral, while the fourth column gives the order at which terms fully nonlocal in the momenta first arise.  When this order is $\ep^0$ we must evaluate the regulated triple-$K$ integral in order to determine the renormalised correlator.  In all other cases, we can determine the renormalised correlator purely from the singularities of the triple-$K$ integral.  Explicit expressions for all such cases are listed in section \ref{subsec:zero_v_scheme}.
\label{sing_table}}
\end{table}

\subsection{Renormalised correlators in $(1,0)$-scheme}
\label{subsec:zero_v_scheme}

In cases (3), (4) and (8), the singularity condition \eqref{e:cond_sigma2} has only $(+--)$ and/or $(---)$ type solutions.  In these cases, the singularities of the regulated triple-$K$ integral involve terms that are only semi- and/or ultralocal in the momenta.  To determine the (fully nonlocal) renormalised correlator then requires a complete evaluation of the regulated triple-$K$ integral including its finite piece of order $\ep^0$.

Here we will focus primarily on the remaining cases, where \eqref{e:cond_sigma2} admits solutions of type $(+++)$ and/or type $(++-)$.
When solutions of these types are present, the regulated triple-$K$ integral has singularities 
that are nonlocal in the momenta, for which there are no corresponding counterterms.
Rather, it is the triple-$K$ integral representation itself that is singular: 
the renormalised 3-point function is given 
by multiplying the regulated triple-$K$ integral through by appropriate positive powers of $\ep$, so as to extract the leading nonlocal singularities in the limit $\ep\rightarrow 0$. 
In cases (5), (6), (7) and (9) an additional complication arises, which is that the  desired nonlocal singularities potentially occur at subleading order in $\ep$ (or even at sub-subleading order in case (9)).
When this occurs, the leading singularities are either ultra- or semilocal, and correspond to the presence of type $(+--)$ and $(---)$ solutions to \eqref{e:cond_sigma2}.  In such instances, one must first remove these leading ultra- or semilocal singularities through the addition of suitable counterterms.

To place this discussion on a more explicit footing, let us now systematically evaluate the divergences of the regulated triple-$K$ integral.  In all cases apart from (3), (4) and (8), we will be able to read off the renormalised 3-point function directly from the leading nonlocal divergence.
A convenient scheme for this computation is $(u,v)=(1,0)$, where the indices of the Bessel functions are preserved.  The individual coefficients in the series expansion of the Bessel functions (the $a_k^\pm(\beta)$ defined below) then have no $\ep$-dependence, making it easy to identify the  overall order of terms.

The Bessel function $K_{\beta}(z)$ has the standard series expansion
\begin{align}\label{Kexpansion}
K_{\beta}(z) = \left\{
\begin{array}{ll}
\displaystyle{\tfrac{1}{2} z^{-\beta} \sum_{k=0}^\infty a^{-}_k(\beta) z^{2k} + \tfrac{1}{2} z^{\beta} \ln z \sum_{k=0}^\infty a^{+}_k(\beta) z^{2k}} & \text{ if } \beta \in \Z^+, \\
\displaystyle{\frac{\pi}{2 \sin(\beta \pi)} \left[ z^{-\beta} \sum_{k=0}^\infty a^{-}_k(\beta) z^{2k} + z^{\beta} \sum_{k=0}^\infty a^{+}_k(\beta) z^{2k} \right]} & \text{ if } \beta \notin \Z,
\end{array}
\right.
\end{align}
where the coefficients $a_k^{\pm}(\beta)$ are
\begin{align}
a^{-}_k(\beta) & = \left\{ 
\begin{array}{ll}
\dfrac{(-1)^k (\beta - k - 1)!}{2^{2k-\beta} k!} & \text{ if } \beta \in \Z^+ \text{ and } k < \beta, \\[2ex]
\dfrac{(-1)^{\beta}}{2^{2 k - \beta} k! (k - \beta)!}  
 \left[ \psi(k-\beta+1)+\psi(k+1) + 2 \ln 2 \right] & \text{ if } \beta \in \Z^+ \text{ and } k \geq \beta, \\[3ex]
\dfrac{1}{2^{2 k - \beta} k! \Gamma(-\beta + k + 1)} & \text{ if } \beta \notin \Z,
\end{array}
\right. \label{e:am}\\
 a^{+}_k(\beta) & = \left\{
\begin{array}{ll}
\dfrac{(-1)^{\beta+1}}{2^{2 k + \beta-1} k! (\beta + k)!} & \text{ if } \beta \in \Z^+, \\[3ex]
 \dfrac{-1}{2^{2 k + \beta} k! \Gamma(\beta + k +1)} & \text{ if } \beta \notin \Z. 
\end{array}
\right.\hspace{5.7cm} \label{e:ap}
\end{align}
Here, $\psi(k)=\Gamma'(k)/\Gamma(k)$ is the digamma function, which for positive integer $k>0$ can be re-expressed 
in terms of the $k$-th harmonic number $H_k=\sum_{n=1}^k n^{-1}$ and the Euler-Mascheroni constant $\gamma_E$ as 
$\psi(k+1)=H_k-\gamma_E$.
When $\beta\in \Z$, the expansion coefficients defined in \eqref{e:am} and \eqref{e:ap} are strictly only valid for $\beta\ge 0$.
Since $K_{-\beta}(z)=K_{\beta}(z)$,  however, we can handle all cases including $\beta<0$ by using $K_{|\beta|}(z)$ in place of $K_\beta(z)$.
We have also pulled out the overall factors in \eqref{Kexpansion} to simplify our later expressions for the renormalised correlators in section \ref{subsec:zero_v_scheme}.  These correlators are only determined up to a finite overall constant of proportionality, to which the terms we have pulled out make a fixed contribution.  Extracting this contribution allows us to simplify our final results, which will be expressed in terms of the expansion coefficients \eqref{e:am} and \eqref{e:ap}.

Writing the regulated triple-$K$ integral as
\[
I_{\alpha+\ep,\{\beta_1, \beta_2,\beta_3\}} = \int_0^\infty \D x\,x^{\alpha+\ep}\prod_i p_i^{\beta_i}K_{|\beta_i|}(p_ix),
\]
the next step is to apply the series expansion \eqref{Kexpansion} to each of the three Bessel functions.
As the $a_k^\pm(\beta)$ coefficients in the Bessel functions are all finite, divergences can only arise from the lower part of the integrals over $x$.  Factoring out all momentum dependence, the only divergent integrals are those of the form
\begin{equation}\label{xintdiv}
{\int_0^{\mu^{-1}}}\D x\, x^{-1 + \epsilon} \ln^n x \:  = \frac{(-1)^n n!}{\epsilon^{n+1}} + O(\epsilon^0),
\end{equation}
where the divergent pieces are independent of $\mu^{-1}$.
The singularities with the highest degree of divergence therefore arise from integrals with the greatest number of logarithms.  The number of logarithms present in a given term corresponds in turn to the number of  coefficients $a^+_k(|\beta_i|)$ for which $\beta_i\in \Z$.
Modulo possible logarithms, the factors of momentum accompanying divergent $x$-integrals of the form \eqref{xintdiv} have the general structure 
\begin{equation}\label{momfactor} 
\sum_{\{k_i\}}\,\Big(\prod_{i=1}^3 a_{k_i}^{\nu_i}(|\beta_i|) p_i^{2k_i+\beta_i+\nu_i|\beta_i|}\Big)
\end{equation}
for some independent choice of signs $\{\nu_i\} \in \pm 1$.  The sum here runs over integer $k_i\ge 0$ such that 
\[\label{xsumcond}
\alpha+ \sum_{i=1}^3 (2k_i +\nu_i|\beta_i| )= -1,
\]  
so as to obtain the appropriate overall power of $x$.
If we denote the sign of each $\beta_i$ by $s_i$ so that $\beta_i=s_i|\beta_i|$, then clearly \eqref{xsumcond} is solved by $\nu_i=s_i\sigma_i$ with $\sum_i k_i = k_{\sigma_1\sigma_2\sigma_3}$ according to \eqref{e:cond_sigma2}.
The factor \eqref{momfactor} can then be re-expressed more conveniently as
\begin{equation} \label{e:F2}
F_{\sigma_1 \sigma_2 \sigma_3} \equiv 
\sum_{\{k_i\}}\,\Big(\prod_{i=1}^3 a_{k_i}^{s_i\sigma_i}(|\beta_i|) p_i^{2k_i+(1+\sigma_i)\beta_i}\Big),
\end{equation}
where the sum runs over all $k_i\ge 0$ such that $\sum_i k_i = k_{\sigma_1\sigma_2\sigma_3}$.
For there to be accompanying logarithms requires both $s_i\sigma_i=+1$ and $\beta_i\in \Z$.  To understand in which of the cases (1)--(9) this occurs, we must introduce one further concept. 

Given a solution $(\sigma_1\,\sigma_2\,\sigma_3)$ of the singularity condition \eqref{e:cond_sigma2}, we will term this solution to be {\it paired} on index $\sigma_1$ if there also exists a solution to \eqref{e:cond_sigma2} of type $(-\sigma_1\,\sigma_2\,\sigma_3)$.  Similarly, the solution is paired on index $\sigma_2$ if there exists a solution of type $(\sigma_1\, {-}\sigma_2\, \sigma_3)$, and it is paired on index $\sigma_3$ if there exists a solution of type $(\sigma_1\,\sigma_2\,{-}\sigma_3)$. 
Thus, the solutions of \eqref{e:cond_sigma2} in cases (1)--(4) are not paired as only a single 
solution is present in each case, while in case (5) the two solutions are both paired on the last index. If we had solutions of type $(++-)$, $(+--)$ and $(---)$, as occurs in one of the subdivisions of case (9), then the first of these solutions is paired on the second index; the second on the first and second indices; and the third on only the first index.

The significance of pairing is as follows.  Given a solution $(\sigma_1\,\sigma_2\,\sigma_3)$ of \eqref{e:cond_sigma2}, for this solution to be paired on index $\sigma_1$ requires the quantity $n$ defined by 
\[
-2n \equiv \alpha+1-\sigma_1\beta_1+\sigma_2\beta_2+\sigma_3\beta_3 = -2(k_{\sigma_1\sigma_2\sigma_3}+\sigma_1s_1|\beta_1|)
\]
to be a non-negative integer.  Thus, if the solution is paired on index $\sigma_1$, then $\beta_1$ must be an integer. 
If instead the solution is not paired, $n$ is either non-integer or else a negative integer.  
In the case where $\sigma_1 s_1=+1$, the solution not being paired on $\sigma_1$ then implies $\beta_1$ is non-integer.  (Recall $k_{\sigma_1\sigma_2\sigma_3}$ is a non-negative integer).  In the remaining case where $\sigma_1 s_1 = -1$, knowing that that solution is not paired on $\sigma_1$ does not tell us anything about  whether or not $\beta_1$ is integer.

To have a logarithm requires both integer $\beta_i$ and also $s_i\sigma_i=+1$.  Tabulating all possibilities as per table \ref{logtable}, we see that if two solutions are paired on an index $\sigma_i$ then we always have a logarithm from the solution with $\sigma_i=s_i$. 
The momentum factor $p_i^{(1+\sigma_i)\beta_i}$ accompanying this log is however always analytic.
On the other hand, if a solution is not paired on some index $\sigma_i$, there are no log contributions, and for the accompanying momentum factor to be non-analytic requires $\sigma_i=+1$.   If $\sigma_i=-1$ the accompanying momentum factor is always analytic, regardless of pairing.

\begin{table}[t]
\begin{center}
\begin{tabular}{|c|c|c|c|c|c|}
\hline
$s_i$ & $\sigma_i$ & Paired & $\beta_i\in \Z$ & $\ln$ present & $p_i^{(1+\sigma_i)\beta_i}$ non-analytic \\ \hline
$+$ & $+$ & $\checkmark$ & $\checkmark$ & $\checkmark$ & $\times$ \\ \hline
$+$ & $+$ & $\times$ & $\times$ & $\times$ & $\checkmark$ \\ \hline
$+$ & $-$ & $\checkmark$ &      $\checkmark$ & $\times$ &    $\times$ \\ \hline
$+$ &$-$ & $\times$ & ? & $\times$ & $\times$ \\ \hline
$-$ & $+$ & $\checkmark$ & $\checkmark$ & $\times$ & $\checkmark$ \\ \hline
$-$ & $+$ & $\times$ & ? & $\times$ & $\checkmark$ \\ \hline
$-$ & $-$ & $\checkmark$ & $\checkmark$ & $\checkmark$ & $\times$ \\ \hline
$-$ & $-$ & $\times$ & $\times$ & $\times$ & $\times$ \\ \hline
\end{tabular}
\end{center}
\caption{For any given solution of \eqref{e:cond_sigma2}, from the sign $s_i$ of $\beta_i$ and whether or not the solution is paired on $\sigma_i$, 
we can deduce whether an order-boosting logarithm is present
as well as whether the accompanying factor of momentum is non-analytic in $p_i^2$.  In this manner one can reconstruct the pole structure and locality properties of the divergent parts of the regulated triple-$K$ integral.\label{logtable}}
\end{table}

With these considerations in place, we can easily understand the order of the leading divergence of the regulated triple-$K$ integral given in table \ref{sing_table}.  From \eqref{xintdiv}, this order is one more than the maximum number of logs that can occur in each case.  The maximum number of logs is in turn given by the maximum number of indices on which any of the solutions present is paired.  Thus, for example, case (7) is only order $\ep^{-1}$ divergent (rather than $\ep^{-2}$, the maximum allowed order when two solutions of \eqref{e:cond_sigma2} are present) because neither solution is paired on any of its indices.
In case (9), on the other hand, the leading divergence can be $\ep^{-3}$ when the $(+--)$ solution (or one of its permutations) is paired on both its first and second indices.

Going through each of the cases (1)--(9) with the aid of table \ref{logtable}, we can now reconstruct all divergences of the regulated triple-$K$ integral and read off the renormalised correlators where possible.  Before proceeding to the complete listing below, let us first run through a few examples. 
In case (5), for instance, both solutions are paired on the third index meaning $\beta_3\in \Z$ and we have one logarithm present.  From table \ref{logtable}, this log is associated with the type $(+++)$ solution if $\beta_3\in \Z^+$; otherwise it is associated with the $(++-)$ solution.  The leading divergence is therefore of order $\ep^{-2}$ and carries a momentum factor of either $F_{+++}$ or $F_{++-}$ according to which solution has the log.  (Note that, after splitting $\ln (p_3 x) = \ln x+ \ln p_3$, only the $\ln x$ part acts to boost the order of the divergence: the remaining $\ln p_3$ piece contributes only to the {\it subleading} $\ep^{-1}$ divergence.)
Examining this leading $\ep^{-2}$ divergence we see that is nonlocal due to the non-analytic factors of momentum associated with the first two indices.  This is indeed as we expect since no counterterms are available for removing divergences: instead we must multiply the regulated triple-$K$ integral by an overall constant of order $\ep^2$ before sending $\ep\rightarrow 0$ to extract the renormalised correlator.

As a second example, let us consider case (9b), the most complicated case, where $(++-)$, $(+--)$, $(-+-)$ and $(---)$ solutions of \eqref{e:cond_sigma2} are present.    Here, each solution is paired on both its first and its second indices.  The number of logarithms associated with each solution then depends on the signs of $\beta_1$ and $\beta_2$.  To have a logarithm requires $\sigma_i s_i = +1$, hence if both $s_1=s_2=+1$ then the $(++-)$ solution has two logarithms (\textit{i.e.}, contributes a factor of $\ln(p_1 x)\ln (p_2 x)$), the $(+--)$ and the $(-+-)$ solution each have only a single logarithm ($\ln (p_1x)$ and $\ln (p_2x)$ respectively), while the $(---)$ solution has none.   The leading divergence is therefore $\ep^{-3} F_{++-}$, however from table \ref{logtable} this is ultralocal as the momentum factors associated with all three indices are analytic.  The subleading divergence at order $\ep^{-2}$ is then semilocal, and only the sub-subleading order $\ep^{-1}$ divergence is nonlocal.
It is this last quantity therefore that is proportional to the renormalised correlator, which may be obtained by removing the leading and subleading divergences through counterterms, multiplying by an overall constant of order $\ep$, then sending $\ep\rightarrow 0$. 
Its momentum dependence, given in \eqref{9bppresult} below,  follows from collecting terms without factors of $\ln x$: for the $(++-)$ solution this is $F_{++-}\ln p_1 \ln p_2$, for the $(+--)$ solution this is $F_{+--}\ln p_1$, {\it etc}.  

We are now ready to list the complete results as follows.  In cases (3), (4) and (8) where it is not possible to determine the renormalised correlator we have instead listed the complete singularity structure of the regulated triple-$K$ integral, which contains only ultralocal or semilocal terms.  In the remaining cases where we provide results for the renormalised correlator, note that these are specified in a  particular choice of renormalisation scheme; when type $(+--)$ or $(---)$ solutions are present we can adjust  the coefficients of ultra- and/or semilocal terms arbitrarily by adding finite counterterms to the action.
The function of momentum $F_{\sigma_1\sigma_2\sigma_3}$ is as defined in \eqref{e:F2}.

\paragraph{Case (1): $\bs{(+++)}$ only} 

\[\label{case1_result}
\lla \O_1(\bs{p}_1)\O_2(\bs{p}_2)\O_3(\bs{p}_3)\rra \propto F_{+++} 
\]

\paragraph{Case (2): $\bm{ (++-)}$ only} 
\[\label{case2_result}
\lla \O_1(\bs{p}_1)\O_2(\bs{p}_2)\O_3(\bs{p}_3)\rra \propto F_{++-} 
\]

\paragraph{Case (3a): $\bm{(+--)}$ only}  
\[
I_{\alpha+\ep,\{\beta_i\}}^{\rm{div}} \propto \ep^{-1} F_{+--}\]

\paragraph{Case (3b): $\bm{(+--)}$ and  $\bm{(-+-)}$ }  

\[
I_{\alpha+\ep,\{\beta_i\}}^{\rm{div}} \propto \ep^{-1} \big(F_{+--}+F_{-+-}\big)\]

\paragraph{Case (3c): $\bm{(+--)}$, $\bm{(-+-)}$ and $\bm{(--+)}$}  

\[
I_{\alpha+\ep,\{\beta_i\}}^{\rm{div}} \propto \ep^{-1} \big(F_{+--}+F_{-+-}+F_{--+}\big)\]

\paragraph{Case (4): $\bm{(---)}$ only }

\[
 I_{\alpha+\ep,\{\beta_i\}}^{\rm{div}}\propto \ep^{-1}F_{---}
\]

\paragraph{Case (5):  $\bm{(+++)}$ and  $\bm{(++-)}$}

\[\label{vppp}
\lla \O_1(\bs{p}_1)\O_2(\bs{p}_2)\O_3(\bs{p}_3)\rra \propto
\left\{ 
\begin{array}{ll}
 F_{+++} & \text{ if } \beta_3\ge 0, \\[1ex]
 F_{++-} & \text{ if } \beta_3<0.
\end{array}
\right.
\]

\paragraph{Case (6a):  $\bm{(++-)}$ and  $\bm{(+--)}$   }

\begin{align}
\lla \O_1(\bs{p}_1)\O_2(\bs{p}_2)\O_3(\bs{p}_3)\rra \propto 
\left\{
\begin{array}{ll}
F_{++-}\ln p_2 + F_{+--} & \text{ if } \beta_2\ge 0, \\[1ex]
F_{++-} + F_{+--} \ln p_2 & \text{ if } \beta_2 < 0.
\end{array}
\right.
\end{align}

\paragraph{Case (6b):  $\bm{(++-)}$, $\bm{(+--)}$ and $\bm{(-+-)}$   }

\begin{align}
\lla \O_1(\bs{p}_1)\O_2(\bs{p}_2)\O_3(\bs{p}_3)\rra \propto 
\left\{
\begin{array}{ll}
F_{++-}\ln p_1 + F_{+--}\ln p_2 +F_{-+-} & \text{ if } \beta_1\ge 0 \text{ and } \beta_2< 0, \\[1ex]
F_{++-}\ln p_2 + F_{+--} +F_{-+-}\ln p_1 & \text{ if } \beta_1<0 \text{ and } \beta_2 \ge 0, \\[1ex]
F_{++-} + F_{+--}\ln p_2 +F_{-+-}\ln p_1 & \text{ if } \beta_1<0 \text{ and } \beta_2 < 0. 
\end{array}
\right.
\end{align}
Note that we cannot have both $\beta_1\ge 0$ {\it and} $\beta_2\ge 0$ here: taking linear combinations of the solutions of \eqref{e:cond_sigma2}, we find $\beta_1+\beta_2= -2k_{++-}+k_{+--}+k_{-+-}$.  In the absence of a $(---)$ solution, we know moreover that $-2k_{---}\equiv\alpha+1-\beta_1-\beta_2-\beta_3 = -2(k_{+--}+k_{-+-}-k_{++-})$ must be such that $k_{---} \in \Z^-$, and hence $\beta_1+\beta_2 = k_{---}-k_{++-}\le 0$.
As we cannot have both $\beta_1\ge 0$ and $\beta_2\ge0$, there are then no double-log contributions to the renormalised correlator even though the $(++-)$ solution is paired on both its first and second indices.  Independently, we  know that such a contribution cannot appear since it would imply the existence of an $\ep^{-3}$ divergence, however this  is forbidden since we have only two different types of solution of \eqref{e:cond_sigma2} ignoring permutations, and hence at most an order $\ep^{-2}$ divergence.

\paragraph{Case (7): $\bm{(++-)}$ and $\bm{(---)}$}

\begin{align}\label{case7result}
\lla \O_1(\bs{p}_1)\O_2(\bs{p}_2)\O_3(\bs{p}_3)\rra &\propto F_{++-}+F_{---}.
\end{align}

\paragraph{Case (8a): $\bm{(+--)}$ and $\bm{(---)}$}

\begin{align}
I_{\alpha+\ep,\{\beta_i\}}^{\rm{div}}
\propto 
\left\{
\begin{array}{ll}
-\ep^{-2}F_{+--}+\ep^{-1}\big(F_{+--}\ln p_1+F_{---}\big) & \text{ if } \beta_1 \ge 0,\\[1ex]
-\ep^{-2}F_{---}+\ep^{-1}\big(F_{+--}+F_{---}\ln p_1\big) & \text{ if } \beta_1<0.
\end{array}
\right.
\end{align}
The relative minus sign between the leading and subleading terms here arises from  \eqref{xintdiv}.

\paragraph{Case (8b): $\bm{(+--)}$, $\bm{(-+-)}$ and $\bm{(---)}$}

\begin{align}
&I_{\alpha+\ep,\{\beta_i\}}^{\rm{div}} \nn\\[1ex]
&\propto 
\left\{
\begin{array}{ll}
-\ep^{-2}\big(F_{+--}+F_{-+-}\big)
+\ep^{-1}\big(F_{+--}\ln p_1+F_{-+-}\ln p_2+F_{---}\big) & \text{ if } \beta_1\ge 0 \text{ and } \beta_2\ge 0, \\[1ex]
-\ep^{-2}\big(F_{+--}+F_{---}\big)
+\ep^{-1}\big(F_{+--}\ln p_1+F_{-+-} +F_{---}\ln p_2 \big) & \text{ if } \beta_1\ge 0 \text{ and } \beta_2< 0, \\[1ex]
-\ep^{-2}\big(F_{-+-}+F_{---}\big)
+\ep^{-1}\big(F_{+--} +F_{-+-}\ln p_2+F_{---}\ln p_1 \big) & \text{ if } \beta_1< 0 \text{ and } \beta_2\ge 0. \\[1ex]
\end{array}
\right.
\end{align}
In this subcase we cannot have both $\beta_1<0$ {\it and} $\beta_2<0$: taking linear combinations of the solutions of \eqref{e:cond_sigma2}, we find $\beta_1+\beta_2 =2k_{---}-k_{+--}-k_{-+-}$.
The absence of a $(++-)$ solution means however that $-2k_{++-}\equiv \alpha+1+\beta_1+\beta_2-\beta_3 = -2(k_{+--}+k_{-+-}-k_{---})$ is such that $k_{++-}\in \Z^-$, and hence $\beta_1+\beta_2 = -k_{++-}+k_{---}>0$.
As we cannot have both $\beta_1<0$ and $\beta+2<0$, there are then no nonlocal double-log contributions to $I_{\alpha+\ep,\{\beta_i\}}^{\rm{div}}$ even though the $(---)$ solution is paired on both first and second indices.  We know independently that such a contribution cannot arise as it would imply the presence of an $\ep^{-3}$ divergence which is forbidden since, discounting permutations, we only have two different types of solution of \eqref{e:cond_sigma2}, and hence at most an order $\ep^{-2}$ divergence.

\paragraph{Case (8c): $\bm{(+--)}$, $\bm{(-+-)}$, $\bm{(--+)}$ and $\bm{(---)}$}

\begin{align}
I_{\alpha+\ep,\{\beta_i\}}^{\rm{div}} 
&\propto 
-\ep^{-2}\big(F_{+--}+F_{-+-}+F_{--+}\big)\nn\\&\quad
+\ep^{-1}\big(F_{+--}\ln p_1 + F_{-+-}\ln p_2 +F_{--+}\ln p_3+ F_{---}\big).
\end{align}
Note in this subcase that all $\beta_i>0$, as can be shown by adding pairwise the $(+--)$, $(-+-)$ and $(--+)$ solutions of \eqref{e:cond_sigma2}.

\paragraph{Case (9a): $\bm{(++-)}$, $\bm{(+--)}$ and $\bm{(---)}$}

\begin{align}\label{9aresult}
\lla \O_1(\bs{p}_1)\O_2(\bs{p}_2)\O_3(\bs{p}_3)\rra 
\propto 
\left\{
\begin{array}{ll}
F_{++-}\ln p_2+F_{+--}\ln p_1+F_{---} 
& \text{ if } \beta_1\ge 0 \text{ and } \beta_2\ge 0, \\[1ex]
F_{++-} +F_{+--}\ln p_1 \ln p_2 +F_{---} 
& \text{ if } \beta_1\ge 0 \text{ and } \beta_2 < 0, \\[1ex]
F_{++-}\ln p_2+F_{+--} +F_{---}\ln p_1 
& \text{ if } \beta_1< 0 \text{ and } \beta_2\ge 0, \\[1ex]
F_{++-} +F_{+--}\ln p_2+F_{---}\ln p_1 
& \text{ if } \beta_1< 0 \text{ and } \beta_2< 0.
\end{array}
\right.
\end{align}

\paragraph{Case (9b): $\bm{(++-)}$, $\bm{(+--)}$, $\bm{(-+-)}$ and $\bm{(---)}$}

\begin{align}\label{9bppresult}
&\lla \O_1(\bs{p}_1)\O_2(\bs{p}_2)\O_3(\bs{p}_3)\rra 
\nn\\[1ex]
&\propto  
\left\{
\begin{array}{ll}
F_{++-}\ln p_1\ln p_2+F_{+--}\ln p_1+F_{-+-}\ln p_2+F_{---} 
& \text{ if } \beta_1\ge 0 \text{ and } \beta_2\ge 0,  \\[1ex]
F_{++-}\ln p_1+F_{+--}\ln p_1\ln p_2 +F_{-+-}+F_{---}\ln p_2 
& \text{ if } \beta_1\ge 0 \text{ and } \beta_2< 0, \\[1ex]
F_{++-}\ln p_2 +F_{+--} +F_{-+-}\ln p_1\ln p_2+F_{---}\ln p_1 
& \text{ if } \beta_1< 0 \text{ and } \beta_2 \ge 0, \\[1ex]
F_{++-} +F_{+--}\ln p_2+F_{-+-}\ln p_1+F_{---}\ln p_1\ln p_2 
& \text{ if } \beta_1< 0 \text{ and } \beta_2< 0.
\end{array}
\right.
\end{align}

\subsection{Non-uniqueness of the triple-$K$ representation and scheme dependence} 
\label{subsec:non-uniqueness}

The homogeneous conformal Ward identities for the regulated correlator are equivalent to the system of equations defining the generalised hypergeometric function of two variables Appell $F_4$ \cite{Bzowski:2013sza, Coriano:2013jba}. 
This system of equations has four solutions in general, but three of these solutions possess singularities in the collinear limit where the momenta satisfy $p_3=p_1+p_2$ (or similar).  Of the four solutions, the only one free from collinear singularities  is the triple-$K$ integral.
For the cases discussed in \cite{Bzowski:2013sza},  for which renormalisation is not required, the triple-$K$ integral is then the unique representation of  the 3-point correlator.

The correlators studied in this paper do however require renormalisation, and the issue of uniqueness of the triple-$K$ representation is consequently more subtle.  Here, the absence of collinear singularities need hold only for the {\it renormalised} correlator obtained after we have sent $\ep\rightarrow 0$.  The {\it regulated} correlator, obtained by solving the regulated homogeneous Ward identities and subtracting divergences with the aid of counterterms, 
must therefore have a finite piece of order $\ep^0$ that is free from collinear singularities (being equal to the renormalised correlator), but also pieces that are of higher order in $\ep$ which vanish in the limit $\ep\rightarrow 0$.  There is no physical reason why these higher-order pieces should be free from collinear singularities, since they make no contribution to the renormalised correlator.  
Thus, given the four general solutions to the regulated homogeneous Ward identities, we should only impose that the finite order $\ep^0$ piece (after subtracting counterterms and multiplying through by any required overall factors of $\ep$) is free from collinear singularities.  This additional freedom renders the triple-$K$ representation non-unique, but the non-uniqueness simply corresponds to our freedom to change the renormalisation scheme by adding finite counterterms to the action.

Let us examine this argument in greater detail.
As per the discussion in \cite{Bzowski:2013sza}, in the present $(1,0)$-scheme the four general solutions of the regulated homogeneous conformal Ward identities take the form 
\begin{equation} \label{e:IIK}
p_1^{\beta_1} p_2^{\beta_2} p_3^{\beta_3} \int_0^\infty \D x \: x^{\alpha+\ep} I_{\pm\beta_1} (p_1 x) I_{\pm\beta_2} (p_2 x) K_{\beta_3} (p_3 x),
\end{equation}
where $I_\beta(x)$ is a modified Bessel function of the first kind. 
As with the triple-$K$ integral, we can split each integral into a finite upper part for which $\mu^{-1}\le x<\infty$, and a lower part for which  
$0\le x < \mu^{-1}$.  Once again, all the divergences as $\ep\rightarrow 0$ arise solely from the lower parts.
From the large-$x$ asymptotic expansions
\begin{equation} \label{e:A}
I_\beta(x) = \frac{1}{\sqrt{2 \pi x}} \,e^x + \ldots, \qquad K_\beta(x) = \sqrt{\frac{\pi}{2 x}} \,e^{-x} + \ldots,
\end{equation}
we see the upper parts are always singular for the collinear momentum configuration $p_3=p_1+p_2$ in any dimension $d\ge 3$.
The only way to eliminate this collinear singularity is to take appropriate linear combinations of the four solutions so that the leading asymptotic behaviours cancel, \textit{i.e.}, by combining the Bessel $I$ to make Bessel $K$ functions, 
\begin{equation}
K_\beta(x) = \frac{\pi}{2 \sin ( \beta \pi)} \left[ I_{-\beta}(x) - I_{\beta}(x) \right].
\end{equation}
The triple-$K$ integral is thus the unique combination with an  upper part that is free from collinear singularities.

Turning now to the lower parts, through a modification of our earlier arguments we easily see that the divergences these contribute are always free from collinear singularities.
First, we recall that Bessel $I$ has the series expansion
\[
I_\beta(z) = z^\beta \,\sum_{k=0}^\infty \frac{1}{2^{2k+\beta}k!\Gamma(\beta+k+1)}\, z^{2k},
\]
valid for any $\beta\notin \Z^-$.
To handle all cases including $\beta\in \Z^-$, it is convenient to choose a different basis for the four general solutions of the homogeneous Ward identities in which all Bessel $I_{-|\beta|}(z)$ are recombined into Bessel $K_\beta(z)=K_{|\beta|}(z)$. 
Our new basis thus consists of the original triple-$K$ integral plus the three  integrals
\begin{align}\label{Idiv1}
I^{(1)}_{\alpha+\ep,\{\beta_i\}} &= p_1^{\beta_1} p_2^{\beta_2} p_3^{\beta_3} \int_0^\infty \D x \: x^{\alpha+\ep} I_{|\beta_1|} (p_1 x) K_{|\beta_2|} (p_2 x) K_{|\beta_3|} (p_3 x),\\ \label{Idiv2}
I^{(2)}_{\alpha+\ep,\{\beta_i\}} &= p_1^{\beta_1} p_2^{\beta_2} p_3^{\beta_3} \int_0^\infty \D x \: x^{\alpha+\ep} K_{|\beta_1|} (p_1 x) I_{|\beta_2|} (p_2 x) K_{|\beta_3|} (p_3 x),\\ \label{Idiv3}
I^{(3)}_{\alpha+\ep,\{\beta_i\}} &= p_1^{\beta_1} p_2^{\beta_2} p_3^{\beta_3} \int_0^\infty \D x \: x^{\alpha+\ep} I_{|\beta_1|} (p_1 x) I_{|\beta_2|} (p_2 x) K_{|\beta_3|} (p_3 x).
\end{align}
To further simplify matters, we observe that
\[
I_{|\beta|}(z) = (-1)^\chi\, z^{|\beta|}\,\sum_{k=0}^\infty a_k^+(|\beta|) z^{2k},
\] 
where the $a_k^+(\beta)$ are as defined in \eqref{e:ap}, where $\chi=1$ if $\beta\notin \Z$ and $\chi=\beta+1$ if $\beta\in\Z^+$.
The divergences  of the three solutions \eqref{Idiv1}-\eqref{Idiv3}  can now be evaluated following the same method we used for the triple-$K$ integral.  In fact, up to an irrelevant constant overall phase arising from the factors of $(-1)^\chi$, the divergences are the same as for the triple-$K$ integral except that we discard logs and set the $a^{-}_k(|\beta|)$ to zero every time we encounter a Bessel $I$ in place of a Bessel $K$.  (Or equivalently, when we have a Bessel $I$, we only obtain a nonzero contribution if $\sigma_i=s_i$ for that index.)
As all these divergences are simply products of momenta raised to various powers, there are consequently never any collinear singularities.

Thus, when the renormalised correlator is given by the {\it finite} part of a solution of the regulated homogeneous conformal Ward identities, the triple-$K$ integral is the unique solution.  When, on the other hand, the renormalised correlator is given by the {\it divergent} part of a solution to the regulated homogeneous Ward identities, there are potentially additional contributions besides the triple-$K$ integral.
These additional contributions encode our freedom to change the renormalisation scheme by adding finite counterterms to the action.  

An example of this is case (7), where we have $(++-)$ and $(---)$ type singularities.  As we saw earlier in  example 13 on page \pageref{Example13label}, the renormalised correlator satisfies the homogeneous conformal Ward identities.  In fact, both the $F_{++-}$ and the $F_{---}$ pieces of the general solution \eqref{case7result} {\it independently} satisfy the homogeneous conformal Ward identities, whose solution is therefore not unique.  (Note the coefficient of the $F_{---}$ term in \eqref{case7result} can be adjusted arbitrarily through the addition of an appropriate counterterm.)
In this case, the renormalised correlator corresponds to the leading $\ep^{-1}$ order divergence of the regulated triple-$K$ integral, which must be multiplied through by an overall constant of order $\ep$.
The non-uniqueness therefore corresponds to the presence of additional solutions of the regulated homogeneous Ward identities of order $\ep^{-1}$.
Collecting together contributions at this order from \eqref{Idiv1}-\eqref{Idiv3}, up to an overall constant of proportionality we obtain
\begin{align}
\left\{ 
\begin{array}{ll}
F_{++-} & \text{ if } s_1=s_2=+1, \\[2ex]
F_{++-}+c F_{---} & \text{ if } s_1=-s_2 \\[2ex]
F_{---} & \text{ if } s_1=s_2=-1. 
\end{array}
\right.\label{case7arb}
\end{align}
(Here, $c$ is an arbitrary constant reflecting the fact that in the case where $s_1=-s_2$, one of the solutions comes from \eqref{Idiv1} and the other from \eqref{Idiv2}.)  
As the contribution from the triple-$K$ integral is proportional to $F_{+++}+F_{---}$ for all values of the signs $s_i$ (see \eqref{case7result}),  $F_{+++}$ and $F_{---}$ are indeed {\it independent} solutions to the homogeneous Ward identities. This uniqueness simply reflects our ability to adjust the coefficient of the $F_{---}$ solution by adding finite counterterms of the $(---)$ type.

\section{Correlators of operators with shadow dimensions: $\Delta$ and $d-\Delta$}\label{sec:shadow}

In this appendix we discuss the relation between correlation functions of operators of dimensions $\Delta$ and $d-\Delta$.
We will assume operators of generic dimensions, \textit{i.e.}, none of the conditions that lead to singularities hold. We also set 
the normalisation of the 2- and 3-point functions to unity.

First, note that under
\be
\Delta \to d- \Delta \quad \Rightarrow \quad \beta = 2 \Delta -d \to -\beta
\ee
It follows that
\be \label{2pt_shadow}
\lla \O_{[d-\Delta]}(\bs{p}) \O_{[d-\Delta]}(-\bs{p}) \rra = \frac{1}{\lla \O_{[\Delta]}(\bs{p}) \O_{[\Delta]}(-\bs{p}) \rra }.
\ee

Moving now to 3-point functions, we note that since $K_{\beta} (x) = K_{-\beta}(x)$, the 3-point functions of correlators of 
$\O_{[\Delta]}$ and $\O_{[d-\Delta]}$ involve the same triple-$K$ integral and thus are simply related to one another.
For example,
\begin{align}
&\lla \O_{[d-\Delta_1]}(\bs{p}_1) \O_{[d - \Delta_2]}(\bs{p}_2) \O_{[d-\Delta_3]}(\bs{p}_3) \rra =\int_0^\infty \D x \: x^\alpha \prod_{j=1}^3 p_j^{-\beta_j} K_{-\beta_j}(p_j x) \nonumber \\
& \quad =\frac{1}{p_1^{2 \beta_1} p_2^{2 \beta_2} p_3^{2 \beta_3}} \int_0^\infty \D x \: x^\alpha \prod_{j=1}^3 p_j^{\beta_j} K_{\beta_j}(p_j x)
=\frac{\lla \O_{[\Delta_1]}(\bs{p}_1) \O_{[\Delta_2]}(\bs{p}_2) \O_{[\Delta_3]}(\bs{p}_3) \rra }{
\prod_{i=1}^3\lla \O_{[\Delta_i]}(\bs{p}_i) \O_{[\Delta_i]}(-\bs{p}_i) \rra}. \label{3pt_shadow}
\end{align}
It was argued in \cite{Klebanov:1999tb,Gubser:2002vv} in the context of AdS/CFT that the CFT with a source for the operator $\O_{[d-\Delta]}$ can be obtained from the CFT with a source for the operator $\O_{[\Delta]}$ by means of a Legendre transform that acts on the sources. It is straightforward to check that  (\ref{2pt_shadow}) and (\ref{3pt_shadow}) can be understood in this fashion. We emphasise however that this holds only for generic dimensions, \textit{i.e.}, when none of the conditions that lead to singularities hold, as it is clear from the discussion of 3-point functions of operators of dimensions one and two in section \ref{subsubsec:1/ep}.

\section{Examples using free fields}\label{sec:freefields}

In this section we use free field computations to check some our results in the main text relating to the two examples $\< \O_{[4]}\O_{[3]}\O_{[3]}\>$ in $d=4$ and $\< \O_{[3]}\O_{[3]}\O_{[3]}\>$ in $d=3$.

\bigskip

{\it \textbullet\ Example 14: 
$d = 4$, $\Delta_1 = 4, \Delta_2 = \Delta_3 = 3$.}

\bigskip

\begin{figure}[t]
\begin{subfigure}{0.4\textwidth}
\includegraphics[width=0.90\linewidth]{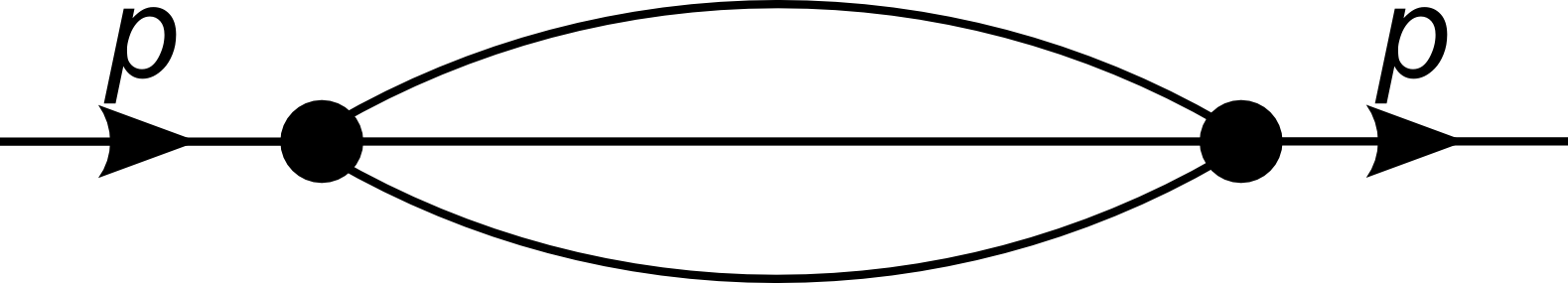}
\end{subfigure}
\hspace{1.5cm}
\begin{subfigure}{0.4\textwidth}
\includegraphics[width=0.90\textwidth]{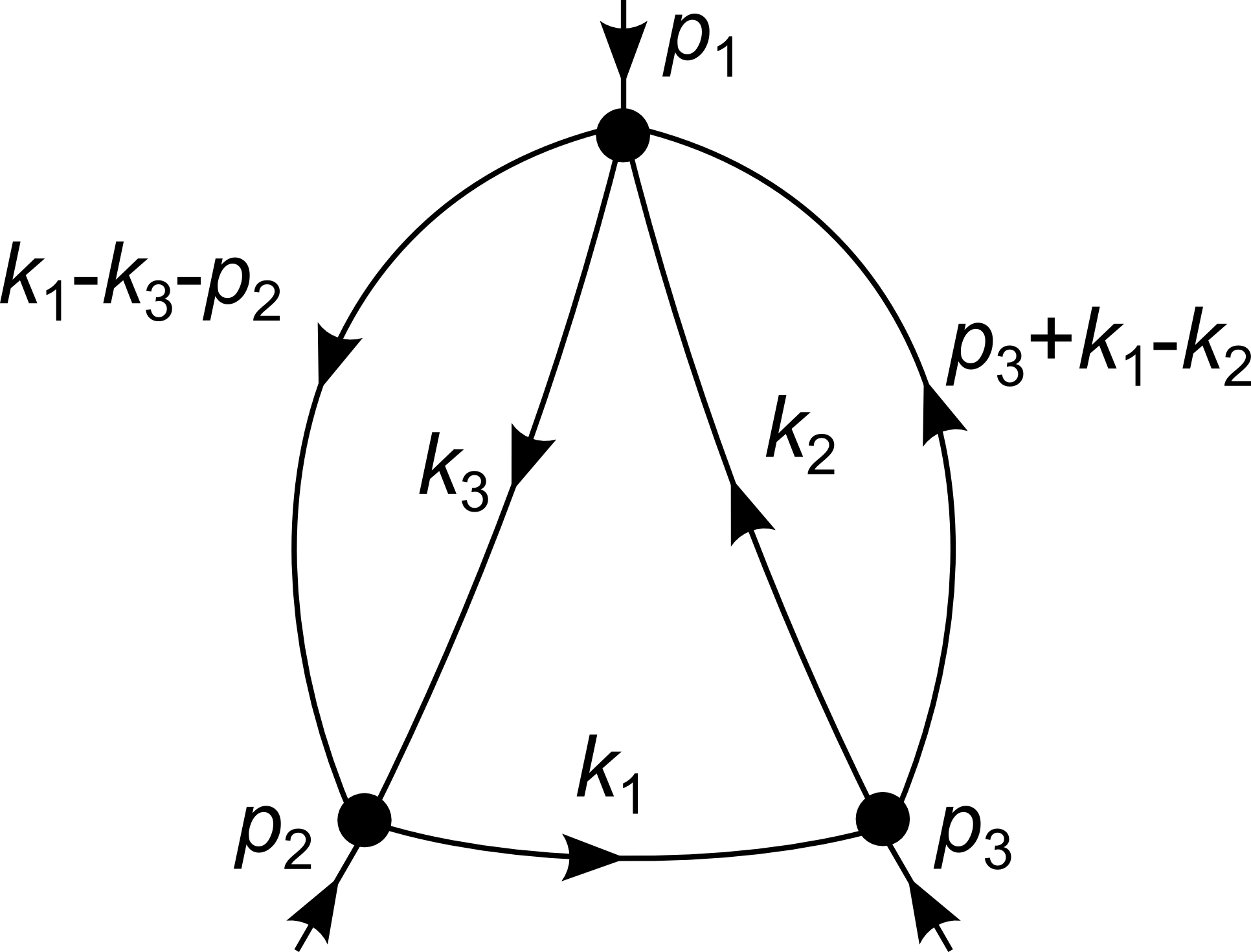}
\end{subfigure}
\centering
\caption{Feynman graphs representing  $\<\O_{[3]}\O_{[3]}\>$  and $\< \O_{[4]} \O_{[3]} \O_{[3]} \>$ for a free scalar $\Phi$ with  $\O_{[3]} =\, \lwick \Phi^3 \rwick$  and $\O_{[4]} =\, \lwick \Phi^4 \rwick$.}
\label{fig:Feynman1}
\end{figure}

The propagator for a single real scalar field $\Phi$ in four dimensions is
\begin{equation}
\< \Phi(\bs{k}) \Phi(\bs{k}') \> = (2 \pi)^d \delta( \bs{k} + \bs{k}' ) \frac{1}{k^2}.
\end{equation}  
In position space, the operators $\O_{[4]}$ and $\O_{[3]}$ can be realised as
\begin{equation}
\O_{[4]}(\bs{x}) =\, \lwick \Phi^4(\bs{x}) \rwick, \qquad\qquad \O_{[3]}(\bs{x}) =\, \lwick \Phi^3(\bs{x}) \rwick.
\end{equation}
Denoting the corresponding sources by $\phi_{[0]}$ and $\phi_{[1]}$, in dimensional regularisation  the canonical dimensions (defined according to the propagator) are 
\begin{align} \label{e:regfree}
\left[ \Phi \right] &= 1 - \frac{\epsilon}{2}, & \left[ \O_{[3]} \right] &= 3 - \frac{3 \epsilon}{2}, & \left[ \O_{[4]} \right] &= 4 - 2 \epsilon, \nn\\
 d &= 4 - \epsilon, &
 \left[ \phi_{[1]} \right] &= 1 + \frac{\epsilon}{2}, &
 \left[ \phi_{[0]} \right] &= \epsilon. 
\end{align}
Up to multiplicity factors, the
2- and 3-point functions are represented by the diagrams presented in figure \ref{fig:Feynman1}. All correlators may be evaluated using the integral
\begin{align} \label{e:i2}
\int \frac{\D^d \bs{k}}{(2 \pi)^d} \frac{1}{k^{2 a} | \bs{p} - \bs{k} |^{2 b}} & = C_{d, a, b} p^{d - 2 (a + b)}, 
\end{align}
where
\[
C_{d, a, b}  = \frac{\Gamma \left(a + b - \frac{d}{2} \right) \Gamma \left(\frac{d}{2} - a\right) \Gamma \left(\frac{d}{2} - b\right)}{(4 \pi)^{d/2} \Gamma(a) \Gamma(b)\Gamma(d - a - b)}.
\]

Immediately, we then find 
\begin{align}\label{2ptreg433free}
\lla \O_{[3]}(\bs{p}) \O_{[3]}(-\bs{p}) \rra_{\text{reg}} & = 6 C_{4-\ep, 1, 1} C_{4 - \ep, 1, \frac{\ep}{2}} p^{2 - 2 \ep} \nn\\
& = - \frac{3 p^2}{256 \pi^4 \epsilon} + \frac{3 p^2}{256 \pi^4} \left[ \ln p^2 + \gamma_E - \ln (4 \pi) - \tfrac{13}{4} \right] + O(\epsilon).
\end{align} 
The counterterm 
\begin{equation} \label{e:Sct2free}
S_{\text{ct}}^{(2)} = a(\epsilon) \int \D^d \bs{x} \: \phi_{[1]} \Box \phi_{[1]} \mu^{-2 \epsilon}
\end{equation}
can be added to the action to yield a finite renormalised 2-point function
\begin{equation}\label{2pt433free}
\lla \O_{[3]}(\bs{p}) \O_{[3]}(-\bs{p}) \rra = \frac{3}{256 \pi^4} p^2 \ln \frac{p^2}{\mu^2}.
\end{equation}
We have chosen subleading terms in the renormalisation constant $a(\epsilon)$ in such a way that the ultralocal portion of the 2-point function vanishes. This choice of renormalisation scheme will simplify subsequent expressions, although other choices of scheme are possible.

The 3-point function is given by the Feynman integral
\begin{equation}\label{433freedef}
\lla \O_{[4]}(\bs{p}_1) \O_{[3]}(\bs{p}_2) \O_{[3]}(\bs{p}_3) \rra_{\text{reg}} = 216 \, I(p_1, p_2, p_3),
\end{equation}
where
\begin{equation} \label{e:toint}
I = \int \frac{\D^d \bs{k}_1}{(2 \pi)^d} \frac{\D^d \bs{k}_2}{(2 \pi)^d} \frac{\D^d \bs{k}_3}{(2 \pi)^d} \frac{1}{k_1^2 k_2^2 k_3^2 (\bs{k}_1 - \bs{k}_2 + \bs{p}_3)^2 (\bs{k}_1 - \bs{k}_3 - \bs{p}_2)^2}.
\end{equation}
After dimensionally regularising to regulate the nested divergences, the integrals over $\bs{k}_2$ and $\bs{k}_3$ can be calculated using \eqref{e:i2} leading to the result
\begin{equation}
I = C^2_{4 - \epsilon, 1, 1} \, \int \frac{\D^{4 - \epsilon} \bs{k}_1}{(2 \pi)^{4 - \epsilon}} \frac{1}{k_1^2 |\bs{k}_1 + \bs{p}_3|^{\epsilon} |\bs{k}_1 - \bs{p}_2|^{\epsilon}}.
\end{equation}
The integral on the right-hand side can be re-expressed as a triple-$K$ integral according to equation (A.3.17) in \cite{Bzowski:2013sza}.  This gives
\begin{equation} \label{e:intfree}
I = \frac{2^{2 + \frac{\epsilon}{2}} C^2_{4 - \epsilon, 1, 1}}{(4 \pi)^{2 - \frac{\epsilon}{2}} \Gamma^2 \left( \frac{\epsilon}{2} \right) \Gamma(3 - 2 \epsilon)} I_{1 - \frac{\epsilon}{2} \{ 2 - \frac{3 \epsilon}{2}, 1 - \epsilon, 1 - \epsilon \}}(p_1, p_2, p_3).
\end{equation}
The divergent part of this expression can then be extracted through the method presented in section \ref{subsubsec:reg}, giving
\begin{align}
& \lla \O_{[4]}(\bs{p}_1) \O_{[3]}(\bs{p}_2) \O_{[3]}(\bs{p}_3) \rra_{\text{reg}} = - \frac{9}{256 \pi^6 \epsilon^2} (p_2^2 + p_3^2) \nn\\
& \qquad+ \: \frac{9}{512 \pi^6 \epsilon} \left[ -  p_1^2 + 3 p_2^2 \ln p_2^2 + 3 p_3^2 \ln p_3^2 + (p_2^2 + p_3^2) \left(-10 + 3 \gamma_E - 3 \ln (4 \pi) \right) \right] + O(\epsilon^0). \label{e:O433Feyn}
\end{align}
The form of the counterterm action is given by \eqref{e:Sct3}, up to factors of the renormalisation scale. Taking into account the choice of the regularisation \eqref{e:regfree} used here, we have
\begin{equation} \label{e:Sct3free}
S_{\text{ct}}^{(3)} = \int \D^{4 - \epsilon} \bs{x} \left[ a_0 \phi_{[0]} \phi_{[1]} \O_{[3]} \mu^{-\epsilon} + (a_1 \phi_{[0]} \phi_{[1]} \Box \phi_{[1]} + a_2 \phi_{[1]}^2 \Box \phi_{[0]}) \mu^{- 3 \epsilon} \right].
\end{equation}
The counterterm contribution following from this action is then
\begin{align} \label{e:ctcontribfree}
\lla \O_{[4]}(\bs{p}_1) \O_{[3]}(\bs{p}_2) \O_{[3]}(\bs{p}_3) \rra_{\text{ct}} & = - a_1 (p_2^2 + p_3^2) \mu^{- 3 \epsilon} - 2 a_2 p_1^2  \mu^{- 3 \epsilon} \nn\\
& \quad 
- \: a_0 \mu^{- \epsilon} \left[ \lla \O_{[3]}(\bs{p_2}) \O_{[3]}(-\bs{p_2}) \rra_{\text{reg}} + \lla \O_{[3]}(\bs{p_3}) \O_{[3]}(-\bs{p_3}) \rra_{\text{reg}} \right].
\end{align}
To cancel the divergences, the counterterm constants must be  
\begin{align}\label{free433cts1}
a_0 & = \frac{9}{2 \pi^2 \epsilon} + a_0^{(0)}, \\
a_1 & = \frac{9}{512 \pi^6 \epsilon^2} + \frac{(-9 + 24 \pi^2
a_0^{(0)})}{2048 \pi^6 \epsilon} + a_1^{(0)}, \\ \label{free433cts2}
a_2 & = -\frac{9}{1024 \: \pi^6 \epsilon} + a_2^{(0)},
\end{align}
where $a_0^{(0)}, a_1^{(0)}, a_2^{(0)}$ are $\ep$-independent undetermined constants. 
(In fact, as we saw in section \ref{sec:anomalies_and_betafns}, $a_0^{(0)}$ and $a_2^{(0)}$ are related to each other by the special conformal Ward identity \eqref{CWI433v4}.) 

From the counterterms we can now read off the beta function and anomaly as follows.  The renormalised source $\phi_{[1]}$ is related to the bare source via
\[
\phi_{[1]}^{\rm{bare}} = \phi_{[1]}+a_0 \mu^{-\ep}\phi_{[0]}\phi_{[1]},
\]
which after inverting yields the beta function
\[
\beta_{\phi_{[1]}} = \lim_{\ep\rightarrow 0}\mu\frac{\partial\phi_{[1]}}{\partial\mu} = a_0^{(-1)}\phi_{[0]}\phi_{[1]} = \frac{9}{2\pi^2}\phi_{[0]}\phi_{[1]}.
\]
Comparing this equation and \eqref{2ptreg433free} with  \eqref{beta433} and \eqref{433reg2pt}, we find
\[\label{c433norm}
c_3^{(-1)} v = \frac{3}{256\pi^4}, \qquad
c_{433} = \frac{27}{256\pi^6}.
\]
From \eqref{e:ctcontribfree}, we also have
\begin{align}
&\mu\frac{\partial}{\partial\mu}
\lla \O_{[4]}(\bs{p}_1) \O_{[3]}(\bs{p}_2) \O_{[3]}(\bs{p}_3) \rra = \lim_{\ep\rightarrow 0}\mu\frac{\partial}{\partial\mu}
\lla \O_{[4]}(\bs{p}_1) \O_{[3]}(\bs{p}_2) \O_{[3]}(\bs{p}_3) \rra_{\rm{ct}} \nn\\
& = \frac{27}{512\pi^6}\Big[
-p_1^2+ p_2^2\ln \frac{p_2^2}{\mu^2}+p_3^2\ln\frac{p_3^2}{\mu^2}+\big(\gamma_E-\ln (4\pi)-\frac{7}{2}+\frac{4}{9}\pi^2a_0^{(0)}\big)(p_2^2+p_3^2)\Big],
\end{align}
which agrees with \eqref{433muderiv} on setting
\[\label{433a0prime}
a'_0 = \frac{27}{1024\pi^6}\,\big(-\gamma_E+4+\ln(4\pi)-\frac{4}{9}\pi^2a_0^{(0)}\big).
\]
The anomaly is then
\[
\mathcal{A}_{433}= \frac{27}{512\pi^6}\,\Big[-p_1^2+\Big(
\gamma_E-\ln(4\pi)-\frac{7}{2}+\frac{4}{9}\pi^2a_0^{(0)}\Big)(p_2^2+p_3^2)\Big],
\]
in accord with \eqref{433anomaly}.  As we saw earlier, only the coefficient of $p_1^2$ is physical, with the remainder of the anomaly depending of the choice of scheme through the constant $a_0^{(0)}$.

Finally, one can evaluate the triple-$K$ integral in \eqref{e:intfree} using
the reduction scheme described in \cite{Bzowski:2013sza, integrals} along with a suitable change of regularisation scheme.  Using \eqref{433freedef} and adding in the counterterm contribution \eqref{e:ctcontribfree} to cancel the divergences, on taking $\ep\rightarrow 0$ we recover our earlier result \eqref{e:O433}, normalised according to \eqref{c433norm}.
The scheme-dependent constant $a_0'$ is as given in \eqref{433a0prime} while 
\begin{align}
a_2' & = - \frac{9}{2048 \pi^6} \,\big( 29 + 6 \log (4 \pi) - 6
\gamma_E \big) - 2 a_2^{(0)}.
\end{align}
The value of $a_1'$ can be retrieved as well, but its expression is longer and not particularly illuminating. As we saw in
section \ref{sec:anomalies_and_betafns}, the scheme-dependent constants $a_0'$
and $a_2'$ are related by \eqref{433a2a0reln}, which followed from the special conformal Ward identity \eqref{CWI433v3}.  In terms of the present calculation, the scheme-dependent constants in \eqref{free433cts1} and \eqref{free433cts2} are therefore related by
\begin{equation}
a_2^{(0)} = \frac{3 (21 - 8 \pi^2 a_0^{(0)})}{4096 \: \pi^6}.
\end{equation}

Notice that throughout the evaluation we have worked consistently with regulated quantities. The procedure presented above highlights the fact that the 3-point function $\< \O_{[4]} \O_{[3]} \O_{[3]} \>$ can be renormalised by adding the counterterms \eqref{e:Sct2free} and \eqref{e:Sct3free} to the regularised action. In particular, the sequence of integrals in \eqref{e:toint} is finite for a small non-zero $\epsilon$, and can in principle be evaluated in any order. While superficially different, the approach we present is however ultimately equivalent to the standard Feynman diagram calculus in which divergences are removed loop by loop.

\begin{figure}[t]
\includegraphics[width=0.80\linewidth]{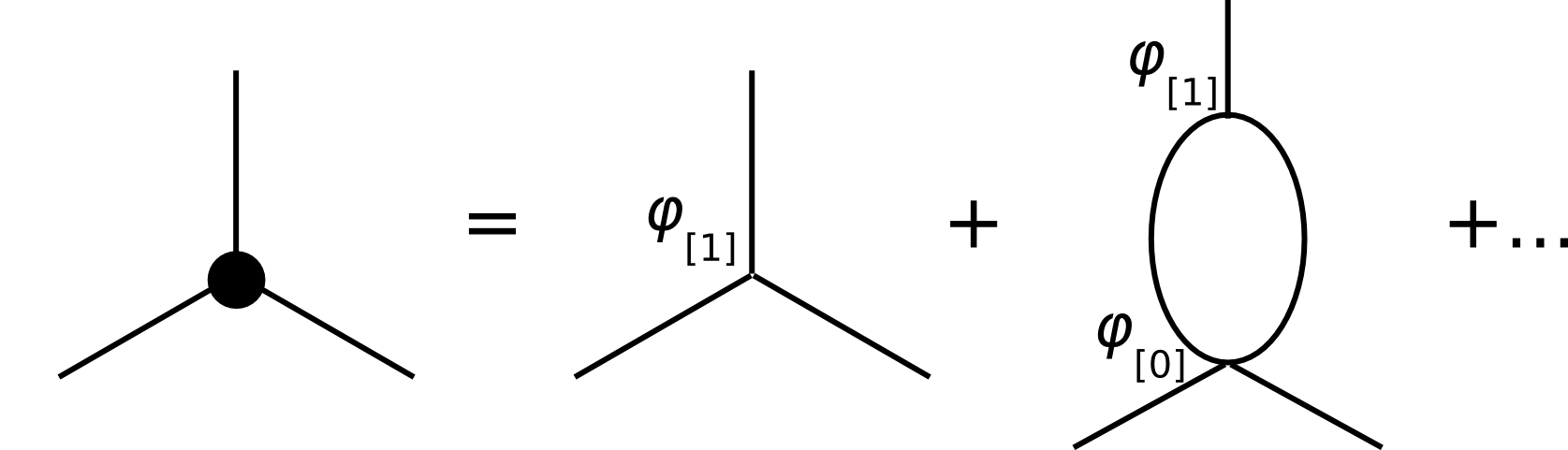}
\centering
\caption{The leading Feynman diagrams contributing to the renormalisation of the $\phi^3$ vertex.}
\label{fig:RenormPhi3}
\end{figure}

From the point of view of Feynman diagrams, the first term in the counterterm action \eqref{e:Sct3free} can be interpreted as the renormalisation of the cubic vertex $\phi^3$. Indeed, after adding to the free field the couplings to the operators $\phi^3$ and $\phi^4$,
the total action is 
\begin{equation}
S = \int \D^{4 - \epsilon} \bs{x} \left[ \frac{1}{2} (\partial \phi)^2 + \phi_{[1]} Z_1 \phi^3 + \phi_{[0]} Z_0 \phi^4 \right],
\end{equation}
where the renormalisation factors $Z_j$ depend on couplings $\phi_{[1]}$ and $\phi_{[0]}$. 
As one can read from \eqref{e:Sct3free},
\begin{equation}
Z_{[1]} = 1 + a_0 \phi_{[0]} \mu^{-\epsilon} + O(\phi_{[0]}^2), \qquad\qquad Z_{[0]} = 1 + O(\phi_{[0]}).
\end{equation}
The renormalisation of the cubic vertex can be then expressed diagrammatically as in figure \ref{fig:RenormPhi3}. The loop integral in the figure is divergent and requires renormalisation.
Evaluating this integral in the dimensionally regulated theory, we find
\begin{equation}
I_{\text{sub}}(\bs{q}) = \int \frac{\D^{4 - \epsilon} \bs{k}}{(2 \pi)^{4 - \epsilon}} \frac{1}{k^2 | \bs{k} - \bs{q} |^2} = C_{4 - \epsilon, 1, 1} q^{-\epsilon}.
\end{equation} 
The divergence as $\epsilon \rightarrow 0$ can be removed by adding an ultralocal counterterm and defining the finite integral
\begin{equation} \label{e:Feynman2ren}
I_{\text{sub}}^{\text{ren}}(\bs{q}) = I_{\text{sub}}(\bs{q}) + \left( - \frac{1}{8 \pi^2 \epsilon} + c_1 \right) \mu^{-\epsilon},
\end{equation} 
where $c_1$ is an arbitrary scheme-dependent constant. It is easy to check that the $\epsilon \rightarrow 0$ limit exists.

This renormalised cubic vertex can now be used in the evaluation of the full 3-point function in figure \ref{fig:Feynman1}. After the renormalisation of the nested divergences has been carried out according to \eqref{e:Feynman2ren}, one can use this expression in \eqref{e:i2}. The corresponding integral reads
\begin{equation}
I' = \int \frac{\D^{4 - \epsilon} \bs{k}_1}{(2 \pi)^{4 - \epsilon}} \frac{1}{k_1^2} I_{\text{sub}}^{\text{ren}}(\bs{k}_1 + \bs{p}_3) I_{\text{sub}}^{\text{ren}}(\bs{k}_1 - \bs{p}_2).
\end{equation}
This integral remains quadratically divergent, but its divergence is purely ultralocal. One can verify this claim by expanding $I_{\text{sub}}^{\text{ren}}$ and comparing with our previous result \eqref{e:O433Feyn}. The logarithmic terms of order $1/\epsilon$ cancel due to the subtraction of the nested divergence in \eqref{e:Feynman2ren} and the divergent part of the integral reads
\begin{align}
216 I' & = \frac{9}{512 \pi^6 \epsilon^2} (p_2^2 + p_3^2) - \frac{9}{2048 \pi^6 \epsilon} \left[ 4 p_1^2 + (1 + 96 c_1 \pi^2 + 6 \log \mu^2) (p_2^2 + p_3^2) \right] + O(\epsilon^0) \nn\\
& = \frac{9}{2048 \pi^6} (p_2^2 + p_3^2) \mu^{-3 \epsilon} \left[ \frac{4}{\epsilon^2} - \frac{1 + 96 c_1 \pi^2}{\epsilon} \right] - \frac{9}{512 \pi^6 \epsilon^2} \mu^{-3 \epsilon} p_1^2 + O(\epsilon^0).
\end{align}
The expression in the last line is ultralocal, with the renormalisation scale $\mu$ ensuring the appropriate dimension.

The conformal 3-point function represented by the Feynman diagram in figure \ref{fig:Feynman1} can thus be computed in the usual perturbative manner, by removing loop divergences at each step of the calculation with the aid of ultralocal counterterms.  
This renormalisability of Feynman diagrams is an important feature of perturbative QFT. 
In the present paper, however,
we achieve the renormalisation of a general CFT 3-point more directly by 
introducing counterterms for the triple-$K$ representation and showing that these counterterms remove all divergences.
At least as far as CFTs are concerned, our approach is the more general since not all CFTs are perturbative. In particular,  there are divergent 3-point functions that cannot be represented by a massless 3-point function of operators in a free field theory of spin-0 or spin-1/2.

\bigskip 
 
{\it\textbullet\, Example 15: $d = \Delta_1 = \Delta_2 = \Delta_3 = 3$.}

\bigskip

Let us now consider  a free scalar field $\Phi$ in $d = 3$ dimensions and evaluate the 3-point function of the operator $\O_{[3]} =\, \lwick \Phi^6 \rwick$.
We will dimensionally regulate in the same fashion as above, so that   
\[\label{e:regfree2}
\left[ \Phi \right] = \frac{1}{2} - \frac{\epsilon}{2},\qquad \left[ \O_{[3]} \right] = 3 - 3 \epsilon, \qquad
\left[ \phi_{[0]} \right] = 2 \epsilon, \qquad d = 3 - \epsilon,
\]
with $\phi_{[0]}$ the source for $\O_{[3]}$ (labelling according to the bare dimensions for brevity).

\begin{figure}[t]
\begin{subfigure}{0.4\textwidth}
\includegraphics[width=1\linewidth]{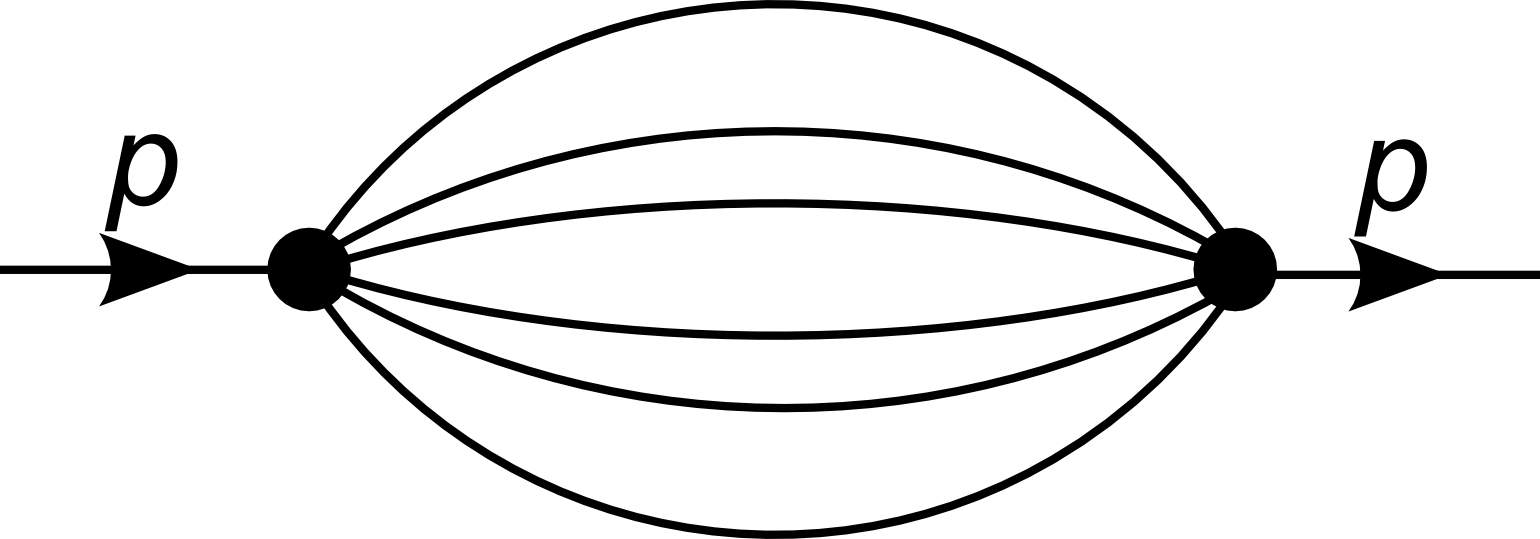}
\end{subfigure}
\hspace{1.5cm}
\begin{subfigure}{0.4\textwidth}
\includegraphics[width=1\textwidth]{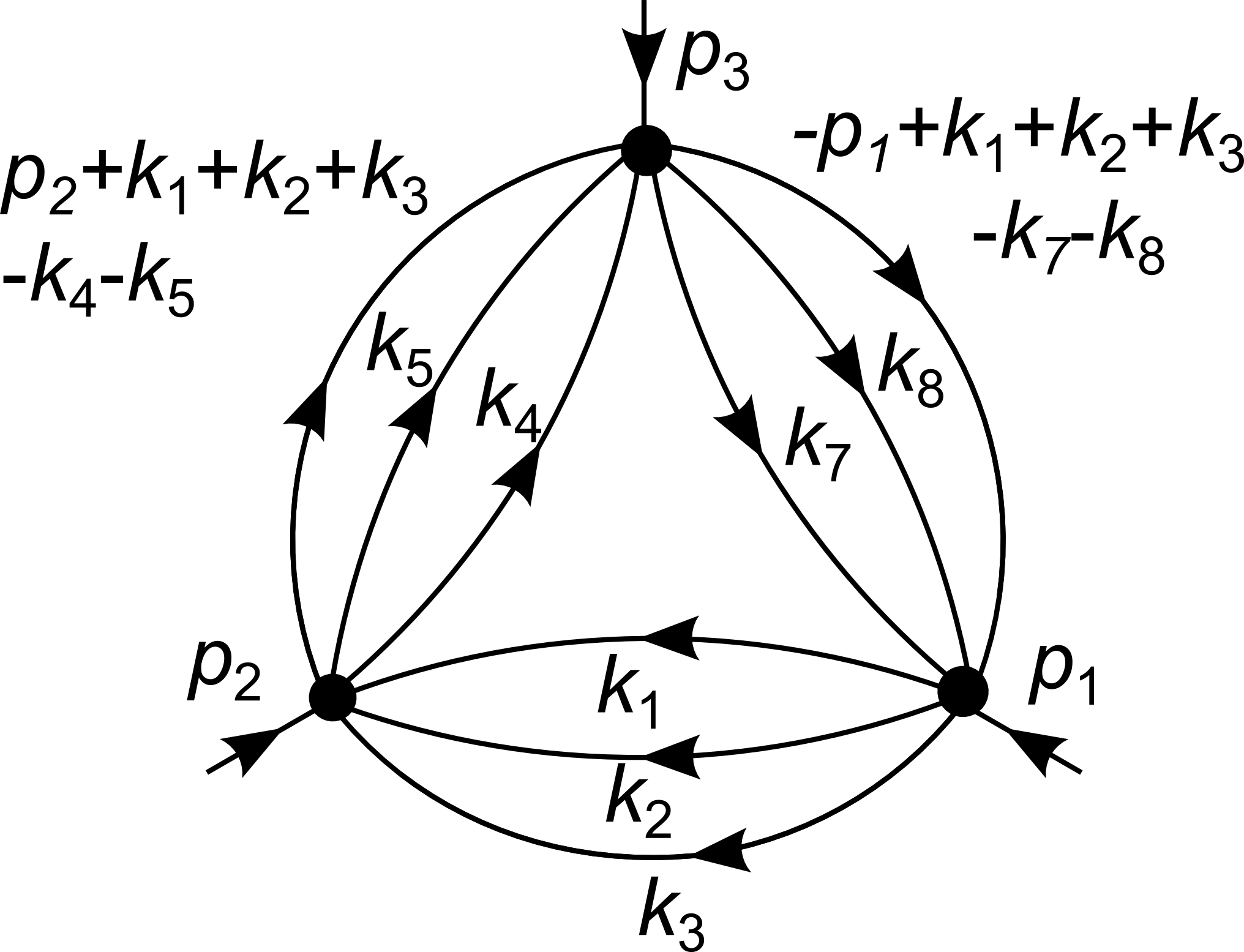}
\end{subfigure}
\centering
\caption{Feynman graphs representing  $\<\O_{[3]}\O_{[3]}\>$ and  $\< \O_{[3]} \O_{[3]} \O_{[3]} \>$  for a free scalar $\Phi$ with 
$\O_{[3]} = \,\lwick \Phi^6 \rwick$. }
\label{fig:Feynman2}
\end{figure}

 The corresponding Feynman diagrams are shown in figure \ref{fig:Feynman2}. 
For the 2-point function, 
\begin{align}
\lla \O_{[3]}(\bs{p}) \O_{[3]}(-\bs{p}) \rra_{\text{reg}} & = 6!\,  p^{3 - 5 \ep} \prod_{j=1}^5 C_{3-\ep, 1, 1 + \frac{1}{2}(j-1)(\epsilon - 1)}  = \frac{15}{1024\, \pi^4}\, p^3 + O(\epsilon),
\end{align} 
with $C_{d, a,b}$ as defined in \eqref{e:i2}. As this result is finite no counterterms are required.

The 3-point function is given by the integral
\begin{equation}
\lla \O_{[3]}(\bs{p}_1) \O_{[3]}(\bs{p}_2) \O_{[3]}(\bs{p}_3) \rra_{\text{reg}} = 1728 \, I(p_1, p_2, p_3),
\end{equation}
where
\begin{equation} \label{e:tointa}
I = \prod_{j=1}^9 \int \frac{\D^d \bs{k}_j}{(2 \pi)^d} \frac{1}{k_j^2} \delta(\bs{p}_2 + \bs{k}_1 + \bs{k}_2 + \bs{k}_3 - \bs{k}_4 - \bs{k}_5 - \bs{k}_6) \delta(- \bs{p}_1 - \bs{k}_1 - \bs{k}_2 - \bs{k}_3 + \bs{k}_7 + \bs{k}_8 + \bs{k}_9).
\end{equation}
A series of 1-loop integrals can then be done by means of the formula \eqref{e:i2} and the result recast as a triple-$K$ integral using equation (A.3.17) of \cite{Bzowski:2013sza}, 
\begin{align}
I &= C^3_{3 - \epsilon, 1, 1} C^3_{3 - \epsilon, 1, \frac{1}{2} (1 + \epsilon)} \, \int \frac{\D^{3 - \epsilon} \bs{k}}{(2 \pi)^{3 - \epsilon}} \frac{1}{k^{2 \epsilon}} |\bs{k} - \bs{p}_1|^{2\epsilon} |\bs{k} + \bs{p}_2|^{2\epsilon} \nn\\
&= \frac{2^{\frac{1}{2}(3 \epsilon - 1)} \pi^{\frac{1}{2}(\epsilon - 3)}}{\Gamma^3(\epsilon) \Gamma(3 - 4 \epsilon)}\, C^3_{3 - \epsilon, 1, 1} C^3_{3 - \epsilon, 1, \frac{1}{2} (1 + \epsilon)} \, I_{\frac{1}{2} - \frac{\epsilon}{2} \{ \frac{3}{2} - \frac{5 \epsilon}{2}, \frac{3}{2} - \frac{5 \epsilon}{2}, \frac{3}{2} - \frac{5 \epsilon}{2} \}}.
\end{align}
In this way we arrive at a representation of the 3-point function in terms of the triple-$K$ integral $I_{\frac{1}{2} \{ \frac{3}{2} \frac{3}{2} \frac{3}{2}\}}$ regulated in the scheme with $(u,v) = (-\frac{1}{2},-\frac{5}{2})$. 
This triple-$K$ integral can be evaluated by starting first in the regularisation scheme $v=0$ for which the Bessel-$K$ functions have half-integral indices and  reduce to elementary functions.  After evaluating the triple-$K$ integral in this scheme, we can then change  to the scheme above with $(u,v) = (-\frac{1}{2}, -\frac{5}{2})$ as described in section \ref{subsubsec:change_scheme}.
The regulated 3-point function thus reads 
\begin{align}
& \lla \O_{[3]}(\bs{p}_1) \O_{[3]}(\bs{p}_2) \O_{[3]}(\bs{p}_3) \rra_{\text{reg}} \nn\\[1ex]
&\qquad = \frac{9}{2^{13} \pi^6 \: \epsilon} (p_1^3 + p_2^3 + p_3^3) + \frac{9}{2^{12} \pi^6} \left[ - p_1 p_2 p_3 + (p_1^2 p_2 + 5 \text{ perms.}) \right.\nn\\
&\qquad\quad  - \: (p_1^3 + p_2^3 + p_3^3) \ln (p_1 + p_2 + p_3) - \frac{5}{2} ( p_1^3 \ln p_1 + p_2^3 \ln p_2 + p_3^3 \ln p_3) \nn\\
& \quad\qquad \left. + \: \frac{1}{12}(86 - 21 \gamma_E + 60 \ln 2 + 21 \ln \pi) (p_1^3 + p_2^3 + p_3^3) \right].
\end{align}

The counterterm action removing the divergence is
\begin{equation} \label{e:SctFeyn2}
S_{\rm{ct}} =  a(\ep)\int \D^{3 - \epsilon} \bs{x}\, \mu^{-2\epsilon}\phi_{[0]}^2 \O_{[3]} ,
\end{equation}
where
\begin{equation}
a = \frac{3}{80 \pi^2  \epsilon} + a^{(0)} + O(\epsilon), 
\end{equation}
with $a^{(0)}$ a scheme-dependent constant independent of $\ep$. The fully renormalised 3-point function then reads 
\begin{align}
\lla \O_{[3]}(\bs{p}_1) \O_{[3]}(\bs{p}_2) \O_{[3]}(\bs{p}_3) \rra &= \frac{9}{2^{12} \pi^6} \left[ - p_1 p_2 p_3 + (p_1^2 p_2 + 5 \text{ perms.}) \right.\nn\\
& \quad \left. - \: (p_1^3 + p_2^3 + p_3^3) \ln \frac{p_1 + p_2 + p_3}{\mu} + a_0' (p_1^3 + p_2^3 + p_3^3) \right],
\end{align}
where
\begin{equation}
a_0' = \frac{3}{2^{14} \pi^6} \left( 19 - 6 \gamma_E + 24 \ln 2 + 6 \ln \pi - 160 \pi^2 a^{(0)} \right).
\end{equation}
This expression matches our previous result \eqref{rencorr333} exactly upon setting 
\begin{equation}
c_{333} = \frac{27}{\sqrt{2^{21}\pi^{15}}}, \qquad b^{(0)} = -\frac{a'_0}{6}.
\end{equation}

As in the previous example, an alternative to the  renormalisation procedure we have just presented would be to proceed via the renormalisation of Feynman diagrams. The counterterm action \eqref{e:SctFeyn2} represents a first quantum correction to the vertex operator $\phi^6$ and  removes the nested subdivergences in diagram \ref{fig:Feynman2}. As previously, the remaining singularity of the diagram then becomes ultralocal. Our renormalisation procedure, however, is more general since it applies to any conformal field theory and does not require a Feynman diagram realisation of the 3-point function.

\section{Triple-$K$ integrals and AdS/CFT}\label{sec:AdS/CFT}

Triple-$K$ integrals appear naturally in the context of AdS/CFT since propagators in Poincar\'{e} coordinates, when transformed to momentum space, are expressible in terms of modified Bessel functions. A scalar 3-point function in the supergravity approximation arises from a cubic interaction term of the bulk action and is usually represented by a Witten diagram as per figure \ref{fig:Witten} (see page \pageref{fig:Witten}).  In this section we will discuss triple-$K$ integrals in a holographic context, and illustrate the holographic renormalisation procedure for the 3-point function of a marginal operator in three dimensions.  

\subsection{Set-up}\label{subsec:setup}

We consider a real scalar field $\Phi$ with a cubic interaction,
\begin{equation}\label{e:act} 
S = \int \D^{d+1} x \sqrt{g} \left[ \frac{1}{2} g^{\mu\nu} \partial_\mu \Phi \partial_\nu \Phi + \frac{1}{2} m^2 \Phi^2 - \frac{\lambda}{3} \Phi^3 \right],
\end{equation}
on a fixed Euclidean AdS background in Poincar\'{e} coordinates,
\begin{equation}
\D s^2 = \frac{1}{z^2} \left[ \D z^2 + \D \bs{x}^2 \right].
\end{equation}
As usual, the mass of the field is parametrised as $m^2 = \Delta ( \Delta - d)$, where $\Delta$ denotes the conformal dimension of the dual CFT operator. Throughout this section we will assume $\Delta > d/2$.  (For cases where $d/2-1\le \Delta < d/2$ see \cite{Klebanov:1999tb} and appendix \ref{sec:shadow}.)

The equation of motion $-\Box_g \Phi + m^2 \Phi = \lambda \Phi^2$ can be solved perturbatively in $\lambda$. For 2- and 3-point functions we only need the first two terms, $\Phi = \Phi_{\{0\}} + \lambda \Phi_{\{1\}} + O(\lambda^2)$, which satisfy
\begin{equation} \label{e:AdSeqns}
( - \Box_g + m^2 ) \Phi_{\{0\}} = 0, \qquad\qquad ( - \Box_g + m^2 ) \Phi_{\{1\}} = \Phi_{\{0\}}^2.
\end{equation}

For the CFT analysis in momentum space we Fourier transform along all directions parallel to the conformal boundary at $z=0$.  Writing the Fourier transform of $\Phi(z, \bs{x})$ as $\Phi(z, \bs{p})$, 
the free field equation \eqref{e:AdSeqns} becomes $L_{d,\Delta}(z, p) \Phi_{\{0\}}(z, \bs{p}) = 0$,
where  
\begin{equation}
L_{d,\Delta}(z, p) = - z^2 \partial_z^2 + (d - 1) z \partial_z + m^2 + z^2 p^2.
\end{equation}
This equation can be solved in terms of modified Bessel functions. 

The equations of motion for $\Phi_{\{n\}}(z,\bs{p})$ with $n>1$ can then be solved in terms of the bulk-to-boundary and bulk-to-bulk propagators. These are uniquely fixed by asymptotic boundary conditions at $z = 0$, together with regularity requirements at $z = \infty$. 

The bulk-to-boundary propagator $\mathcal{K}_{d, \Delta}$ is defined by
\begin{equation}
\left\{ \begin{array}{l}
L_{d, \Delta}(z, p) \mathcal{K}_{d, \Delta}(z, p) = 0, \\[2ex] 
\lim_{z \rightarrow 0} [z^{-(d - \Delta)} \mathcal{K}_{d, \Delta}(z, p)] = 1, \\[2ex]
\mathcal{K}_{d, \Delta}(\infty, p) = 0. \end{array} \right.
\end{equation}
while the bulk-to-bulk propagator $\mathcal{G}_{d, \Delta}$ solves
\begin{equation}
\left\{ \begin{array}{l}
L_{d, \Delta}(z, p) \mathcal{G}_{d, \Delta}(z, p; \zeta) = \zeta^4 \delta(z - \zeta), \\[2ex]
\lim_{z \rightarrow 0} [z^{-(d - \Delta)} \mathcal{G}_{d, \Delta}(z, p; \zeta)] = 0, \\[2ex] 
\mathcal{G}_{d, \Delta}(\infty, p; \zeta) = 0.
\end{array} \right.
\end{equation}
The unique solutions to these equations are
\begin{equation} \label{e:bulktobnd}
\mathcal{K}_{d, \Delta}(z, p) = \frac{2^{\frac{d}{2} - \Delta+1}}{\Gamma \left( \Delta - \frac{d}{2} \right)} p^{\Delta - \frac{d}{2}} z^{\frac{d}{2}} K_{\Delta - \frac{d}{2}}(p z)
\end{equation}
for the bulk-to-boundary propagator and
\begin{equation} \label{e:bulktobulk}
\mathcal{G}_{d, \Delta}(z, p; \zeta) = \left\{ \begin{array}{ll}
(z \zeta)^{d/2} I_{\Delta - \frac{d}{2}}(p z) K_{\Delta - \frac{d}{2}}(p \zeta) & \text{for } z \leq \zeta, \\
(z \zeta)^{d/2} I_{\Delta - \frac{d}{2}}(p \zeta) K_{\Delta - \frac{d}{2}}(p z) & \text{for } z > \zeta, \end{array} \right.
\end{equation}
for the bulk-to-bulk propagator. The solution to the equations of motion \eqref{e:AdSeqns}, with the boundary value of $\Phi_{(0)}$ set to $\phi_{0}$, are then
\begin{align}
\Phi_{\{0\}}(z, \bs{p}) & = \mathcal{K}_{d, \Delta}(z, p) \phi_0(\bs{p}), \label{e:Phi0} \\
\Phi_{\{1\}}(z, \bs{p}) & = \int_0^\infty \frac{\D \zeta}{\zeta^{d+1}} \mathcal{G}_{d, \Delta}(z, p; \zeta) \int \frac{\D^d \bs{k}}{(2 \pi)^d} \mathcal{K}_{d, \Delta}(\zeta, k) \mathcal{K}_{d, \Delta}(\zeta, |\bs{p} - \bs{k}|) \phi_{(0)}(\bs{k}) \phi_{0}(\bs{p} - \bs{k}) \nonumber \\
& = \int \frac{\D^d \bs{k}}{(2 \pi)^d} \phi_{0}(\bs{k}) \phi_{0}(\bs{p} - \bs{k}) \int_0^\infty \frac{\D \zeta}{\zeta^{d+1}} \mathcal{G}_{d, \Delta}(z, p; \zeta) \mathcal{K}_{d, \Delta}(\zeta, k) \mathcal{K}_{d, \Delta}(\zeta, | \bs{p} - \bs{k} |) \label{e:Phi1}
\end{align}
provided the integral converges. A diagrammatic representation of solution \eqref{e:Phi1} is presented in figure \ref{fig:Witten}. 

\subsection{3-point functions}\label{subsec:hol3pt}

\begin{figure}[t]
	\centering
	\begin{subfigure}[ht]{0.315\textwidth}
		\centering
		\includegraphics[width=1.00\textwidth]{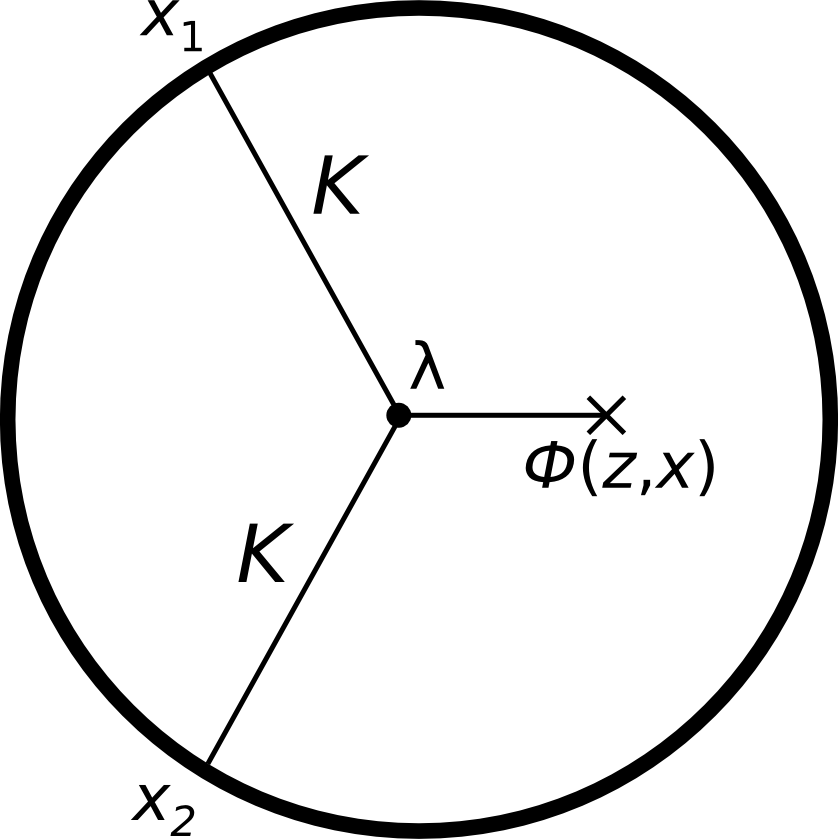}\vspace{2mm}					
	\end{subfigure}
	\qquad\qquad\qquad
	\begin{subfigure}[ht]{0.345\textwidth}
		\centering
		\includegraphics[width=1.00\textwidth]{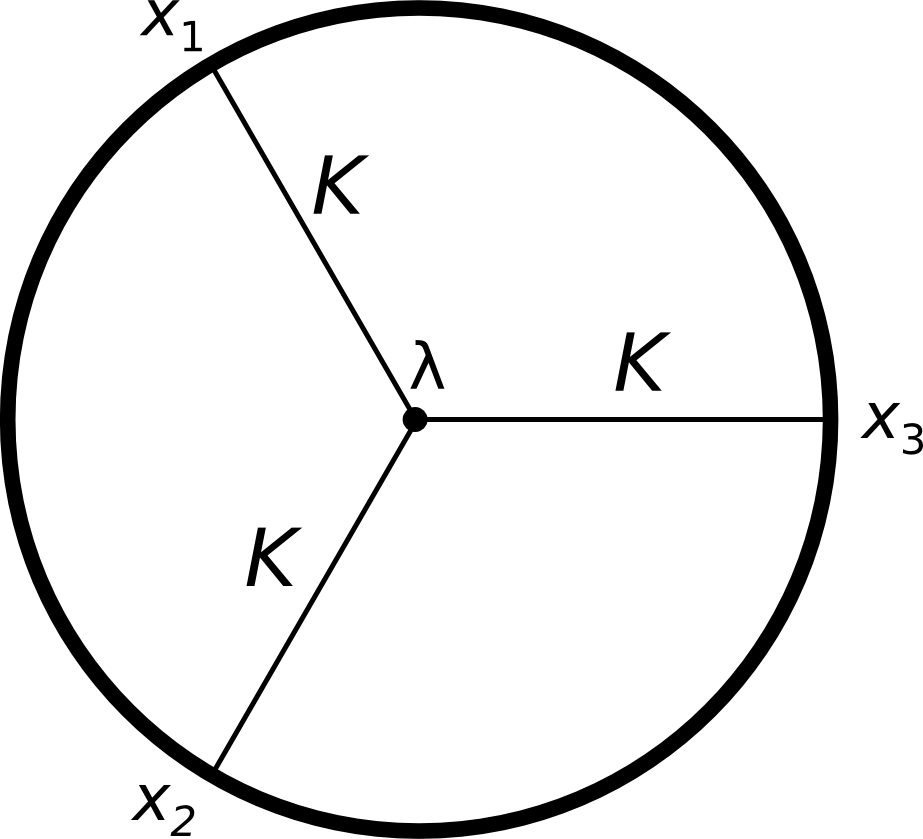}\vspace{2mm}
	\end{subfigure}
	\caption{
	(a) Witten diagram for the evaluation of the scalar field $\Phi$ at a point $x$ in the bulk; (b) taking $x$ to a point $x_3$ on the boundary, the diagram now represents a 3-point function.
	}\label{fig:Witten}
\end{figure}

The 1-point function in the presence of sources for the operator $\O$ dual to the bulk scalar $\Phi$ reads \cite{deHaro:2000xn, Skenderis:2002wp}
\begin{equation} \label{e:1pt}
\< \O \>_s = - (2 \Delta - d) \Phi_{(\Delta)} + X[\phi_0],
\end{equation}
where $\Phi_{(\Delta)}$ denotes the coefficient of $z^{\Delta}$ in the near-boundary expansion of $\Phi$, and $X[\phi_0]$ is a functional whose contribution to correlation functions is at most local.

In order to extract the 3-point function, we need to identify the piece of $\Phi$ which depends quadratically on the source $\phi_0$. This piece is given by \eqref{e:Phi1}, after evaluating the integral on the right-hand side.  When this integral diverges, we can  introduce a cut-off at $z=\delta$, 
\begin{equation} \label{e:iInt}
\mathcal{I}^{\delta}_{d, \Delta}(z, \bs{p}, \bs{k}) = \int_\delta^\infty \frac{\D \zeta}{\zeta^{d+1}} \mathcal{G}_{d, \Delta}(z, p; \zeta) \mathcal{K}_{d, \Delta}(\zeta, k) \mathcal{K}_{d, \Delta}(\zeta, | \bs{p} - \bs{k} | ).
\end{equation}
The 3-point function of $\O$ then follows from this integral, with any divergences that may be present removed by holographic renormalisation of the supergravity on-shell action. (A complete example of this procedure for a marginal operator will be presented shortly in section \ref{secD4}; for a related discussion of holographic renormalisation for irrelevant operators see \cite{vanRees:2011fr,vanRees:2011ir}.)
From \eqref{e:1pt}, we  have
\begin{align} \label{e:AdS3pt}
\lla \O(\bs{p}_1) \O(\bs{p}_2) \O(\bs{p}_3) \rra & = - \lambda (2 \Delta - d) \lim_{\delta \rightarrow 0} \left[ \mathcal{I}^{\delta}_{d, \Delta}(z, \bs{p}_1, \bs{p}_2) + \mathcal{I}^{\delta}_{d, \Delta}(z, \bs{p}_1, \bs{p}_3) + \mathcal{I}^{\delta}_{\text{ct}}(z) \right]_{(\Delta)}  \nn\\
& \qquad\qquad + \left. \frac{\delta^2 X(\bs{p}_1)}{\delta \phi_0(-\bs{p}_2) \delta \phi_0(-\bs{p}_3)} \right|_{\phi_0 = 0},
\end{align}
where $\mathcal{I}^{\delta}_{\text{ct}}$ is a suitable counterterm and $(\ldots)_{(\Delta)}$ denotes the coefficient of $z^{\Delta}$ in the near-boundary expansion. As we will 
now show, the first part of \eqref{e:AdS3pt} can be re-written as a triple-$K$ integral.

Firstly, the piecewise form of the bulk-to-bulk propagator in  \eqref{e:bulktobulk} splits the integral \eqref{e:iInt} into two regions: a near-boundary region $\zeta \leq z$ and an inner region $\zeta > z$. Denoting the corresponding integrals as $\mathcal{I}^{\delta,<}_{d, \Delta}$  and  $\mathcal{I}^{>}_{d, \Delta}$, we then have
\begin{equation}
\mathcal{I}^{\delta}_{d, \Delta} = \mathcal{I}^{\delta,<}_{d, \Delta} +  \mathcal{I}^>_{d, \Delta},
\end{equation}
where only the near-boundary integral depends on the regulator $\delta$.

In the near-boundary region $\zeta \leq z$, the integral reads
\begin{equation} \label{e:iBnd}
 \mathcal{I}^{\delta,<}_{d, \Delta} = z^{\frac{d}{2}} K_{\Delta - \frac{d}{2}}(p z) \int_{\delta}^{z} \D \zeta \: \zeta^{-\frac{d}{2} - 1} I_{\Delta - \frac{d}{2}}(\zeta p) \mathcal{K}_{d, \Delta}(\zeta, k) \mathcal{K}_{d, \Delta}(\zeta, | \bs{p} - \bs{k} |).
\end{equation}
As we have discussed, the integral will diverge as $\delta \rightarrow 0$ and these divergences can be removed by holographic counterterms.  
To compute the 3-point function we now only need to extract the coefficient of $z^{\Delta}$.
By power expanding the integrand, one finds that an appropriate term exists only if an independent choice of signs can be found such that
\begin{equation} \label{e:AdScond}
\frac{d}{2} \pm \beta \pm \beta \pm \beta = -2 k, \qquad \beta = \Delta - \frac{d}{2},
\end{equation}
where $k$ is a non-negative integer. This is exactly our fundamental condition \eqref{e:condition} for all $\beta_j = \beta$: when satisfied, the near-boundary integral produces a contribution to the 3-point function. While such a contribution is local 
(see the discussion in section \ref{subsec:non-uniqueness}),
it is crucial in order for the 3-point function to have the correct symmetry properties, as we will see in the next section.

Consider now a contribution to the 3-point function from the inner region $\zeta > z$. Since the expansion of the bulk-to-bulk propagator reads
\begin{equation}
\mathcal{G}_{d, \Delta}(z, p; \zeta) = \frac{z^{\Delta}}{2 \Delta - d}\, \mathcal{K}_{d, \Delta}(\zeta, p) + O(z^{\Delta + 2}) \text{ for } \zeta > z,
\end{equation}
the integral gives
\begin{equation} \label{e:iInter}
\mathcal{I}^>_{d, \Delta} = \frac{z^{\Delta}}{2 \Delta - d} \int_{z}^{\infty} \D \zeta \: \zeta^{-\frac{d}{2} - 1} \mathcal{K}_{d, \Delta}(\zeta, p) \mathcal{K}_{d, \Delta}(\zeta, k) \mathcal{K}_{d, \Delta}(\zeta, | \bs{p} - \bs{k} |) + O(z^{\Delta + 2}).
\end{equation}
When the expression \eqref{e:bulktobnd} for the bulk-to-boundary propagator is substituted, this integral is proportional to a triple-$K$ integral with a lower cut-off. 
To extract the coefficient of $z^{\Delta}$ for the complete right-hand side, we simply have to strip off the overall prefactor of $z^\Delta$ then evaluate the $z$-independent piece of the integral.  To find this $z$-independent piece, it is tempting to send $z\rightarrow 0$ leaving us with a genuine triple-$K$ integral.   We  know, however, that when \eqref{e:AdScond} is satisfied this triple-$K$ integral diverges, since \eqref{e:AdScond} is equivalent to the singularity condition 
\eqref{e:condition} with all $\beta_j = \beta$.

Thus, provided the condition \eqref{e:AdScond} is \emph{not} satisfied, the contribution to the 3-point function from the near-boundary part of the integral \eqref{e:iBnd} vanishes, while the contribution from the inner region \eqref{e:iInter} reduces to a finite triple-$K$ integral upon sending $z\rightarrow 0$. 
(Diagrammatically, we can think of this as moving the internal point in the Witten diagram to the boundary as shown in figure \ref{fig:Witten}.)  As the local functional $X[\phi_0]$ in \eqref{e:AdS3pt} moreover vanishes, the complete correlation function is then given by this triple-$K$ integral,
\begin{equation} \label{e:gen3pt}
\lla \O(\bs{p}_1) \O(\bs{p}_2) \O(\bs{p}_3) \rra = - 2 \lambda \left( \frac{2^{\frac{d}{2} - \Delta + 1}}{\Gamma \left( \Delta - \frac{d}{2} \right)} \right)^3 I_{\frac{d}{2} - 1 \{\Delta - \frac{d}{2}, \Delta - \frac{d}{2}, \Delta - \frac{d}{2} \}}(p_1, p_2, p_3),
\end{equation}
as follows by expanding the propagators in \eqref{e:iInter}.  This triple-$K$ integral is finite, although in some cases it may be necessary to use analytic continuation to define its precise value (as in example 3 on page \pageref{Example3}).
 
If, on the other hand, the condition \eqref{e:AdScond} holds, one still expects to obtain the non-local part of the correlation function from the inner region in \eqref{e:iInter}.  This non-local contribution corresponds to the finite order $z^0$ piece of the integral as $z\rightarrow 0$, and so is equivalent to a triple-$K$ integral up to local terms.  (The overall correlator therefore receives local contributions from both \eqref{e:iBnd} and \eqref{e:iInter}.) We will illustrate this case with an example in the following section.

Notice however that the procedure of holographic renormalisation is \emph{not} equivalent to shifting the $\alpha$ and $\beta$ parameters in the triple-$K$ integral in \eqref{e:gen3pt}.  Instead, holographic regularisation amounts to the introduction of a cut-off on the integration variable in the triple-$K$ integral; 
in the complete holographic renormalisation scheme, one then has to include additional local contributions from \eqref{e:iBnd} and the functional $X[\phi_0]$ in \eqref{e:AdS3pt}.

\subsection{Marginal operator in $d = 3$}\label{subsec:marginal}
\label{secD4}

To illustrate the general discussion above, we now discuss the complete holographic renormalisation of the 3-point function for a marginal operator in $d=3$ dimensions. This case satisfies the condition \eqref{e:AdScond} with a single plus sign and $k = 0$.

The near-boundary expansion for the solution to the equations of motion reads
\begin{align}
\Phi & = ( \phi_{\{0\}(0)} + \lambda \phi_{\{1\}(0)}) + z^2 ( \phi_{\{0\}(2)} + \lambda \phi_{\{1\}(2)}) + z^3 ( \phi_{\{0\}(3)} + \lambda \phi_{\{1\}(3)} ) \nn\\
& \qquad + 2 \lambda \ln z \left[ \psi_{(0)} + z^2 \psi_{(2)} + z^3 \psi_{(3)} \right] + O(z^4, \lambda^2), \label{e:expPhi0}
\end{align}
where we have labelled the $\phi$ coefficients in this expansion with round brackets to indicate the power of the radial variable $z$ and curly brackets to denote the power of the coupling constant $\lambda$.  
(As we will not need such an expansion for the $\psi$ coefficients, however, we will omit the curly bracket label for these variables.)

The boundary source is then $\phi_0 = \phi_{(0)} = \phi_{\{0\}(0)} + \lambda \phi_{\{1\}(0)} + O(\lambda^2)$, although to begin with we will switch off all the subleading coefficients by setting $\phi_{\{n\}(0)}=0$ for $n>0$.  (Later, we will see however that these subleading contributions must be reintroduced, and so we will retain them explicitly in the following.)
The `vev' coefficient $\phi_{(3)}=\phi_{\{0\}(3)} + \lambda \phi_{\{1\}(3)} + O(\lambda^2)$, with the equations of motion implying that $\phi_{\{0\}(3)}$ is linearly dependent on the source $\phi_0$, while $\phi_{\{1\}(3)}$ has a quadratic dependence, {\it etc}. 
All the remaining coefficients can be expressed locally in terms of $\phi_{(3)}$ and $\phi_0$, {\it e.g.,}
\begin{align}
\psi_{(0)} & = \frac{1}{6} \phi_{\{0\}(0)}^2, \nn \\
\phi_{\{0\}(2)} & = \frac{1}{2} \partial^2 \phi_{\{0\}(0)}, \nn\\
\phi_{\{1\}(2)} & = \frac{1}{2} \partial^2 \psi_{(0)} + \frac{1}{2} \phi_{\{0\}(0)} \partial^2 \phi_{\{0\}(0)} + \frac{1}{2} \partial^2 \phi_{\{1\}(0)}, \nn\\
\psi_{(2)} & = \frac{1}{2} \partial^2 \psi_{(0)}, \nn\\
\psi_{(3)} & = - \frac{1}{3} \phi_{\{0\}(0)} \phi_{\{0\}(3)}. \label{e:psi3adscft}
\end{align}

To regulate the action \eqref{e:act} we impose a cutoff $z \geq \delta$. The divergent part of the regulated action is then
\begin{align}
S_{\text{div}} & = \frac{\lambda}{6} \int_{z \geq \delta} \D z \D^3 \bs{x} \sqrt{g} \Phi^3 - \frac{1}{2} \int_{z = \delta} \D^3 \bs{x} \sqrt{g} g^{zz} \Phi \partial_z \Phi, \nn\\
& = - \int_{z = \delta} \D^3 \bs{x} \sqrt{\gamma_z} \left[ \frac{1}{2} ( \phi_{\{0\}(0)} + \lambda \phi_{\{1\}(0)}) \partial^2 ( \phi_{\{0\}(0)} + \lambda \phi_{\{1\}(0)}) + \right.\nn\\
& \qquad\qquad \left. + \: \lambda \left( \frac{1}{9}  \phi_{\{0\}(0)}^3 + \frac{1}{2} z^2  \phi_{\{0\}(0)}^2 \partial^2  \phi_{\{0\}(0)} + \frac{1}{3} z^2 \ln z \:  \phi_{\{0\}(0)}^2 \partial^2  \phi_{\{0\}(0)} \right) \right], \label{e:Sreg}
\end{align}
where $\gamma_z$ is the induced metric on a slice of constant $z$, \textit{i.e.}, $(\gamma_z)_{ij} = z^{-2} \delta_{ij}$. It is easy to check that these divergent terms can be repackaged into a local functional of the bulk field, allowing us to write the following counterterms,
\begin{equation} \label{e:Sct}
S_{\text{ct}} = \int_{z = \delta} \D^3 \bs{x} \sqrt{\gamma_z} \left[ \frac{1}{2} \Phi \Box_z \Phi + \lambda \left( \frac{1}{9} \Phi^3 + \frac{1}{3} \Phi^2 \Box_z \Phi \right) \right],
\end{equation}
where $\Box_z$ is the Laplacian for the metric $(\gamma_z)_{ij}$ on the slice of constant $z$. When these counterterms are added to the regulated action, the variation of $S_{\text{sub}} = S_{\text{reg}} + S_{\text{ct}}$ gives
\begin{equation}
\frac{\delta S_{\text{sub}}}{\delta \Phi} = - 3 \left( \phi_{\{0\}(3)} + \lambda \phi_{\{1\}(3)} \right) + \lambda \left( \frac{4}{3} \phi_{\{0\}(0)} \phi_{\{0\}(3)} + 2 \phi_{\{0\}(0)} \phi_{\{0\}(3)} \ln \delta \right).
\end{equation}
The logarithmically divergent piece cancels against the functional derivative of the bulk field with respect to the source when we compute the 1-point function,
\begin{align}
\langle \mathcal{O} \rangle_s
& = \frac{1}{\sqrt{g_{(0)ij}}} \frac{\delta S_{\text{ren}}}{\delta \phi_{0}} = \lim_{\delta \rightarrow 0} \frac{1}{\sqrt{\gamma_\delta}} \int \D^3 \bs{x} \sqrt{\gamma_\delta} \frac{\delta \Phi}{\delta \phi_{0}} \frac{\delta S_{\text{sub}}}{\delta \Phi} \nonumber \\
& = - 3 \left( \phi_{\{0\}(3)} + \lambda \phi_{\{1\}(3)} \right) + \frac{4}{3} \lambda \phi_{\{0\}(0)} \phi_{\{0\}(3)}, \label{e:Os}
\end{align}
leading, as expected, to \eqref{e:1pt} with a specific non-vanishing $X[\phi_0]$.

By taking a single derivative of the above expression with respect to the source $\phi_0 = \phi_{\{0\}(0)}$ we obtain the holographic 2-point function
\begin{equation}
\lla \mathcal{O}(\bs{p}) \mathcal{O}(-\bs{p}) \rra = 3 \left[ \mathcal{K}_{3,3}(z, p) \right]_{(3)} = p^3,
\end{equation}
where $\left[ \mathcal{K}_{3,3}(z, p) \right]_{(3)}$ denotes the coefficient of $z^3$ in the near-boundary expansion of the bulk-to-boundary propagator, as follows from \eqref{e:expPhi0}.

We are now in position to evaluate the 3-point function as given in \eqref{e:AdS3pt}.  All propagators are elementary functions (\textit{e.g.}, $\mathcal{K}_{3,3}(z, p) = e^{-z p} (1 + z p)$) allowing exact computations to be performed.
Evaluating the triple-$K$ integral with a cut-off in \eqref{e:iInt}, we find
\begin{align} \label{e:AdSdiv}
\mathcal{I}^{\delta}_{d, \Delta}(z, p_j) & = \frac{1}{9} \left[ 1 + 3 \ln \left( \frac{z}{\delta} \right) \right] \mathcal{K}_{3,3}(z, p_1) - \frac{1}{12} z^2 \left[ p_1^2 + 3(p_2^2 + p_3^2) \right] + \: \nn\\
& \qquad - \: \frac{1}{9} z^3 \left[ p_1 p_2 p_3 - ( p_1^2 p_2 + 5 \text{ perms.} ) + (p_1^3 + p_2^3 + p_3^3) \ln \left( (p_1 + p_2 + p_3) z \right) + \right.\nn\\
& \qquad\qquad\qquad\qquad \left. + \: (\gamma_E - 1) (p_1^3 + p_2^3 + p_3^3) - \tfrac{2}{3} (p_2^3 + p_3^3) \right].
\end{align}
Naively this integral, and hence the 3-point function, is divergent as $\delta\rightarrow 0$.  However, via \eqref{e:Phi1}, this divergence in \eqref{e:iInt} leads to a corresponding divergent contribution to $\Phi_{\{1\}}$, and one can check that this divergent contribution satisfies the homogeneous free field equations.  It can therefore be cancelled by turning on a subleading order $\lambda$ contribution to the source, namely $\phi_{\{1\}(0)}$, since this also obeys the homogeneous free field equations and contributes to $\Phi_{\{1\}}$.
For consistency, we should then regard the full expansion $\phi_{(0)} = \phi_{\{0\}(0)} + \lambda \phi_{\{1\}(0)} + O(\lambda^2)$ as the source $\phi_0$ for the dual operator $\O$, rather than just the leading piece $\phi_{\{0\}(0)}$ as earlier when $\phi_{\{1\}(0)}$ was switched off.

To cancel the divergence in this fashion requires setting 
\begin{equation} \label{e:redef}
\phi_0 = \phi_{(0)} = \phi_{\{0\}(0)} - \frac{1}{9} \lambda \left[ 1 - 3 \ln \left( \delta \mu \right) \right] \phi_{\{0\}(0)}^2 + O(\phi_{\{0\}(0)}^3),
\end{equation}
where $\mu$ is a renormalisation scale introduced on dimensional grounds. Equation \eqref{e:Os} does not change to this order in $\lambda$, but $\phi_{\{1\}(3)}$ -- and hence \eqref{e:AdS3pt} -- receives an additional contribution cancelling the divergence. The final holographic 3-point function then reads
\begin{align} \label{e:AdSfinal}
 \lla \O(\bs{p}_1) \O(\bs{p}_2) \O(\bs{p}_3) \rra &= -\frac{2}{3} \lambda \,\Big[- p_1 p_2 p_3 + ( p_1^2 p_2 + 5 \text{ perms.} )  \nn\\
& \quad  - \: (p_1^3 + p_2^3 + p_3^3) \ln \left( \frac{p_1 + p_2 + p_3}{\mu} \right) + (1-\gamma_E) (p_1^3 + p_2^3 + p_3^3) \Big].
\end{align}
Note that the local functional $X[\phi_0]$ in \eqref{e:AdS3pt} makes a contribution of $4(p_2^3 + p_3^3)/9$ to this expression: this contribution is crucial for the final 3-point function to be symmetric under any permutation of momenta.

From the perspective of the dual CFT, the redefinition of the source \eqref{e:redef} introduces a beta function. Identifying $\phi_{\{0\}(0)}$ as the bare source (independent of $\mu$) and $\phi_0 = \phi_{(0)}$ in \eqref{e:redef} as the renormalised source, then
\begin{equation} \label{e:AdSbeta}
\beta_{\phi_0} = \mu\frac{\partial \phi_{0}}{\partial \mu} = \frac{1}{3} \lambda \phi_{0}^2 + O(\phi_{0}^3).
\end{equation}
These results are in complete agreement with our earlier discussion in example 8 on page \pageref{example8}.
The form of the 3-point function in \eqref{e:AdSfinal} agrees with \eqref{rencorr333} on setting the theory-dependent normalisation constant to
\begin{equation}
c_{333} = - 2 \lambda \left( \frac{\pi}{2} \right)^{-3/2}
\end{equation}
and the scheme-dependent constant $a^{(0)} = (\gamma_E-1)/6$.  Moreover, with the beta function as in \eqref{e:AdSbeta}, the Callan-Symanzik equation \eqref{CSeqn333} is satisfied.  Further discussion of the Callan-Symanzik equation in a holographic context may be found in \cite{vanRees:2011ir}.

\bibliographystyle{jhep}
\bibliography{cwis2015}

\end{document}